\documentclass[fleqn,usenatbib]{mnras}

\usepackage[T1]{fontenc}

\DeclareRobustCommand{\VAN}[3]{#2}
\let\VANthebibliography\thebibliography
\def\thebibliography{\DeclareRobustCommand{\VAN}[3]{##3}\VANthebibliography}

\usepackage{graphicx}	
\usepackage{amsmath}	
\usepackage{amssymb}	

\usepackage{newtxtext,newtxmath}

\newcommand{\mdot}{$\mathrm{M}_\odot$}
\newcommand{\zdot}{$\mathrm{Z}_\odot$}
\newcommand{\healpix}{\textsc{healpix}}
\newcommand{\treecol}{\textsc{treecol}}

\title[Rapid self-enrichment of star clusters]{Star clusters forming in a low metallicity starburst -- rapid self-enrichment by (very) massive stars}

\author[Lah\'en et al.]{
Natalia Lah\'en$^{1}$\thanks{E-mail: nlahen@mpa-garching.mpg.de},
Thorsten Naab$^{1}$ and
Dorottya Sz\'ecsi$^{2}$
\\
$^{1}$Max-Planck-Institute f\"{u}r Astrophysik, Karl-Schwarzschild-Stra{\ss}e 1, D-85740 Garching, Germany\\
$^{2}$Institute of Astronomy – Faculty of Physics, Astronomy and Informatics – Nicolaus Copernicus University, Grudzi\k{a}dzka 5, 87-100 Toru\'n, Poland\\
}

\date{Accepted XXX. Received YYY; in original form ZZZ}

\pubyear{2024}

\begin{document}
\label{firstpage}
\pagerange{\pageref{firstpage}--\pageref{lastpage}}
\maketitle

\begin{abstract}

Stellar winds of massive ($\gtrsim9\,\mathrm{M_\odot}$) and very massive ($\gtrsim100\,\mathrm{M_\odot}$) stars may play an important role in the metal-enrichment during the formation of star clusters. With novel high-resolution hydrodynamical \textsc{griffin}-project simulations, we investigate the rapid recycling of stellar wind-material during the formation of massive star clusters up to $M_\mathrm{cluster}\sim2\times10^5\,\mathrm{M_\odot}$ in a low-metallicity dwarf galaxy starburst. The simulation realises new stars from a stellar initial mass function (IMF) between $0.08\,\mathrm{M_\odot}$ and $\sim400\,\mathrm{M_\odot}$ and follows stellar winds, radiation and supernova-feedback of single massive stars with evolution tracks. Star clusters form on timescales less than \mbox{$\sim5$ Myr}, and their supernova-material is very inefficiently recycled. Stellar wind-material, however, is trapped in massive clusters resulting in the formation of stars self-enriched in Na, Al, and N within only a few Myr. Wind-enriched (second population) stars can be centrally concentrated in the most massive clusters ($\gtrsim10^4\,\mathrm{M_\odot}$) and the locked wind-material increases approximately as $M_\mathrm{cluster}^{2}$. These trends resemble the characteristics of observed second population stars in globular clusters. We fit scaling relations to the log-normal distributed wind-mass fractions and extrapolate to possible globular cluster progenitors of \mbox{$M_\mathrm{cluster}=10^7\,\mathrm{M_\odot}$} to investigate whether a dominant second population could form. This can only happen if the IMF is well sampled, single massive stars produce at least a factor of a few more enriched winds e.g. through a top-heavy IMF, and a significant fraction of the first population (unenriched) stars is lost during cluster evolution. 

\end{abstract}




\section{Introduction}

The majority of present-day globular clusters (GCs) have been observed to host multiple populations (MPs) of stars, identified through variations in the stellar light-element composition (\citealt{2001A&A...369...87G, 2009A&A...505..139C, 2015AJ....149...91P, 2017A&A...601A.112P, 2019MNRAS.487.3815M}; see e.g. \citealt{2015MNRAS.454.4197R, 2016EAS....80..177C, 2018ARA&A..56...83B, 2019A&ARv..27....8G} for comprehensive reviews). The stars in GCs can be divided into normal, field-like stars called the first population (1P) and to anomalous, light-element peculiar second population (2P). The presence of MPs is often surveyed by contrasting certain abundance ratios against each other: for instance, N, Na, Al and Si are typically enhanced when C, O and Mg are depleted. In extreme cases, the abundances of Mg and K are anticorrelated \citep{2011ApJ...740...60C, 2015ApJ...801...68M}. Heavy element abundances, on the other hand, show fairly uniform distributions and only small variations across GC stars \citep{2019MNRAS.487.3815M}, with iron-complex GCs being the exception \citep{2015MNRAS.450..815M, 2015AJ....150...63J}.

The spread in the abundance ratios and the fraction of stars in the 2P of GCs correlate with GC mass and anticorrelate with metallicity \citep{2009A&A...505..139C, 2009A&A...505..117C, 2010A&A...516A..55C, 2017MNRAS.464.3636M, 2020MNRAS.491..515M}. Tentative correlation with cluster age has been reported based on the fact that almost all ancient massive clusters host a significant number of 2P stars while young, relatively massive clusters do not \citep{2016MNRAS.460.1869C, 2017MNRAS.468.2482L, 2017MNRAS.468.3150M, 2018MNRAS.473.2688M, 2020MNRAS.491..515M}. The radial distribution of the 2P stars within GCs is often centrally concentrated \citep{2011A&A...527L...9K, 2011A&A...525A.114L, 2016MNRAS.463..449S, 2019ApJ...884L..24D} while some GCs have a centrally concentrated 1P or a radially uniform mix of 1P and 2P stars \citep{2014ApJ...791L...4D, 2015ApJ...804...71L, 2023MNRAS.520.1456L}. The absence of a central concentration in 2P might either be the result of dynamical mixing \citep{2013MNRAS.429.1913V} or there being no primordial segregation between 1P and 2P. Star  clusters also lose mass through gas expulsion and while orbiting their host galaxies \citep{2010A&A...516A..73D, 2010MNRAS.409..305L, 2015MNRAS.452..924K, 2019MNRAS.482.5138B, 2016MNRAS.458.1450W}, with complex impact on the MPs depending e.g. on the galactic orbit and the initial radial distribution of the stars \citep{2016A&A...587A..53K, 2020MNRAS.491..515M}.

The variety of models that can at least partly explain the abundance and scaling relations of MPs is arguably among the most diverse sets of proposed solutions to an open problem in modern astrophysics \citep{2018ARA&A..56...83B, 2019A&ARv..27....8G}. A comparison of the enhanced and depleted elements with the products of nucleosynthesis at increasingly high temperatures \citep{1999A&A...347..572A} indicates such abundance variations may be related to nuclear burning within intermediate mass or massive stars \citep{1989ATsir1538...11D, 1993PASP..105..301L}. As the nuclear burning products are released into the surroundings of a massive star, they pollute the local environment. If the ejection process is not energetic enough to expel the material out of a star-forming region, the material may be locked into newly forming stars. This so-called self-enrichment will lead to chemical variations within the population of stars that formed spatially and temporally correlated. Self-enrichment by the first formed stellar population is especially favoured  as the origin of the ubiquitous MPs because the peculiar abundance ratios are observed across a range of present-day stellar masses and evolutionary stages \citep{1998MNRAS.298..601C, 2001A&A...369...87G, 2003A&A...402..985Y}, which is hard to reconcile only with internal evolution of all those stars.

From a stellar nucleosynthesis point of view, to produce the Na-O anticorrelation observed in the majority of GCs, the Ne-Na chain requires a temperature of tens of MK \citep{1993PASP..105..301L, 2007A&A...470..179P}. The Al-Mg anticorrelation observed in a limited set of GCs would be obtained through the Mg-Al chain, operating efficiently at temperatures above $70$ MK \citep{2007A&A...470..179P}. The less frequently observed Si and K enhancement would require more than $80$ MK and $150$ MK, respectively \citep{2017A&A...608A..28P}. The most popular polluters suggested today include intermediate mass asymptotic giant branch (AGB) stars \citep{1981ApJ...245L..79C, 2001ApJ...550L..65V, 2010MNRAS.407..854D}, fast-rotating massive stars \citep{2004ApJ...612L..25N, 2006A&A...448L..37M, 2006A&A...458..135P, 2007A&A...464.1029D}, massive binary stars \citep{2009A&A...507L...1D, 2010MNRAS.407..277S, 2013MNRAS.436.2398B}, very massive \citep{2018A&A...615A.119V} and supergiant stars \citep{2018A&A...612A..55S, 2019ApJ...871...20S}, supermassive stars \citep{2014MNRAS.437L..21D, 2018MNRAS.478.2461G, 2023A&A...673L...7C} and stellar mergers \citep{2020MNRAS.491..440W}. Tidal disruption events \citep{2016MNRAS.458..127K} have also been suggested as a possible source for strong N-enhancements observed in compact high-redshift objects (e.g. \citealt{2023A&A...677A..88B}) that may be progenitors to present-day GCs. Many of the proposed scenarios require dilution of the polluting material with the unenriched (i.e. non-polluted) interstellar medium (ISM) either to produce the proper abundance correlations or to alleviate the so-called mass budget problem \citep{2007A&A...470..179P}. The latter describes the discrepancy between the significant, often dominant, fraction of 2P stars per cluster and the fact that stellar populations typically return only a small fraction of their mass for recycling. As pointed out by \citet{2015MNRAS.454.4197R} and \citet{2018ARA&A..56...83B} among others, none of the proposed scenarios have been able to alone explain all aspects of the ubiquitous MP phenomenon; it may turn out that multiple sources and evolutionary scenarios need to be invoked together \citep{2010MNRAS.407..277S, 2012ApJ...750L..14C, 2014ApJ...795L..28C, 2015MNRAS.449.3333B, 2019MNRAS.485.4311J}.

Numerical simulations have been used to address a variety of the scenarios that lead to the formation of chemically anomalous stars within GCs. The formation of the 2P polluted by AGB winds in hydrodynamical simulations \citep{2008MNRAS.391..825D, 2011MNRAS.412.2241B, 2019MNRAS.486.2570B, 2019MNRAS.489.3269C, 2021MNRAS.500.4578M} is among the most commonly studied scenarios. Semianalytic hydrodynamical models of self-enrichment and gas retainment through stellar winds of low-metallicity massive stars have been investigated in \citet{2019ApJ...871...20S}. Such detailed enrichment simulations typically neglect modelling the formation of the 1P and instead impose the polluting component as a single age pre-existing population evolving e.g. within a gaseous background. Self-pollution by massive OB-star winds within a star-forming molecular cloud has been self-consistently modelled by \citet{2021ApJ...922L...3L} using a detailed stellar wind prescription but at solar metallicity, incompatible with GC-forming environments, and with no attention to chemical enrichment. \citet{2019MNRAS.486.1146H} likewise followed the star cluster formation process within giant molecular clouds using sink particles, but painted the stellar enrichment over the cluster stars only in post-processing. Monte Carlo simulations \citep{2021MNRAS.502.4290V, 2022MNRAS.517.4768H} and N-body simulations have been used to probe the dynamical evolution of MPs \citep{2015MNRAS.450.1164H, 2017MNRAS.472...67H, 2019MNRAS.487.5535T, 2024A&A...681A..45L} but again rely on assumptions of the initial distribution of the 1P and 2P stars. So far, the formation of star clusters in the GC mass range, the development of MPs, and their complex variety of correlated abundance ratios have not been investigated in hydrodynamical simulations that realise single stars in an evolving galactic star forming environment. 

Here we assess the feasibility of the coeval cluster formation and self-enrichment by following star formation and stellar feedback from galactic to sub-parsec scales throughout a dwarf galaxy starburst. We simulate how the interaction of two dwarf galaxies composed of gas, stars and dark matter leads to compression of dense, high-pressure gas in a starburst phase where the star formation rate (SFR) exceeds the quiescent phase by two orders of magnitude. We follow the formation and evolution of massive star clusters with more than $10^5$ stars by applying pre-computed and tabulated stellar-evolutionary models to stellar particles that represent single stars. The stars release chemically enriched material through stellar winds and supernovae, which we trace element-by-element and channel-by-channel throughout the galactic ISM. This allows us to follow when and how a chemically enriched second population of stars forms, including their exact chemical composition. We then compare our theoretical results to observations of present-day GCs, which may have formed in such conditions when the Universe was still young \citep{2023ApJ...945...53V, 2024arXiv240103224A}. Because MPs are most ubiquitous in GCs with sub-solar metallicity, we concentrate on stellar enrichment processes in a galactic environment with a low metallicity of only $1\%$ of the solar value. 

\begin{figure*}
\includegraphics[width=\textwidth]{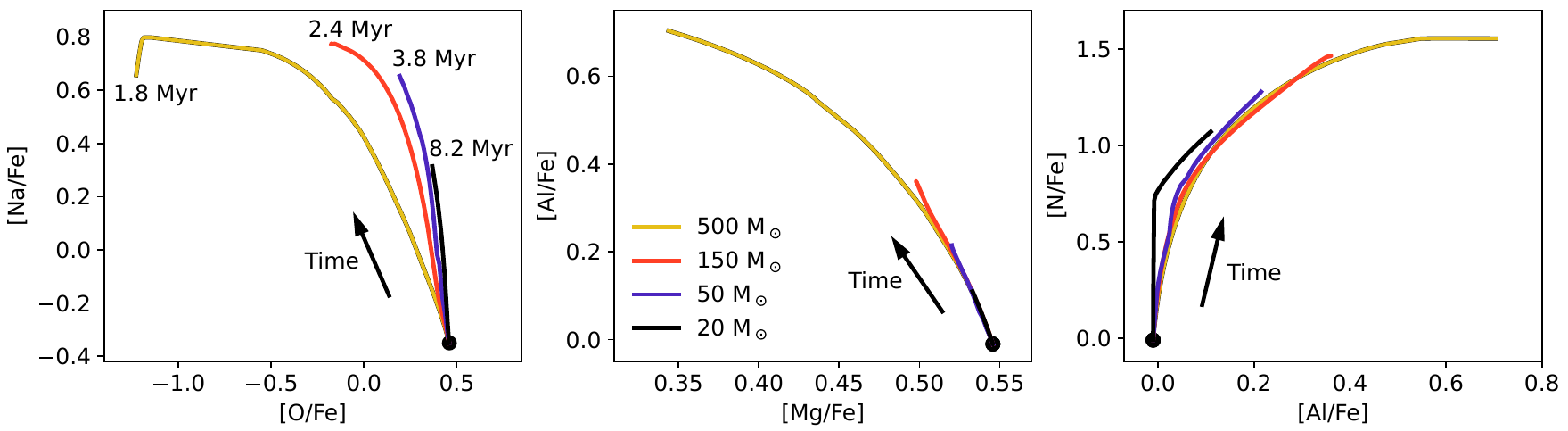}
\caption{The evolution of the abundance ratios [Na/Fe], [O/Fe], [Al/Fe], [Mg/Fe] and [N/Fe] on the surface of the star in four BoOST stellar models with initial masses of \mbox{$20$ \mdot{}}, \mbox{$50$ \mdot{}}, \mbox{$150$ \mdot{}} and \mbox{$500$ \mdot}. The black dots indicate the zero age main sequence and the abundance ratios get more extreme values as the stars evolve. The models describe the stellar evolution until core-helium exhaustion, for a total of \mbox{1.8 Myr}, \mbox{2.4 Myr}, \mbox{3.8 Myr} and \mbox{8.2 Myr} in order of increasing mass. The stellar wind that is injected into the interstellar medium surrounding each star is assumed to have the composition of the stellar surface. \label{fig:tracks}}
\end{figure*}

For hydrodynamical simulations we employ the \textsc{sphgal} code \citep{2014MNRAS.443.1173H, 2016MNRAS.458.3528H, 2017MNRAS.471.2151H, 2019MNRAS.483.3363H}, supplemented with a fully realised stellar initial mass function (IMF; \citealt{2023MNRAS.522.3092L}) and evolutionary tracks for massive stars \citep{2022A&A...658A.125S}. The simulations are part of the Galaxy Realizations Including Feedback From INdividual massive stars\footnote{\url{https://wwwmpa.mpa-garching.mpg.de/~naab/griffin-project}} (\textsc{griffin}) project, geared to address outstanding challenges in galaxy formation and evolution \citep{2017ARA&A..55...59N}. The methodology has been successfully used to study the formation of stars, star clusters and globular clusters \citep{2019ApJ...879L..18L, 2020ApJ...891....2L, 2020ApJ...904...71L, 2022MNRAS.509.5938H, 2023MNRAS.522.3092L, 2023MNRAS.526.1408S} and their observational properties \citep{2022MNRAS.514.4560L}, as well as the CII-emission and H$_2$ distribution in the ISM \citep{2022ApJ...934..115B, 2022MNRAS.512.4736S} in low-metallicity dwarf galaxies. 

The article is structured as follows. Section \ref{section:simulations} introduces the simulation methodology and initial conditions of the hydrodynamical simulations. In Section \ref{section:results} we discuss the results, starting with general star formation properties of the dwarf galaxy starburst and continuing to detailed investigation of chemical enrichment within the young massive star clusters formed during the starburst. We discuss the implications of our results for the formation of self-enriched MPs in GCs in Section \ref{section:discussion}. Conclusions are given in Section \ref{section:summary}. In the following, we will use the terms enriched or second population (2P) for the chemically peculiar GC stars, and unenriched or first population (1P) for the GC stars that have a chemical composition equivalent to the field stars.

\section{Simulations}\label{section:simulations}

For simulations we use the \textsc{sphgal} code \citep{2014MNRAS.443.1173H, 2016MNRAS.458.3528H, 2017MNRAS.471.2151H, 2019MNRAS.483.3363H} that has been modified from \textsc{gadget-3} \citep{2005MNRAS.364.1105S} to introduce improvements to the performance of the smoothed particle hydrodynamics (SPH) method. Modifications to star formation and stellar feedback routines to e.g. account for single massive stars have been outlined in \citet{2014MNRAS.443.1173H, 2016MNRAS.458.3528H, 2017MNRAS.471.2151H}, \citet{2019MNRAS.483.3363H}, \citet{2020ApJ...891....2L} and \citet{2023MNRAS.522.3092L}. Here we give a brief description of the star formation and stellar feedback models most relevant for the current study and refer the reader for further details to the references listed above. For SPH we adopt 100 neighbours.

\subsection{Star formation}\label{section:SF}

Non-equilibrium cooling and heating processes in the low-temperature ISM (\mbox{$<3\times 10^4$ K}) are accounted for using a chemical network \citep{2016MNRAS.458.3528H} that includes six chemical species (H$_2$, H$_+$, H, CO, C$_+$, O) and free electrons. At temperatures exceeding \mbox{$>3\times 10^4$ K}, we use the equilibrium metal-dependent cooling rates of \citet{2009MNRAS.393...99W}. 

Star formation is implemented as in \citet{2023MNRAS.522.3092L} with a few minor modifications. We employ a Jeans mass
dependent star formation threshold and sample the gas mass into individual stars after a gravitational collapse time (local dynamical timescale) has elapsed. The local Jeans mass is computed as
\begin{equation}
   M_\mathrm{J} = \frac{\pi^{5/2}c_\mathrm{s}^3}{6G^{3/2}\rho^{1/2}}, 
\end{equation}
where $c_\mathrm{s}$ is the sound speed, $G$ is the gravitational constant, and $\rho$ is the density of the gas particle. Gas particles for which the Jeans mass crosses below a threshold of half the SPH-kernel mass, \mbox{$0.5 M_\mathrm{kernel}= 200$ \mdot{}}, are turned into \textit{reservoir} particles and decoupled from hydrodynamics. The reservoir particles interact with their surroundings only through gravity, while they remain inert for a duration given by the local dynamical time scale $t_\mathrm{dyn}=(4\pi G \rho)^{-1/2}$ measured for the gas at the time when the particles crossed the star formation threshold. During this inert time, the particles are also decoupled from any nearby stellar feedback: they do not receive metal-enriched material once the gravitational collapse has been triggered. 

After the dynamical time of a reservoir particle has passed, we sample the mass of the particle immediately into single stars randomly according to the \citet{2001MNRAS.322..231K} IMF between 0.08 \mdot{} and 500 \mdot. If the sampled mass exceeds the particle mass by more than 0.08 \mdot, we search for neighbouring reservoir particles to account for the missing mass. Compared to \citet{2023MNRAS.522.3092L} where we used a fixed search radius of 1 pc, we now use the local Jeans length at the time of conversion as the search radius, given by
\begin{equation}
   R_\mathrm{J} = \left( \frac{3}{4\pi} \frac{M_\mathrm{J}}{\rho} \right) ^{1/3}.
\end{equation}
The Jeans length provides a natural proxy for the size of the region that undergoes gravitational collapse. The radius is allowed to vary between 0.3 pc and 5 pc, limited by the size of the gravitational softening kernel and the size of typical molecular clouds. In case there is not enough mass to satisfy the sampled stellar mass, the last value is discarded and a new random mass is drawn. The sampling and neighbour search is continued until the mass of the particle is successfully sampled, one reservoir particle at a time. As a result, the simulation realises massive (\mbox{$> 9$ \mdot}) and very massive (\mbox{$> 100$ \mdot}) stars only in conditions where the local mass reservoir is large enough to satisfy the stellar mass through strict conservation of mass. As shown in \citet{2023MNRAS.522.3092L}, this results in a realistic relation between the mass of star clusters and the most massive stars they can host, while star formation in the galactic field can only result in relatively low-mass stars.

\subsection{Stellar feedback}\label{section:feedback}

Mass is conserved regardless of particle type throughout star formation and stellar feedback. This is especially important in the case of the chemical elements, as we are interested in the propagation of the elements originating from different enrichment channels. 
The simulations trace the mass of 13 elements: H, He, N, C, O, Si, Al, Na, Mg, Fe, S, Ca and Ne. In addition to the uniform initial metallicity of $0.016$ \zdot{}, we follow the chemical enrichment through stellar winds of massive stars and AGB stars, as well as core-collapse and pair-instability supernovae (SNe). In the current study we investigate stellar enrichment during the formation of star clusters in a dwarf galaxy starburst mainly over a time scale of a few Myr, therefore we focus on the elements originating from stellar winds of massive stars and supernovae.

\begin{figure*}
\includegraphics[width=\textwidth]{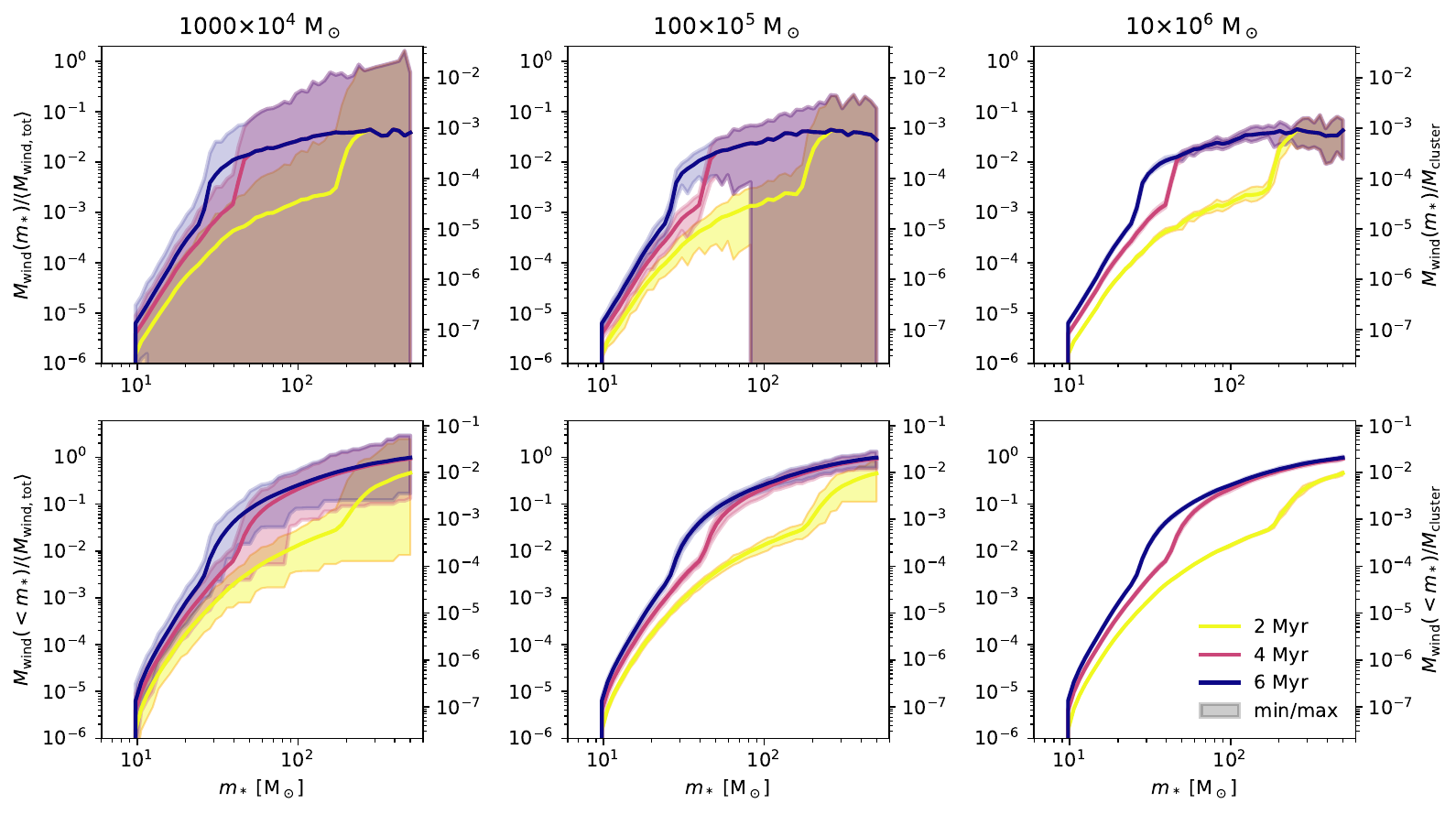}
\caption{Stellar wind-material produced by $9$--$500$ \mdot{} stars in a randomly sampled stellar population in $10^4$ \mdot{} (left), $10^5$ \mdot{} (middle) and $10^6$ \mdot{} (right) star clusters, computed using the BoOST stellar tracks for massive stars. The wind production is binned ($\sim0.04$ dex per bin) according to stellar mass, each cluster has a uniform age, and we show the wind-material produced until cluster age of 2 Myr (yellow), 4 Myr (red) and 6 Myr (blue). Each cluster has been randomly populated with an IMF between 0.08 \mdot{} and 500 \mdot{}, and we sample in total $10^7$ \mdot{} of stars by Monte Carlo sampling the $10^4$ \mdot{}, $10^5$ \mdot{} and $10^6$ \mdot{} clusters 1000, 100 and 10 times, respectively, to showcase the effect of stochastic sampling. The solid lines show the mean wind-mass of each Monte Carlo set at a given cluster age, and the shaded regions show the minimum and maximum value in each mass bin to indicate the range of most extreme values with respect to the mean. The top row shows the wind production at a given stellar mass, and the bottom row shows the cumulative distribution of wind-material with increasing stellar mass. The wind-masses are shown normalized to the total produced wind-material by the time the cluster has aged to \mbox{25 Myr} (left y-axis) and to the mass of the respective cluster (right y-axis). Very massive stars ($>100$ \mdot) dominate the mass-loss during the first couple of Myr, 90\% of the wind-material is released by $>50$ \mdot{} stars already by cluster age of 6 Myr, and the impact of stochasticity on the minimum and maximum wind-mass released per mass bin reduces with increasing cluster-mass. \label{fig:theor_output}}
\end{figure*}

All stars in the simulation are considered individually while computing the interstellar radiation field through far-ultraviolet radiation (FUV, \mbox{$6$--$13.6$ eV}). The radiation field at the location of each gas particle is attenuated using the \treecol{} algorithm \citep{2012MNRAS.420..745C} along 12 incoming directions, divided evenly around the particle using \healpix{} \citep{2011ascl.soft07018G}. In addition, massive stars create H$\,$II regions through photoionising radiation propagated using a Str\"omgren type approximation that iteratively searches for ionization equilibrium. Details of how radiation is implemented in the simulation can be found in \citet{2016MNRAS.458.3528H} and \citet{2017MNRAS.471.2151H}.

We follow the evolution of single massive and very massive stars until the end of the core helium-burning phase using the Bonn Optimized Stellar Tracks (BoOST, \citealt{2022A&A...658A.125S}). We use the low-metallicity (\mbox{$Z=0.016$ \zdot}) IZw18-grid that contains 1856 slowly rotating stellar models with 608 timestamps each, logarithmically spaced between \mbox{9 \mdot{}} and \mbox{498.4 \mdot{}}. As described in \citep{2023MNRAS.522.3092L}, each massive star is given an FUV luminosity, a photoionising photon rate, and stellar wind properties such as the outflow rate, chemical composition and velocity, all of which evolve throughout the stellar lifetimes. The choice of applying the BoOST library of massive-stellar wind yields was based on the fact that this library provides a comprehensive list of elements, including Na, Mg and Al, that are under wide-spread scrutiny in GC-studies. Fig. \ref{fig:tracks} gives examples of how some key abundance ratios on the stellar surface, and hence in the stellar wind, evolve during the stellar lifetime (see Fig. 2 in \citealt{2023MNRAS.522.3092L} for more examples). The stellar winds of up to \mbox{$\sim 10^{-3}$ \mdot{} yr$^{-1}$} are injected radially outward assuming a momentum conserving interaction between the stellar wind and the ISM as described in \citet[see their Section 2.2]{2023MNRAS.522.3092L}.

Lower-mass stars ($0.8$--\mbox{$9$ \mdot}) are provided with FUV luminosities using the Geneva stellar tracks at $Z=0.0004\sim 0.02$ \zdot \citep{2019A&A...627A..24G} and the \textsc{BaSeL} spectral library at corresponding $Z\sim 0.01$ \zdot{} \citep{1997A&AS..125..229L, 1998A&AS..130...65L, 2002A&A...381..524W}\footnote{Because of the lower mass limit of \mbox{1.7 \mdot} in the $Z=0.0004$\mbox{$\sim 0.02$ \zdot{}} tables, we approximate the fluxes between \mbox{$0.8$--$1.7$ \mdot{}} by scaling up the equivalent values at $Z=0.002\sim 0.1$ \zdot{} from \citet{2013A&A...558A.103G}.}. For completeness, for the lowest mass stars (0.08--\mbox{0.8 \mdot}) we scale the lowest tabulated FUV luminosities downward according to $L\propto M^{3.5}$. 

Once a massive star in our simulation reaches the end of its lifetime (according to \citealt{2019A&A...627A..24G}), it either dies as a core-collapse SN, a pair-instability SN (PISN) explosion, or in collapse to a black hole. We assume that stars with initial masses between \mbox{8 \mdot{}} and \mbox{40 \mdot{}} explode by releasing $10^{51}$ erg of thermal energy\footnote{For discussion of the impact of more detailed explodability and explosion energies see e.g. \citet{2021MNRAS.501.5597G} and \citet{2023arXiv231011495S}.} and metal-enriched material interpolated using the metallicity dependent yield tables of \citet{2004ApJ...608..405C}. PISNe occur for stars with helium-core masses between $65$--\mbox{$133$ \mdot{}} \citep{2002ApJ...567..532H}. In our implementation, we check the helium-core mass of the BoOST stellar models at core-helium exhaustion. In the adopted models at \mbox{$Z=0.016$ \zdot}, PISN explosions occur in stars with initial mass between $107.2$--\mbox{$203.4$ \mdot{}}. As differences between zero and non-zero metallicity PISN models have been shown to be small \citep{2014A&A...566A.146K, 2017ApJ...846..100G}, the PISN explosion energies between $5\times10^{51}$--\mbox{$9\times10^{52}$ erg} and the chemical yields are interpolated logarithmically according to the zero-metallicity massive star (helium-core) models of \citet{2002ApJ...567..532H}\footnote{\url{https://2sn.org/DATA/HW01/}}. The total mass released in the PISN explosion equals the mass of the progenitor star at helium-exhaustion, with a combined chemical composition given by the helium-core model of \citet{2002ApJ...567..532H} and the rest of the star as given by the surface composition of the BoOST model. Because there is no remnant, the stellar particle is removed from the simulation upon the explosion.

Lower-mass stars with initial masses between $0.5$--$8$ \mdot{} are assumed to end their lives in an AGB phase, releasing a single burst of enriched wind-material with yields interpolated from \citet{2010MNRAS.403.1413K}. In practice, the simulation of $65$ Myr, and especially the starburst of $<10$ Myr, concern too short timescales to allow for a significant AGB contribution.

The stellar winds and SN injections are implemented using the \healpix{} algorithm to distribute the energy and mass around the stars into 12 equal-sized pixels, each containing $8\pm 2$ gas particles. Further details can be found in \citet{ 2019MNRAS.483.3363H}, \citet{2023MNRAS.522.3092L} and references therein.

\subsubsection{Stochastic wind output of an idealized stellar population}

A few representative examples of stellar tracks selected from the BoOST models are shown here in Fig. \ref{fig:tracks}, in \citet[their Figs. 2 and 3]{2022A&A...658A.125S} and in \citet[their Fig. 2]{2023MNRAS.522.3092L}. As clustered star formation and feedback are the focus of the current study, we illustrate the output of an idealized stellar population through stellar winds in Fig. \ref{fig:theor_output}. We have compiled the averaged wind output of three Monte Carlo cluster populations with total mass of \mbox{$10^7$ \mdot{}} composed of 1000, 100 and 10 uniform-age star clusters with masses of $10^4$ \mdot{}, $10^5$ \mdot{} and $10^6$ \mdot{}, respectively. Each cluster has been populated with a stochastically sampled \citet{2001MNRAS.322..231K} IMF between $0.08$--$500$ \mdot. The wind output of stars between \mbox{9 \mdot{}} and \mbox{500 \mdot{}} in each single cluster has been integrated to acquire the wind-mass released by the cluster by the time it reaches a stellar age of \mbox{2 Myr}, \mbox{4 Myr} and \mbox{6 Myr}. Fig. \ref{fig:theor_output} shows the mean and mimimum/maximum wind output of the three Monte Carlo sets of clusters ($10^4$ \mdot{} to $10^6$ \mdot{}) binned by stellar mass and the respective cumulative output with increasing stellar mass, i.e. the contribution of cluster stars up to mass $m_*$. The wind output is shown as normalized both to the total wind output of the averaged cluster sample until cluster age of 25 Myr and to the cluster-mass to indicate the relative contribution of different mass bins.

First we note in Fig. \ref{fig:theor_output} the dominating role of massive stars already very early on in the evolution of a uniform-age cluster. More than $40\%$ of all wind-material has already been released by the time the cluster reaches 2 Myr in age, with the major contribution from very massive ($>100$ \mdot) stars. Stars more massive than \mbox{$50$ \mdot{}} produce more than $90\%$ of all wind-material in the cluster. 

The second aspect emphasized in this simple Monte Carlo experiment is the impact of stochasticity on the output of the star clusters. For lower-mass star clusters (e.g. the $10^4$ \mdot{} sample) the variation of stellar masses at the high-mass end produces a significant scatter in the integrated wind output. Depending on the distribution of very massive stars, the maximum wind output of a given stellar mass bin can exceed the mean value of the sample by more than an order of magnitude. The lower limit of zero in the $10^4$ \mdot{} and $10^5$ \mdot{} samples indicates no stars in those bins in some of the randomly populated clusters and hence a zero contribution to the total wind yield. As can be seen in Fig. \ref{fig:theor_output}, the wind output (and, consequently, the total energy output) of massive stars is highly susceptible to stochasticity even in relatively massive star clusters. The star formation and feedback methods introduced here and in \citet[][see also \citealt{2021PASJ...73.1036H}]{2023MNRAS.522.3092L} have been specifically tailored to capture some of this stochasticity in a physically motivated manner. The aim is to improve upon traditional galaxy scale simulations that implement stellar feedback either using stellar population particles or assuming always a fully sampled IMF.

\subsection{Detection of star clusters}\label{section:clusters}

In the dwarf starburst simulation, we follow the formation of hundreds of star clusters across the mass range from a few 10 \mdot{} up to more than $10^5$ \mdot. Star clusters are identified using the structure finding algorithms \textsc{friend-of-friends} and \textsc{subfind} \citep{2001MNRAS.328..726S, 2009MNRAS.399..497D} included in \textsc{gadget-3}. For constructing the \textsc{friend-of-friends} catalogue we use a linking length of \mbox{0.2 pc}. Once \textsc{subfind} is used to find gravitationally bound substructures in the \textsc{friend-of-friends} catalogue, we consider gravitationally bound star clusters containing at least 50 stars in our final analysis. This corresponds to a typical minimum cluster-mass of $\sim25$ \mdot{} at our adopted input IMF. Unless stated otherwise, we analyse the cluster data as recorded in the final snapshot of the simulation.

\begin{figure*}
\includegraphics[width=\textwidth]{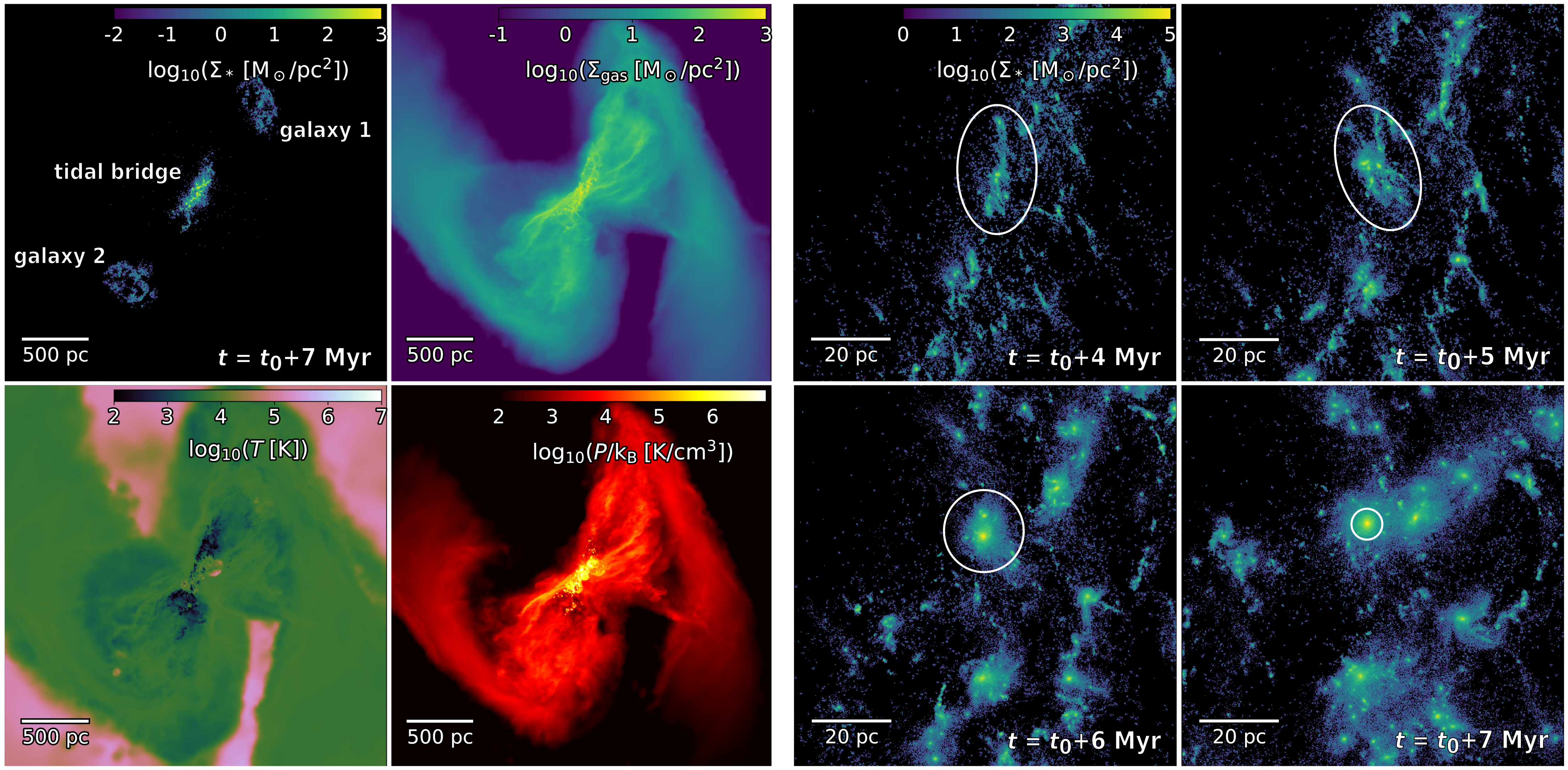}
\caption{\textit{Left half:} the surface density of stars formed during the simulation (\textit{top left}), the gas surface density (\textit{top right}), the gas temperature (\textit{bottom left}) and the thermal gas pressure (\textit{bottom right}) at the end of the off-nuclear dwarf galaxy starburst. The quantities have been computed within $\pm 500$ pc of the midplane. The most massive star clusters in the tidal bridge exceed $10^{5}$ \mdot. The old stellar disks of the progenitor dwarf galaxies (not shown) coincide with the concentrations of new stars labeled as galaxy 1 and galaxy 2 in the top left image. The image resolution is $\sim6$ pc. \textit{Right half:} the evolution of the stellar surface density within a \mbox{100 pc} patch in the central starburst region, shown in \mbox{1 Myr} steps across the starburst from left to right, top to bottom. The sequence also follows the formation of the most massive star cluster in the simulation (\mbox{$\sim 2\times 10^5$ \mdot}), circled in each panel. The image resolution is $\sim0.2$ pc. The timestamps in the panels indicate time with respect to $t_0$, which is the epoch when the starburst begins as defined in Fig. \ref{fig:sfr}. \label{fig:2dmap}}
\end{figure*}

\subsection{Initial conditions}\label{section:ICs}

We follow the procedure in \citet[and references therein]{2020ApJ...891....2L} in setting up the initial conditions for the dwarf galaxy encounter. The initial conditions consist of two identical compact gas rich dwarf galaxies selected from the simulation suite presented in \citet{2023MNRAS.522.3092L}, set up with methods of \citet{2005MNRAS.364.1105S}. The compact disk configuration is otherwise identical to that used in \citet{2017MNRAS.471.2151H} and \citet{2022MNRAS.509.5938H} except for the initial metallicity which we set to $\sim0.016$ \zdot. The initial dwarf galaxies are chosen right before star formation begins in the isolated configuration of \citet{2023MNRAS.522.3092L}. The initial disk scale lengths are 0.73 kpc, the disk gas fraction is 66\%, and the virial mass and total baryon mass are $2\times 10^{7}$ \mdot{} and $6\times 10^{7}$ \mdot, respectively. Gas particles and the old pre-existing disk stars have initially a $4$ \mdot{} resolution, dark matter is represented as $6.8\times10^3$ \mdot{} particles, and gravitational softening lengths are 0.1 pc for stars and gas and 62 pc for dark matter. Compared to \citet{2020ApJ...891....2L}, we reduce the approach velocities of the parabolic encounter by a factor of four to allow a strong tidal interaction between the dwarf galaxies already during the first passage. In addition, the gas distribution is more centrally concentrated in the compact dwarf galaxy models compared to the initial conditions of \citet{2020ApJ...891....2L}. These modifications shift the starburst observed during and after the second encounter after $\sim170$ Myr of simulation time in \citet{2020ApJ...891....2L} to the first passage between the two interacting galaxies already at $\sim60$ Myr. As a result, the galaxies experience an off-nuclear starburst of up to \mbox{0.5 \mdot{} yr$^{-1}$} along the tidal bridge between the galactic disks, ensuring as enhanced star and star cluster formation in as unenriched ISM conditions as possible. We continue the simulation until the most massive star clusters have formed the majority of their stellar mass $\sim 7$ Myr into the starburst, corresponding to a total simulation time of $\sim65$ Myr.

\section{Results}\label{section:results}

\subsection{Star formation in the dwarf galaxy starburst}

A visualization of the off-nuclear starburst and the related gas quantities are given in Fig. \ref{fig:2dmap}. The galactic disks can be identified as concentrations of young stars in the top right and bottom left quadrants of the first panel, connected with the star forming tidal bridge. Similarly vigorous off-nuclear star formation is common in galactic interactions both in observations \citep{1999AJ....118.1551W, 2001AJ....122.2969H, 2011ApJ...731...93M} and simulations \citep{2010ApJ...715L..88K, 2010ApJ...720L.149T, 2022MNRAS.514..265L} and qualitatively similar star forming configurations have been reported in interacting dwarf galaxies \citep{2021ApJ...912...89K, 2023ApJ...944..160M}. The tidal bridge is forming the majority of new stars in the simulated galaxies, and the star clusters in the bridge are more than two orders of magnitude more massive than those formed in the identical galaxies run in isolation in \citet{2023MNRAS.522.3092L}.

Fig. \ref{fig:2dmap} also shows a sequence of star formation within the central $\sim100$ pc of the off-nuclear starburst. The filamentary gas structures form stars and star clusters, some of which accumulate hierarchically into more massive clusters. The formation of the most massive cluster in the simulation with a final bound mass of \mbox{$2\times 10^5$ \mdot{}} can be followed \mbox{$\sim 15$ pc} above the image centre. Cluster sub-structure and tidal features, resolved with single stars, are visible in and around the massive clusters.

\begin{figure}
\includegraphics[width=\columnwidth]{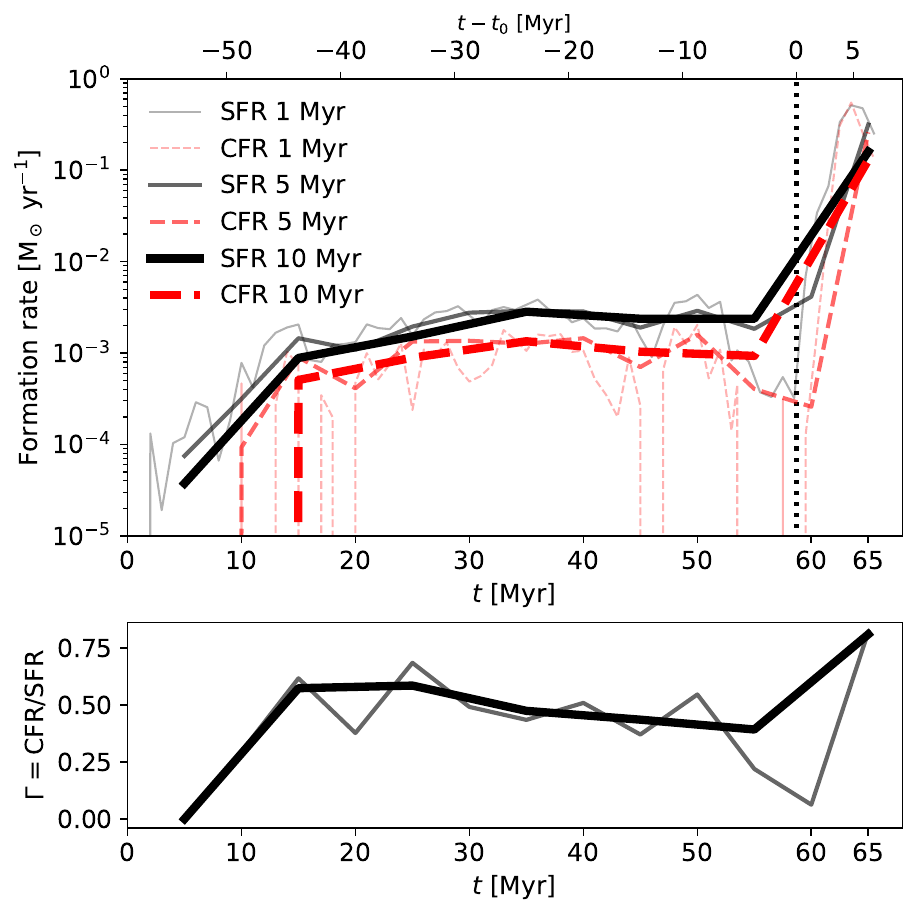}
\caption{Top: star formation rate (black solid lines) and the cluster formation rate (red dashed lines, $M_\mathrm{cluster}>100$ \mdot) of the interacting dwarf galaxies. The rates have been averaged in bins of 1 Myr (thin lines), 5 Myr (thicker lines) and 10 Myr (thickest lines). The vertical dotted line indicates the start of the starburst, $t_0$. Bottom: the cluster formation efficiency CFR/SFR using the 5 and 10 Myr averaged rates in the top panel. \label{fig:sfr}}
\end{figure}

Fig. \ref{fig:sfr} shows the SFR during the dwarf galaxy interaction. The SFR is computed as the sum of stars formed in the past 1 Myr, 5 Myr or 10 Myr, the latter estimating the timescale over which e.g. observed SFRs based on ultraviolet bands tend to smooth the star formation history \citep{2012ARA&A..50..531K}. The system enters into a starburst shortly after the galaxies start interacting for their first encounter. The most extreme SFRs exceed the quiescent phase by more than two orders of magnitude. We identify the starburst as the duration when the SFR is elevated compared to the early quiescent stages of the interaction and the isolated counterpart (see Fig. 9 in \citealt{2023MNRAS.522.3092L}). In the following, we define \mbox{$t_0=58$ Myr} as the start of the starburst and centre the time axis at this epoch when discussing the progression of the simulation. The $t_0$-centreed simulation time is indicated in Fig. \ref{fig:sfr} as well. 

In addition to the SFR, Fig. \ref{fig:sfr} shows the cluster formation rate (CFR), computed as the sum of all clusters more massive than \mbox{$100$ \mdot{}} with mean stellar age younger than \mbox{1 Myr}, \mbox{5 Myr} or \mbox{10 Myr} identified according to Section \ref{section:clusters}, divided by the respective time interval of 1 Myr, 5 Myr or 10 Myr. We also show in Fig. \ref{fig:sfr} the cluster formation efficiency $\Gamma=\mathrm{CFR/SFR}$, computed over the averaging intervals of 5 Myr and 10 Myr. Comparing the SFR and the CFR, we observe an enhancement in clustered star formation during the starburst, in agreement with the $\Gamma$ -- SFR surface density correlation found in \citet{2020ApJ...891....2L} and various surveys of star forming galaxies (see e.g. the discussion in \citealt{2020SSRv..216...69A}). This correlation is in contrast with the isolated dwarf galaxy simulations e.g. in \citet{2023MNRAS.522.3092L} and \citet{2024A&A...681A..28A} as well as recent observations of dwarf galaxies (e.g. \citealt{2023MNRAS.519.3749C, 2023ApJ...949..116C}) where no clear correlation is typically found. The observed dwarf galaxy populations do include dwarf starbursts (e.g. all three galaxies analysed in \citealt{2023ApJ...949..116C}) as well as galaxies with very little star formation (e.g. minimum 1--10 Myr $\Sigma_\mathrm{SFR}$ in \citealt{2023MNRAS.519.3749C} is less than $10^{-3}$ \mdot yr$^{-1}$ kpc$^{-2}$). It remains unclear whether the undetected $\Gamma$--SFR surface density correlation across dwarf galaxy populations is real or the result of observational limitations and low number statistics.

\begin{figure}
\includegraphics[width=\columnwidth]{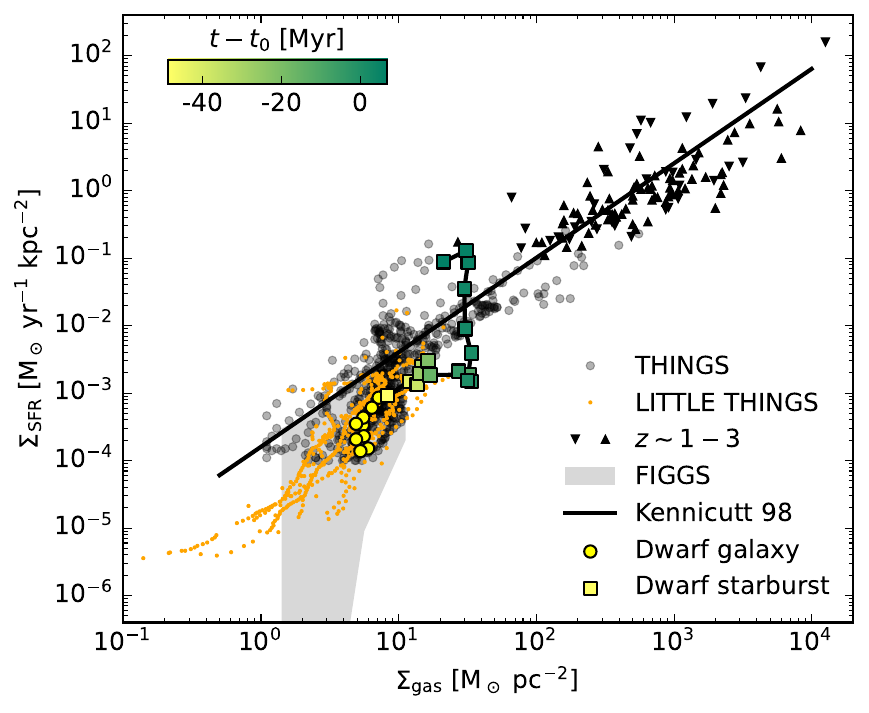}
\caption{The relation between star formation rate surface density and gas surface density in the star forming regions of the simulated dwarf galaxies. The yellow circles show one of the progenitor dwarf galaxies run in isolation over 500 Myr in 50 Myr steps and the squares coloured by time and connected by lines show the progression of the dwarf galaxy merger in 5 Myr steps before the starburst ($-48\,\mathrm{Myr}<t-t_0<0\,\mathrm{Myr}$) and in 1 Myr steps during the starburst ($0 \,\mathrm{Myr}<t-t_0<7\,\mathrm{Myr}$). The star formation rates are always computed over past 10 Myr and averaged over \mbox{400 pc} pixels, to match the typical averaging scale in the reference observations. The observed data points are from the THINGS survey (gray dots, \citealt{2008AJ....136.2782L}), the LITTLE THINGS survey (orange dots, \citealt{2015ApJ...805..145E}), 5-95 percentile range of the dwarf galaxies in the FIGGS survey (shaded gray, \citealt{2015MNRAS.449.3700R}) and high-z star forming galaxies (up and down pointing triangles, respectively, from \citealt{2013ApJ...768...74T} and \citealt{2010MNRAS.407.2091G}). The black diagonal line shows the observed fit $\Sigma_\mathrm{SFR}=1.6\times 10^{-4} \Sigma_\mathrm{gas}^{1.4}$ from \citet{1998ApJ...498..541K} that accounts for helium and metals by a correction factor of 1.36. \label{fig:KS}}
\end{figure}

The environmental dependence of the star formation rate is illustrated in Fig. \ref{fig:KS}. We show the star formation rate surface density $\Sigma_\mathrm{SFR}$ as a function of total gas surface density $\Sigma_\mathrm{gas}$ in the merging dwarf galaxy system and in one of the progenitor dwarf galaxies run in isolation for reference. The data have been binned in \mbox{400 pc} pixels and the data shown are averages over pixels that have had star formation within the past \mbox{10 Myr}. The observational reference data are from various surveys targeting dwarf galaxies and more massive star forming galaxies in the local Universe \citep{2008AJ....136.2782L, 2015ApJ...805..145E, 2015MNRAS.449.3700R} as well as star forming massive galaxies at higher redshifts \citep{2013ApJ...768...74T, 2010MNRAS.407.2091G}. The isolated dwarf galaxy and the initial stages of the dwarf galaxy interaction match the range of values observed in local systems that host low levels of star formation. As the starburst begins, first the $\Sigma_\mathrm{gas}$ increases, followed by increasing amounts of star formation up to \mbox{0.1 M$_\odot$ yr$^{-1}$ kpc$^{-2}$}. The most extreme starburst phase exceeds the normal star-forming relation indicated in the figure and extends toward the transition between the local star forming galaxies and higher redshift galaxies. Once the starburst subsides, the $\Sigma_\mathrm{SFR}$ and $\Sigma_\mathrm{gas}$ loop back down toward more quiescent values.

\subsubsection{Formation of star clusters}

\begin{figure}
\includegraphics[width=\columnwidth]{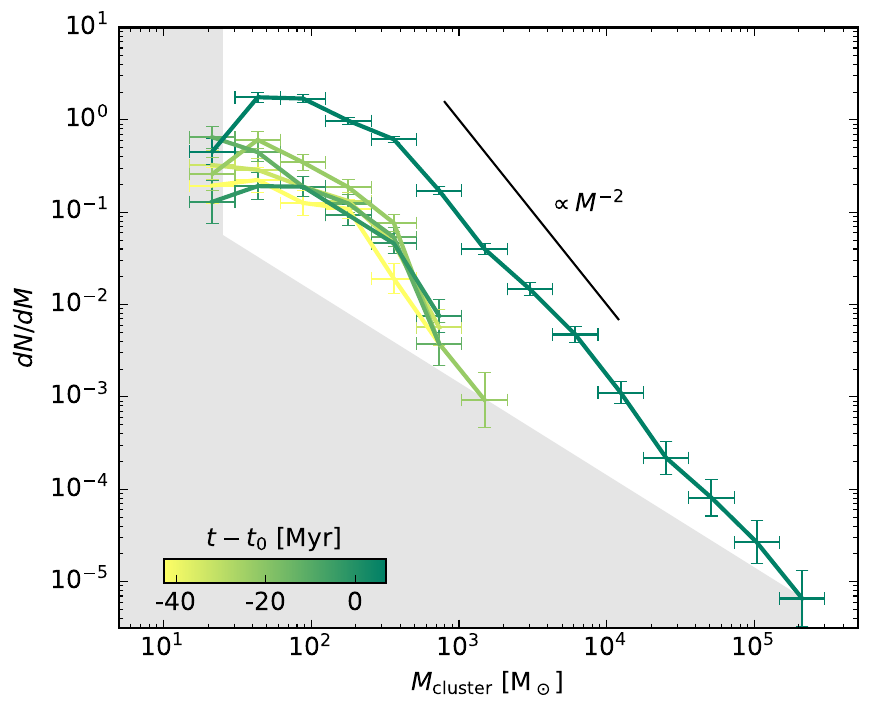}
\caption{The cluster mass function of clusters younger than 10 Myr, starting at simulation time  $t-t_0=-43$ Myr in 10 Myr steps until $t-t_0=7$ Myr. The errorbars indicate the Poisson error and the width of the mass bins. The shaded area shows the region below the average cluster-mass limit of $\sim25$ \mdot{} and below the limit of one cluster per mass bin. The line indicates a power-law relation of $dN/dM\propto M_\mathrm{cluster}^{-2}$. \label{fig:CMF}}
\end{figure}

As we already found in \citet{2020ApJ...891....2L} using 3D cluster data and in \citet{2022MNRAS.514.4560L} using mock observations, the increase in SFR results in the formation of more massive star clusters. Fig. \ref{fig:CMF} shows the star cluster mass function ($\mathrm{CMF}=dN/dM$) throughout the simulation in 10 Myr steps for star clusters with mean stellar age less than 10 Myr. Only the final snapshot, which represents the system after the starburst, includes young clusters more massive than \mbox{$\sim10^3$ \mdot}. The high-mass end of the CMF drops off the traditional power-law slope of $-2$ in the early stages when the SFR is low, similar to the CMFs found in simulated dwarf galaxies in \citet{2022MNRAS.509.5938H} and \citet{2024A&A...681A..28A}, while at enhanced SFR during the starburst (Fig. \ref{fig:sfr}) the power-law slope is in agreement with $-2$ from \mbox{$10^3$ \mdot{}} up to more than \mbox{$10^5$ \mdot}.

\begin{figure}
\includegraphics[width=\columnwidth]{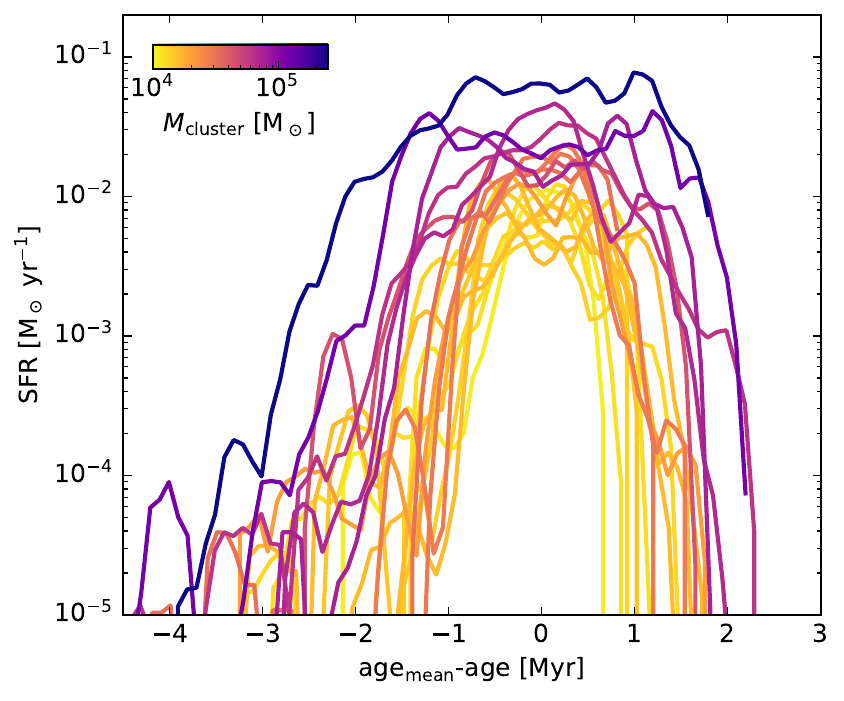}
\caption{The star formation histories of all clusters more massive than \mbox{$10^4$ \mdot}, colour-coded by the cluster-mass. Zero age corresponds to the mean stellar age of each cluster and simulation time progresses to the right. All clusters form on a timescale of less than 10 Myr and the majority of the stellar mass forms over a timespan of 2--3 Myr. \label{fig:cluster_SFR}}
\end{figure}

Massive star clusters form hierarchically within interconnected filaments along the tidal bridge (see Fig. \ref{fig:2dmap}). Fig. \ref{fig:cluster_SFR} shows the individual star formation histories of all clusters more massive than $10^4$ \mdot. The time axis has been shifted by the mean stellar age of each cluster, so that positive values correspond to the stars formed closer to the end of the simulation.
The clusters form rapidly over timescales of less than 10 Myr across the mass range, with the majority of stars forming in a 2--3 Myr timespan. The standard deviation around the mean value spans from 0.7 Myr to 2 Myr. The peak SFR and the width of the star formation profile increase with cluster-mass. Many of the clusters show multiple bursts of star formation. The increasingly high SFRs in massive clusters increases the chance of massive stars to form and the prolonged star formation histories allow the ejected material of those massive stars to be captured in regions of ongoing star formation. On the other hand, the period of star formation is short compared to the lifetime of the progenitors of core-collapse SNe ($\sim 5$ Myr), thus only PISN ejecta would be able to contribute to self-enrichment on short enough timescale.

\subsubsection{Initial stellar mass function}

\begin{figure}
\includegraphics[width=\columnwidth]{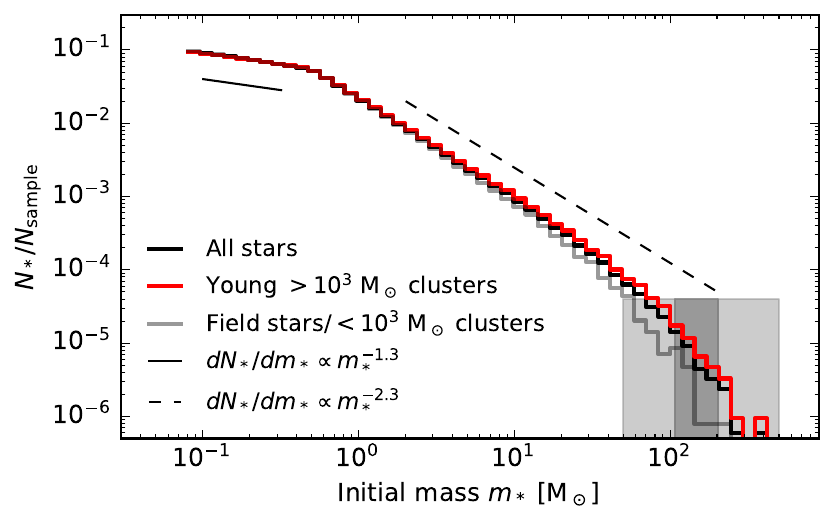}
\caption{The initial masses of stars realised by the end if the simulation at $t-t_0\sim7$ Myr. All stars are shown in black, stars in clusters more massive than $10^3$ \mdot{} in red, and stars in clusters less massive than \mbox{$10^3$ \mdot{}} and stars that are not bound to clusters combined in gray. The binned number counts have been normalized to the total number of stars in each sample. The thin solid and dashed lines indicate the input Kroupa IMF slopes of $-1.3$ and $-2.3$ used in sampling the stellar masses during the simulation. The shaded regions show the initial mass range where stars die either as direct collapse black holes (light gray) or where the stars explode as PISNe leaving no remnant (dark gray). \label{fig:IMF}}
\end{figure}

Fig. \ref{fig:IMF} shows the initial masses of all stars realised in the simulated dwarf galaxy interaction. We also show the initial masses of all stars in clusters more massive than \mbox{$10^3$ \mdot}, compared to the combined mass function of lower-mass clusters and field stars that are not bound to any cluster. We compare the realized IMF to the IMF shape of $dN_*/dm_*\propto m^{-1.3}$ below 0.5 \mdot{} and $dN_*/dm_*\propto m^{-2.3}$ above 0.5 \mdot{} that was used to randomly sample the initial masses. In the high-mass end, where we have to group particles to conserve mass, the realized stellar masses are limited by the local sampling reservoir (reservoir particles, Section \ref{section:SF}). The histograms have been normalized to the total number of stars in each sample to ease comparison. In contrast to the corresponding isolated compact dwarf galaxy simulation in \citet{2023MNRAS.522.3092L}, where we had only a handful of stars close to \mbox{100 \mdot{}} and only one star in the PISN mass range in the entire simulation sample, we now see almost the entire IMF range realised during the starburst. Until the end of the simulation, more than $\sim 16\,000$ massive stars have formed in the system, out of which $120$ had masses in excess of \mbox{$100$ \mdot}. 682 massive stars have already died: there have been 22 PISN explosions and 95 stars have died as direct collapse black holes with final black hole masses in the range from $38.5$ \mdot{} to $273$ \mdot. The most massive star realised in the simulation had an initial mass of $379.4$ \mdot.

Comparing the initial masses of stars in the field and low-mass clusters to the stars in massive clusters, the mass function in the massive clusters follows better the high-mass slope of $m_*^{-2.3}$ up to tens of solar masses. Very high masses above 200--300 \mdot{} are only present in the most massive clusters that reach more than \mbox{$10^4$ \mdot{}}. Conversely, regions corresponding to lower star formation activity exhibit a more pronounced deficit in initially massive and very massive stars. This difference shows the impact of the local mass conservation and the Jeans length dependent mass reservoir in the IMF sampling, which ensure that we allow massive stars to be realised strictly only when there has been enough star forming gas available within the region where stars are being spawned. There is a deficiency of initially very massive stars compared to a fixed power-law IMF across the simulation, therefore a fully sampled IMF is not guaranteed even in the most intense star forming regions that host the most massive clusters of the low-metallicity galaxies. 

\begin{figure*}
\includegraphics[width=\textwidth]{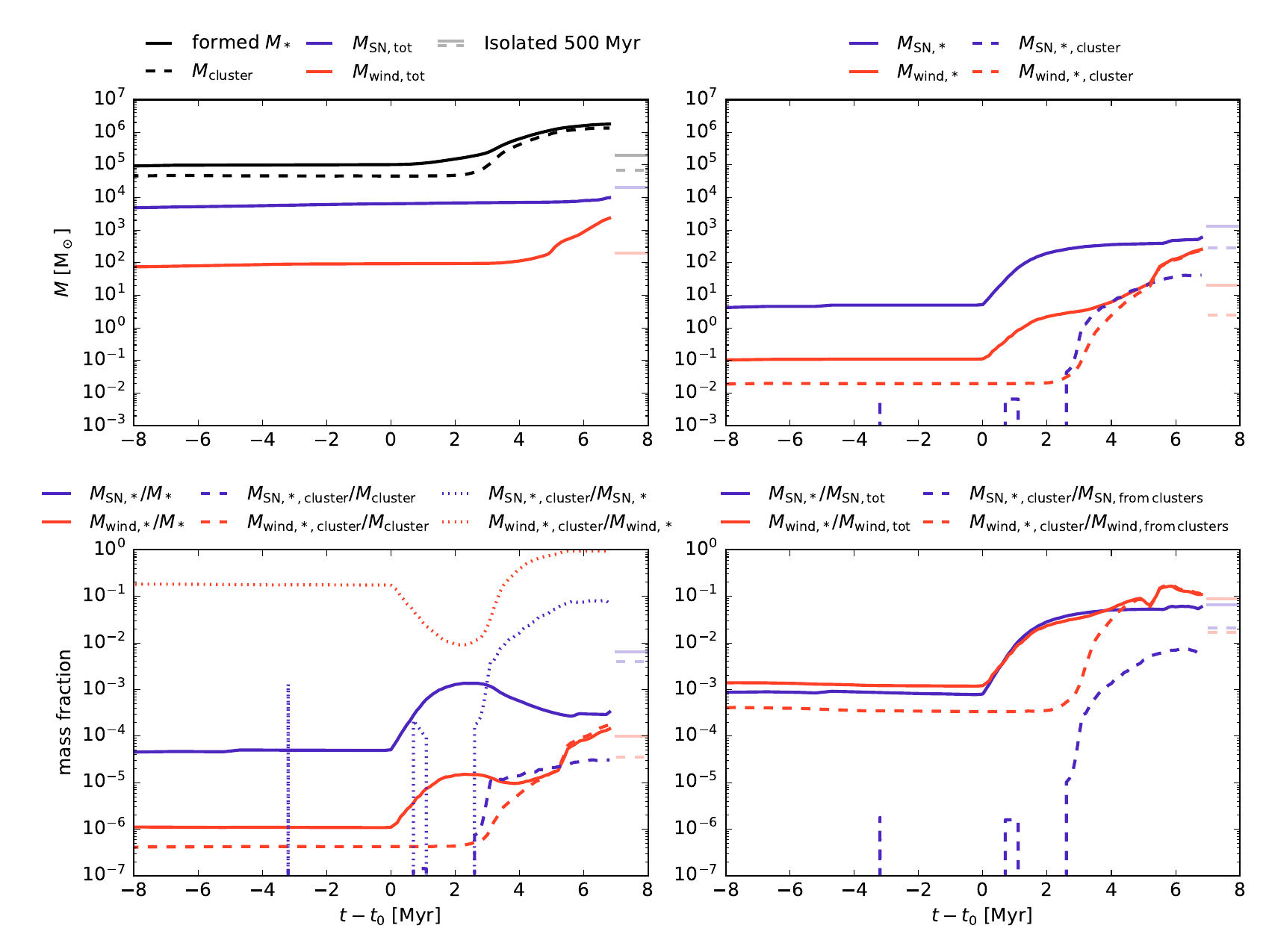}
\caption{The integrated wind and SN-enrichment during the dwarf starburst. Solid lines show the values for all stars in the simulation and dashed lines show the equivalent for stars in bound clusters; blue lines are for SN-material and red lines are for wind-material. The light coloured bars on the right of each panel show the equivalent values for a dwarf galaxy run for 500 Myr in isolation in \citet{2023MNRAS.522.3092L}. The labels in the legends are described in more detail in the text. The time axis has been shifted by the start of the starburst $t_0$ as indicated in Fig. \ref{fig:sfr}. \textit{Top left:} total formed stellar mass, bound star cluster mass, produced SN-mass and produced wind-mass. \textit{Top right:} wind and SN-material recycled in stars and bound star clusters. \textit{Bottom left:} wind and SN-material recycled in stars and star clusters relative to the total mass in stars, star clusters and the overall recycled mass. \textit{Bottom right:} wind and SN-material recycled in stars and star clusters relative to the produced wind and SN-material from all stars and cluster clusters. $\sim 10\%$ of the wind and SN-material is recycled in new stars ($M_\mathrm{wind,*,cluster}/M_\mathrm{wind,tot}\sim M_\mathrm{SN,*,cluster}/M_\mathrm{SN,tot}\sim 10\%$) and the rest ($\sim 90\%$) resides in the gaseous phase. 91\% of the wind-material that is recycled in new stars remains predominantly in star clusters ($M_\mathrm{wind,*,cluster}/M_\mathrm{wind,*}= 91\%$), while only $\sim8\%$ of the recycled SN-material is in cluster stars ($M_\mathrm{SN,*,cluster}/M_\mathrm{SN,*}=7\%$). \label{fig:ejecta}}
\end{figure*}

\subsection{Enrichment by massive stars}

The 13 individual elements (Section \ref{section:feedback}) released through stellar winds and SN explosions are traced throughout the starburst. Fig. \ref{fig:ejecta} shows the integrated total and relative masses of wind and SN-material across the starburst. From left to right and top to bottom we show

\begin{itemize}
    \item $M_*$ \& $M_\mathrm{cluster}$: cumulative mass of new stars and star clusters
    \item $M_\mathrm{SN}$ \& $M_\mathrm{wind}$: cumulative mass of wind and SN-material
    \item $M_\mathrm{wind,*}$ \& $M_\mathrm{SN,*}$: wind and SN-material locked in new stars
    \item $M_\mathrm{wind,*,cluster}$ \& $M_\mathrm{SN,*,cluster}$: wind and SN-material locked in bound star clusters 
    \item $M_\mathrm{wind,*}/M_*$ \& $M_\mathrm{SN,*}/M_*$: wind and SN-material locked in new stars in relation to total stellar mass 
    \item $M_\mathrm{wind,*,cluster}/M_\mathrm{cluster}$ \& $M_\mathrm{SN,*,cluster}/M_\mathrm{cluster}$: wind and SN-material locked in bound star clusters in relation to total cluster-mass 
    \item $M_\mathrm{wind,*,cluster}/M_\mathrm{wind,*}$ \& $M_\mathrm{SN,*,cluster}/M_\mathrm{SN,*}$: wind and SN-material locked in bound star clusters in relation to total locked in stars
    \item $M_\mathrm{wind,*}/M_\mathrm{wind,tot}$ \& $M_\mathrm{SN,*}/M_\mathrm{SN,tot}$: wind and SN-material locked in new stars in relation to the total produced material
    \item $M_\mathrm{wind,*,cluster}/M_\mathrm{wind\,from\,clusters}$ \& \newline$M_\mathrm{SN,*,cluster}/M_\mathrm{SN\,from\,clusters}$: wind and SN-material locked in bound star clusters in relation to the produced material 
\end{itemize}
The last two quantities have been computed in \mbox{0.1 Myr} steps from the wind and SN-material locked in cluster stars at any given moment, divided by the total wind and SN-material produced by massive stars in total. For comparison, we also indicate in Fig. \ref{fig:ejecta} the final values in a single dwarf galaxy analysed in \citet{2023MNRAS.522.3092L}, measured after 500 Myr of isolated evolution.

As the starburst ramps up, massive stars form in increasing numbers, releasing equivalently increasing amounts of wind-material as they evolve. Meanwhile, the total mass of SN-material is only starting to build up Myrs into the starburst. This is partly because the majority of the most short-lived stars die in a direct collapse without a SN (e.g. $87\%$ of $>40$ \mdot{} stars that live $<4$ Myr), with the PISN progenitors being the exception. SN-material still always dominates the mass returned by stars in the simulation, however by the end of the starburst the integrated winds have narrowed the gap down to only a factor of 4. The wind and SN-material recycled in all new stars is enhanced by two orders of magnitude during the starburst, amounting to approximately $11\%$ of the wind-mass and $6\%$ of the SN-mass produced in total by the time the starburst is over. As indicated in Fig. \ref{fig:ejecta}, the value of approximately $10\%$ of recycled feedback material is similar to that accumulated in stars of the isolated compact dwarf galaxy after 500 Myr of evolution in \citet{2023MNRAS.522.3092L}. The dwarf starburst can therefore recycle wind-material significantly faster than  either of the progenitor galaxies evolved in isolation. Winds are more prone to be locked in clustered stars compared to SN-material: by the end of the starburst, $91\%$ of the wind-material recycled in stars resides in clusters, compared to the corresponding value of only $8\%$ for all SN-material. The inefficient recycling of SN-material in clusters is also seen directly in $M_\mathrm{SN,*,cluster}/M_\mathrm{SN,\,from\,clusters}$, which shows only $0.6\%$ of the SN-mass produced by cluster stars as recycled in the clusters compared to the value of $\sim 11\%$ for wind-material.

In total, wind and SN-material account for $\sim10^{-4}$ and a few $10^{-4}$, respectively, of the total stellar mass formed by the end of the starburst. The difference between the two feedback channels is much smaller than what was found in the case of isolated dwarf galaxies in \citet{2023MNRAS.522.3092L}, where the SN-material in stars was built up gradually over 500 Myr through cooling and mixing within the galactic ISM. As can be seen in the top left panel of Fig. \ref{fig:ejecta}, the spatially clustered, rapidly evolving star formation during the starburst occurs on a shorter time scale than what would be required to cool and recycle the SN-material \textit{within} the region that is producing it. This is demonstrated in the bottom panels, where the SN-enrichment within star clusters is an order of magnitude less efficient compared to the overall SN recycling. Even though the starburst has both produced and locked in stars a similar amount of SN-material ($\sim 10^4$ \mdot) compared to the isolated counterpart at \mbox{500 Myr}, the ten-fold larger stellar mass in the starburst results in a ten-fold smaller contribution of SN-material to the total stellar mass. Meanwhile, because the winds are released continuously with lower injection energies, they are easier to capture in the gas \textit{within} the same star formation region. The formation of wind-enriched stars is dominated by enrichment from and to the star clusters, as can be seen in the converging lines for all stars and clusters in Fig. \ref{fig:ejecta}. We will investigate this \textit{self-enrichment} scenario within star clusters on a cluster-by-cluster level in the following Sections.

\begin{figure}
\includegraphics[width=\columnwidth]{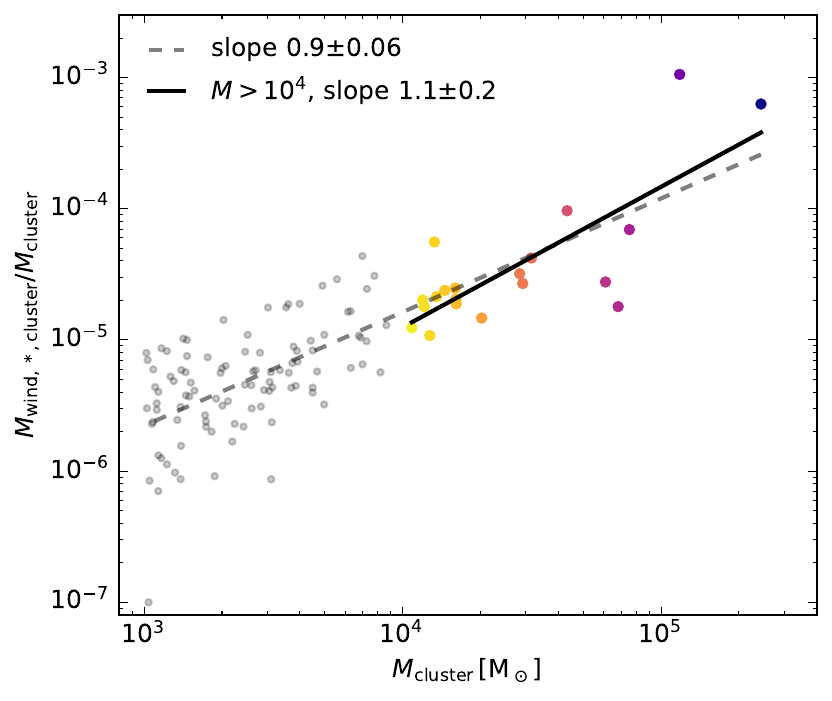}
\caption{The wind-mass fraction in bound star clusters more massive than \mbox{$10^3$ \mdot},  measured within $2r_\mathrm{1/2}$ at the end of the simulation. One single cluster with no wind-material has been placed at $10^{-7}$. The lines show the best-fit power-law relations $\propto M_\mathrm{cluster}^\alpha$ to all clusters (dashed) and to clusters more massive than \mbox{$10^4$ \mdot}, and the power-law slopes are indicated in the legend. Clusters more massive than \mbox{$10^4$ \mdot{}} have been colour-coded as in Fig. \ref{fig:cluster_SFR}. The wind-mass fraction, sourced mainly through self-enrichment, scales approximately linearly with cluster-mass and hence $M_\mathrm{wind,*,clusters}\propto M_\mathrm{cluster}^{2}$. \label{fig:mean_enr}}
\end{figure}

\subsection{Chemical enrichment within star clusters}

\subsubsection{Cluster-averaged wind-mass fraction}

To address whether chemical enrichment through stellar winds and SNe may result in self-enrichment during the formation of star clusters, we take a detailed look at the population of bound star clusters formed in the dwarf starburst. We focus on massive star clusters from $10^3$ \mdot{} to $>10^5$ \mdot{}, which host the majority of the massive and very massive stars (Fig. \ref{fig:IMF}) that are the sources of enrichment in the simulation. In the two following Sections, we consider the properties of the clusters as averaged over bound stars in each cluster at the end of the simulation, as determined in Section \ref{section:clusters}.

We show in Fig. \ref{fig:mean_enr} the mass fraction of wind-material in each cluster measured within twice the half-mass radius $r_{1/2}$. The correlation of the wind-mass fraction with cluster-mass indicates enhanced self-enrichment via stellar winds in massive star clusters. More massive star clusters form in more dense star-cluster-forming environments, which in our implemented star formation routines (Section \ref{section:SF}) leads to the formation of more massive stars both in stellar mass and in number. Since the scaling of the wind-mass fraction is linear or super-linear with cluster-mass, it follows that the wind-material recycled in cluster stars scales as $M_\mathrm{wind,*,clusters}\propto M_\mathrm{cluster}^{2}$. The more massive clusters therefore produce and/or capture increasing amounts of wind-material per unit cluster-mass. 

For comparison, self-enrichment in star-forming molecular clouds of varying initial density has also been investigated by \citet{2021ApJ...922L...3L} in high-resolution cloud-scale simulations. Unfortunately the models concerned only solar metallicity, and chemical composition of the winds of their massive stars (up to $100$ \mdot) was not a focus. Nevertheless, their higher initial densities resulted in higher efficiency of star formation while the new stars also captured increasing amounts of wind-material, in agreement with our findings in Fig. \ref{fig:mean_enr}. Their total wind mass fractions ranged from $10^{-4}$ to a few times $10^{-3}$ in their clusters with masses between $2.8\times 10^4$ \mdot{} and $8.5\times 10^4$ \mdot{}, indicating a super-linear relation between the wind-mass fraction and the cluster-mass as well. Given their assumption of a fully sampled Kroupa IMF, they attributed the increased efficiency of wind-capture to two factors: the enhanced trapping of the wind-bubbles in higher density gas (see the theoretical arguments in \citealt{2021ApJ...914...89L} and \citealt{2021ApJ...914...90L}), and the longer duration of star formation that can continue for multiple free-fall times in the densest star-forming regions. \citet{2016A&A...587A..53K}  have also argued that the compactness of the cluster, defined as $C_5=(M_\mathrm{cluster}/10^5$ \mdot)$/(r_{1/2}/\mathrm{pc})$, may govern whether the cluster gas can be retained against expulsion due to stellar winds.  Using the definition of cluster compactness from \citet{2016A&A...587A..53K}, we recover values between 0.1--4 for clusters more massive than $10^4$ \mdot{} with half-mass radii between 0.16--1 pc (see Fig. \ref{fig:stamps}). \citet{2016A&A...587A..53K} report very similar values for young massive star clusters from \citet{2014MNRAS.445..378B}, as well as present day GCs that are known to host chemically peculiar stars but should have been more compact in the past.  For lower-mass clusters, the compactness increases from 0.001 to 0.4 with increasing mass, similar to those reported for open clusters in \citet{2016A&A...587A..53K} where no multiple populations are found. \citet{2017ApJ...835...60W} and \citet{2017MNRAS.470..977L} have further shown that stellar winds can radiatively cool on a rapid timescale of a couple Myrs if the cluster is compact enough. For us, an additional component is the non-universal high-mass end of the IMF, where the less massive clusters simply form less massive stars.

\begin{figure}
\includegraphics[width=\columnwidth]{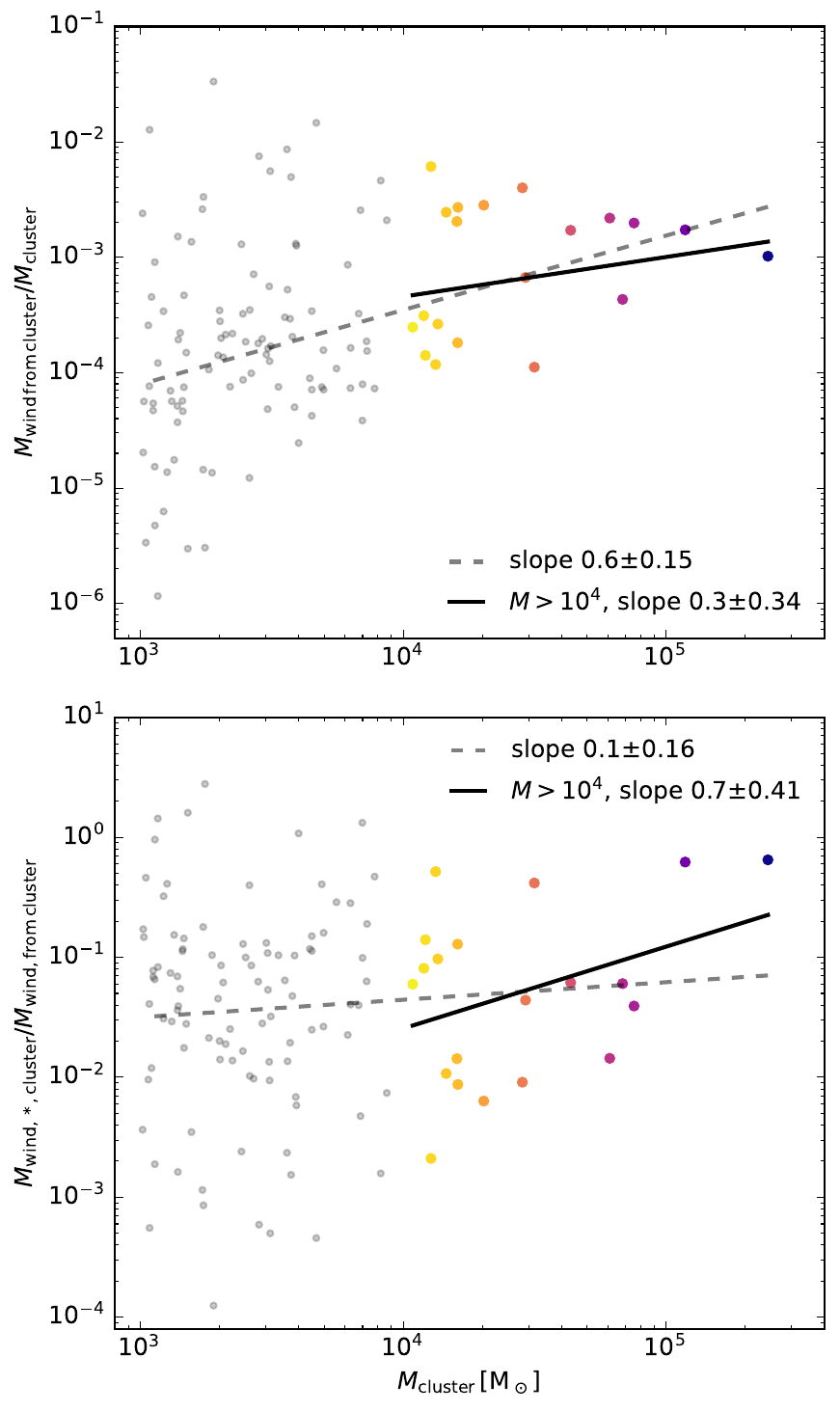}
\caption{\textit{Top:} the relative wind-mass produced by clusters more massive than $10^3$ \mdot, measured in the final snapshot as the integrated wind-output of all massive stars bound in a cluster related to the cluster mass. \textit{Bottom:} the recycled wind-fraction in each cluster in the top panel, computed as the wind-material locked in cluster stars divided by the wind-material produced by the cluster. The lines in both panels show the best-fit power-law relations $\propto M_\mathrm{cluster}^\alpha$ to all clusters (dashed) and to clusters more massive than \mbox{$10^4$ \mdot}, and the power-law slopes are indicated in the legends. Clusters more massive than \mbox{$10^4$ \mdot{}} have been colour-coded as in Fig. \ref{fig:cluster_SFR}. Even though the majority of the clusters are very young (mean stellar age $<5$ Myr), the produced mass-estimate is a lower limit as it excludes escaped stars as well as stars that have exploded as PISN since the PISN events leave no remnants. Due to the lower-limit of total produced mass in the top panel, the retained ratios, especially in the higher cluster-mass range, are upper limits. \label{fig:mean_ret}}
\end{figure}

\begin{figure}
\includegraphics[width=\columnwidth]{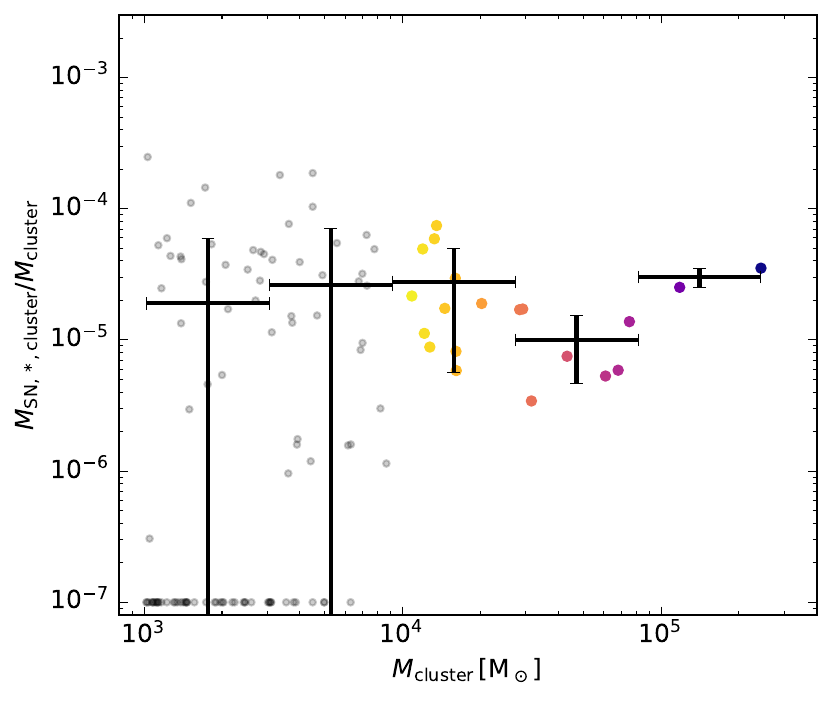}
\caption{The SN-mass fraction in clusters more massive than $10^3$ \mdot, measured within $2r_{1/2}$. Clusters with no feedback material have been placed at $10^{-7}$ and clusters more massive than \mbox{$10^4$ \mdot{}} have been colour-coded as in Fig. \ref{fig:cluster_SFR}.
The errorbars show the linear mean and standard deviation of the binned SN-mass fraction to guide the eye at low cluster-masses where many of the clusters have not been enriched by SNe. The SN-mass fraction remains flat with cluster-mass, reflecting SN-material that is uniformly mixed across the galaxy.
 \label{fig:mean_SN}}
\end{figure}

To disentangle whether the clusters simply produce more wind-material or whether the deeper cluster potential is able to capture more wind-material, we inspect in Fig. \ref{fig:mean_ret} the wind-mass \textit{produced} and \textit{recycled} in the clusters shown in Fig. \ref{fig:mean_enr}. The relative wind-production is the ratio between the integrated wind-mass produced by all massive stars in the cluster, divided by the cluster mass. The recycled wind-mass fraction has been computed as the ratio between the wind-mass locked in the cluster stars and the wind-mass produced by the clusters. The wind-mass produced by each massive star bound in each cluster has been integrated over the stellar lifetime, assuming that the star was always bound to the cluster in which it resides in the last snapshot. The wind-production may represent lower limits, since massive stars may have escaped and some have already exploded as PISN, leaving no trace of how much wind they produced. Tracing back all stars that contributed to the wind-enrichment would require detailed tracing of all of the clusters, which is beyond the scope of the current study. To minimize the impact of cluster evolution, we analyse the clusters immediately at the end of the starburst, within a few Myr after their formation. If the recycled fraction is above unity, the cluster has been enriched by stars that are not (anymore) bound to the cluster. The large scatter in the both panels is driven by the stochastic IMF.

The relative wind-production scales moderately with cluster-mass ($M_\mathrm{wind\,from\,clusters}/ M_\mathrm{cluster}\propto M_\mathrm{cluster}^{0.3-0.6}$), therefore the total wind-production has a slightly super-linear scaling with cluster mass ($M_\mathrm{wind\,from\,clusters}\propto M_\mathrm{cluster}^{1.3-1.6}$). This reflects the fact that higher mass clusters are better guaranteed to host a larger number of massive stars, with increasingly high upper mass limit in their IMF. If all wind-material would be captured as efficiently in all clusters, variations in the cluster-to-cluster wind-production alone would produce a too weak scaling in the wind-mass per cluster compared to the one measured in Fig. \ref{fig:mean_enr}.

The recycled fraction, likewise, scales as $M_\mathrm{wind,*,cluster}/M_\mathrm{wind\,from\,clusters}\propto M_\mathrm{cluster}^{0.1\mathrm{-}0.7}$, with a steeper correlation toward larger cluster masses. If the recycled fraction would be constant, the correlation of wind-mass fraction with cluster mass in Fig. \ref{fig:mean_enr} could be expected to emerge directly from variations in the cluster IMF. However, because the recycled fraction has a positive correlation with cluster mass, especially at high cluster masses, there seems to be a contribution of enhanced wind-trapping with increasing cluster-mass.  The most massive clusters here have the highest concentration $C_5$, indicating that the most massive clusters should be most resistant against gas expulsion and result in high recycling efficiencies in Fig. \ref{fig:mean_ret}. The strongly super-linear wind-mass -- cluster-mass relation $M_\mathrm{wind,*,clusters}\propto M_\mathrm{cluster}^{2}$ in Fig. \ref{fig:mean_enr} is therefore likely driven by the combination of increasing number of massive stars and more efficient capture of the wind-material.

\subsubsection{Cluster-averaged SN-mass fraction}

The mass fraction of SN-material in cluster stars is shown in Fig. \ref{fig:mean_SN}. The SN-fraction remains flat throughout the cluster-population analysed at the end of the simulation. This means that the total SN-material locked in cluster-stars scales only linearly with cluster-mass, $M_\mathrm{SN,*,cluster}\propto M_\mathrm{cluster}$. Since the SN production is only slowly increasing by the time the starburst is already over, the clusters do not have the time to self-enrich by SNe during their formation. This is also demonstrated in Fig. \ref{fig:cluster_SFR}, where the duration of the star formation in even the most massive clusters is only \mbox{5 Myr} or so, barely enough for the first core-collapse SNe to occur. The only possible SN contributor on such short timescales would be the PISNe in clusters massive enough to host very massive stars. PISNe are, however, more energetic and extremely rare in the simulation compared to normal SNe as there is one PISN for every 188 SN events.  Instead of local self-enrichment, the flat SN-mass fraction in Fig. \ref{fig:mean_SN} illustrates mixing of small amounts of SN-material throughout the galactic ISM: the clusters form during the starburst out of a gas mixture uniformly polluted by traces of SN-material.

The lower-mass clusters that host SN-enriched stars show a large scatter in mean SN-mass. This is partly driven by the adopted SN injection scheme where each SN is on average distributed over a fixed number of $\sim96$ gas particles. The SN-enriched stars occur in clumps of at least $4$ \mdot{} whenever a SN-enriched gas particle is sampled into stars, increasing the contribution of one such clump in low-mass star clusters.

\begin{figure}
\includegraphics[width=\columnwidth]{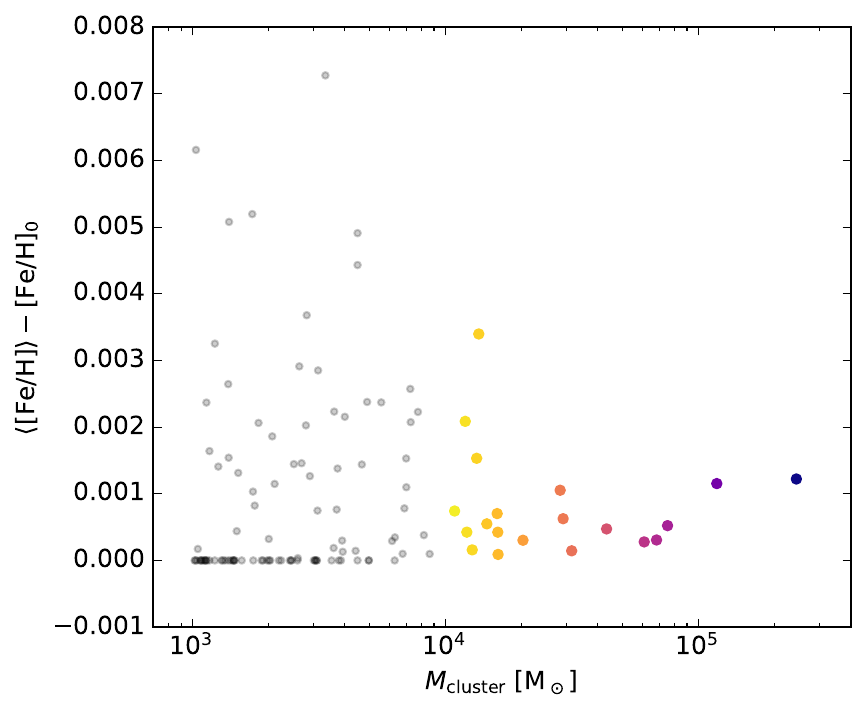}
\caption{The mean abundance ratios of [Fe/H] for stars bound to clusters more massive than $10^3$ \mdot, measured within $2r_\mathrm{1/2}$. The material used to compute the ratio for each star is composed of the SN and wind-mass locked in the stars, diluted by the rest of the stellar mass with the initial ISM composition. The mean [Fe/H] have been shifted by the [Fe/H] ratio of the unenriched ISM composition at the beginning of the simulation. Clusters more massive than \mbox{$10^4$ \mdot{}} have been colour-coded as in Fig. \ref{fig:cluster_SFR}.\label{fig:FeH}}
\end{figure}

\subsubsection{Heavy-element enrichment}\label{section:fe_per_h}

One of the key features of bona fid\'e GCs are their fairly uniform heavy element abundances across the stars of a given cluster, indicating no significant SN-enrichment during their formation. Recent spectral studies have however shown that some GCs do in fact host stars that have small spreads e.g. in their iron abundances, especially in the first, chemically normal populations \citep{2023A&A...669A..19L, 2023ApJ...958...31M}. When there is a spread in [Fe/H], its presence in the first population reflects the initial ISM composition, rather than self-enrichment. This observational interpretation is in agreement with our result of uniform SN-content and little to no SN self-enrichment even within the most massive star clusters simulated here.

Fig. \ref{fig:mean_SN} shows that our simulated clusters do, on average, contain a small amount of SN products. To assess the heavy-metal enrichment in our simulated clusters, Fig \ref{fig:FeH} shows the mean [Fe/H] abundance ratio for cluster stars within $2 r_{1/2}$ of the centre of each cluster. The abundances have been shifted by the initial galactic ISM composition to highlight the enrichment by heavy elements, or the lack thereof, compared to the initial composition. All the analysed clusters have been enriched in total by less than \mbox{0.01 dex} in iron compared to the unenriched ISM. The majority of the stars in all clusters have formed out of an ISM composition that does not include any SN products produced during the simulation. The mean [Fe/H] enrichment for most of the low-mass clusters is practically zero even when essentially all of the clusters contain some form of feedback material, because the stellar wind composition is never significantly enriched in heavy elements. 

The SN-material is concentrated in up to 1\% of the stars in each cluster. These stars can have highly enhanced [Fe/H] values of up to \mbox{1.4 dex} due to limitations of the SN-injection method that does not include diffusion of metals between particles. Due to these SN-enriched stars, the total [Fe/H] standard deviation of the stars in a given cluster is up to \mbox{0.07 dex} (not shown in Fig. \ref{fig:FeH} for clarity), even though the mean [Fe/H] is always very low. For comparison, the maximum star-to-star spread is in observed iron-normal GCs $\Delta[\mathrm{Fe/H}]\sim 0.2$--$0.3$ \citep{2023A&A...669A..19L, 2023ApJ...958...31M}.  As noted above, the [Fe/H] spread in our simulated clusters is driven by low levels of SN-enrichment. The total star-to-star [Fe/H] spread from winds alone reaches at most \mbox{$\Delta[\mathrm{Fe/H}]\sim0.01$ dex} with respect to the unenriched ISM and the 99-percentile of $\Delta[\mathrm{Fe/H}]$ from wind-enrichment is within \mbox{$\sim0.005$ dex} in all clusters more massive than $10^3$ \mdot{}.

\begin{figure}
\includegraphics[width=\columnwidth]{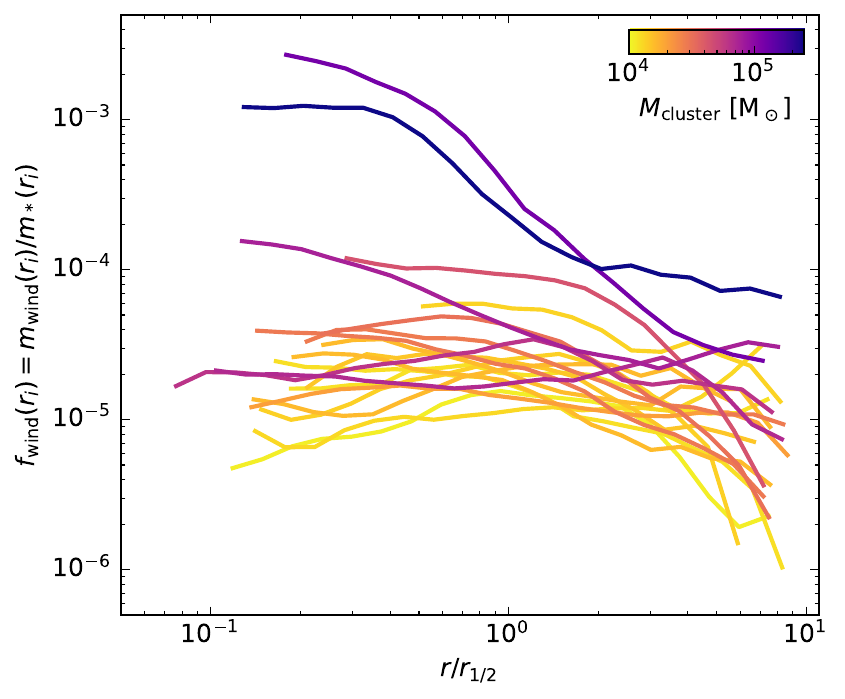}
\caption{The radially binned wind-mass fraction in bound cluster stars in all clusters more massive than $10^4$ \mdot. The fraction $f_\mathrm{wind}(r_i)$ is computed as the ratio between the wind mass and the stellar mass in bin $r_i$. The radii are scaled by the half-mass radii of the clusters and lines are colour-coded by cluster-mass. Clusters up to a few \mbox{$10^4$ \mdot{}} have relatively flat wind-mass fraction profiles, while more massive clusters show increasingly centrally concentrated enrichment profiles. \label{fig:radial_enr}}
\end{figure}

\subsubsection{Radial distribution of enriched stars}

Fig. \ref{fig:radial_enr} shows the azimuthally averaged radial profile of the star-by-star wind-mass fraction in clusters more massive than \mbox{$10^4$ \mdot}. The profiles show a transition from a flat wind-mass fraction profile at low cluster-masses to a centrally concentrated profile as the cluster-mass approaches $10^5$ \mdot. The average central wind-mass fractions are up to two orders of magnitude enhanced compared to the outer regions of the clusters. \citet{2021ApJ...922L...3L} argued that the efficiency of wind-recycling was essentially determined by the gas-density in the star-forming regions, which governs the ability of winds to be trapped while the star formation continues, possibly for multiple free-fall times. Our centrally enhanced wind-mass profiles may be the result of prolonged star formation histories in the central regions of the most massive clusters as indicated in Fig. \ref{fig:cluster_SFR}. However, with our locally limited IMF, the most massive stars are formed in the central regions of the most massive clusters, hence they also deposit their winds in those locations.

\begin{figure*}
\includegraphics[width=0.9\textwidth]{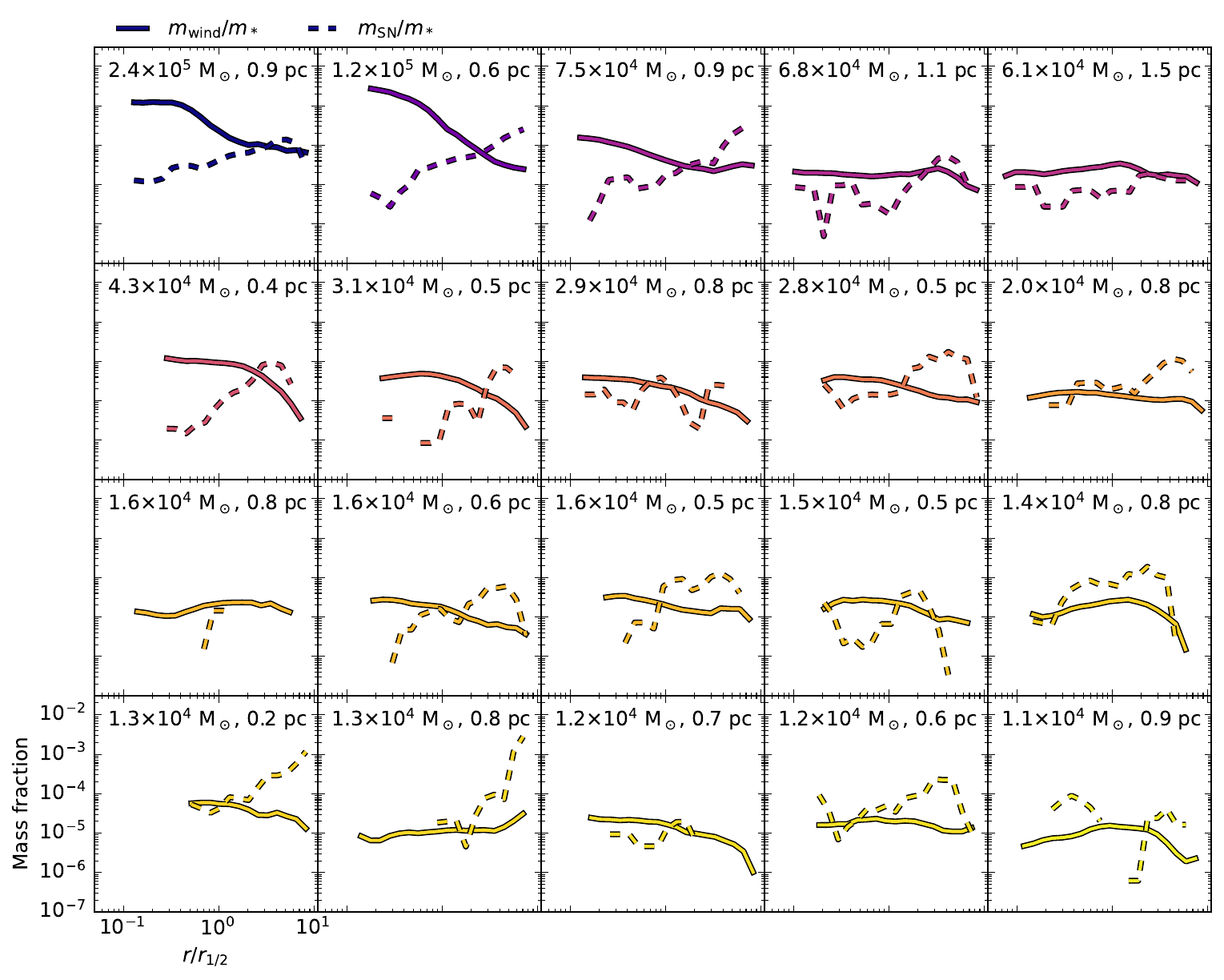}
\caption{The radial wind (solid) and SN (dashed) mass fractions of the bound cluster stars in the 20 most massive clusters. The lines use the same colour scale as in Fig.  \ref{fig:radial_enr} and the cluster-mass and half-mass radius $r_{1/2}$ are indicated at the top of each panel. In most of the clusters the radial SN-mass fraction is flat or increases with radius, as opposed to the wind-mass fraction which shows increasingly centrally concentrated profiles toward higher cluster-masses. The wind-mass fraction and the SN-mass fraction are often anti-correlated in clusters that have centrally concentrated wind-mass profiles. \label{fig:stamps}}
\end{figure*}

Fig. \ref{fig:stamps} shows a comparison between the radial wind-mass fraction and SN-mass fraction in the 20 most massive clusters from Fig. \ref{fig:radial_enr}. The wind material that originates from self-enrichment is often centrally concentrated or flat because star formation occurs preferentially in cluster centres during the late stages of cluster assembly and because the central region can capture the ejected wind material. On the other hand, the SN-material that does not have an origin within the clusters themselves is typically either flat or centrally depressed. The most centrally concentrated wind fraction mass profiles also coincide with centrally depressed SN-mass fraction profiles. 

In observed GCs, the chemically peculiar stars are often \citep{2019ApJ...884L..24D} but not always \citep{2023MNRAS.520.1456L} found to be more centrally concentrated than the unenriched stars. Central concentration should be a natural outcome of the self-enrichment scenario if the enriched stars form and release their winds in the deepest parts of the gravitational well of the forming massive cluster. The central radii of a massive star cluster could also be expected to be the only region able to retain the enriched material in the presence of the energetic massive stars that produce the enriched stellar winds. The enrichment properties of our simulated clusters seem to agree with the observations that show a centrally concentrated second population.

\begin{figure}
\includegraphics[width=\columnwidth]{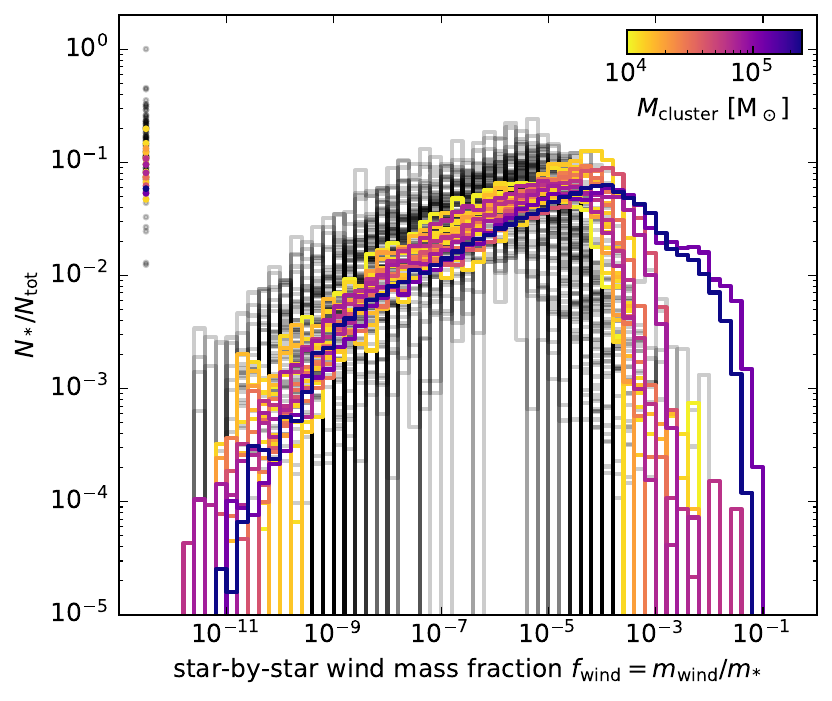}
\caption{The mass fraction of wind-material captured in bound cluster stars, measured within $2r_\mathrm{1/2}$ in clusters more massive than $10^3$ \mdot. Stars with very little ($<10^{-12}$) or no wind-material have been collected in a set of points at $10^{-12}$. Clusters more massive than $10^4$ \mdot{} have been colour-coded as in Fig. \ref{fig:cluster_SFR}. The peak of the $f_\mathrm{wind}$ distribution and the largest $f_\mathrm{wind}$ value per cluster shift to higher values with increasing cluster-mass. \label{fig:wind_fraction}}
\end{figure}

\subsubsection{Chemical abundances of enriched populations of stars}

A detailed look at the wind-material across cluster stars is given in Fig. \ref{fig:wind_fraction} where we show the distribution of stellar wind-mass fractions $f_\mathrm{wind}$ in stars within $2r_\mathrm{1/2}$ of each cluster. As reflected in the wind-mass fractions in Fig. \ref{fig:mean_enr}, the peak of the wind-mass distribution shifts toward higher $f_\mathrm{wind}$ with increasing cluster-mass. Consequently, the most wind-enriched stars in each cluster also shift towards higher $f_\mathrm{wind}$, with the most enriched stars consisting of 10\% of wind-material. The width of the distribution increases with cluster-mass, which we will discuss in more detail below. In the massive clusters, typically 10\% of the stars have not been enriched by winds at all. The fraction of unenriched stars increases with decreasing cluster-mass, as lower-mass clusters host on average lower-mass stars that evolve slower and release less winds compared to the most massive stars in the more massive clusters.

\begin{figure*}
\includegraphics[width=0.9\textwidth]{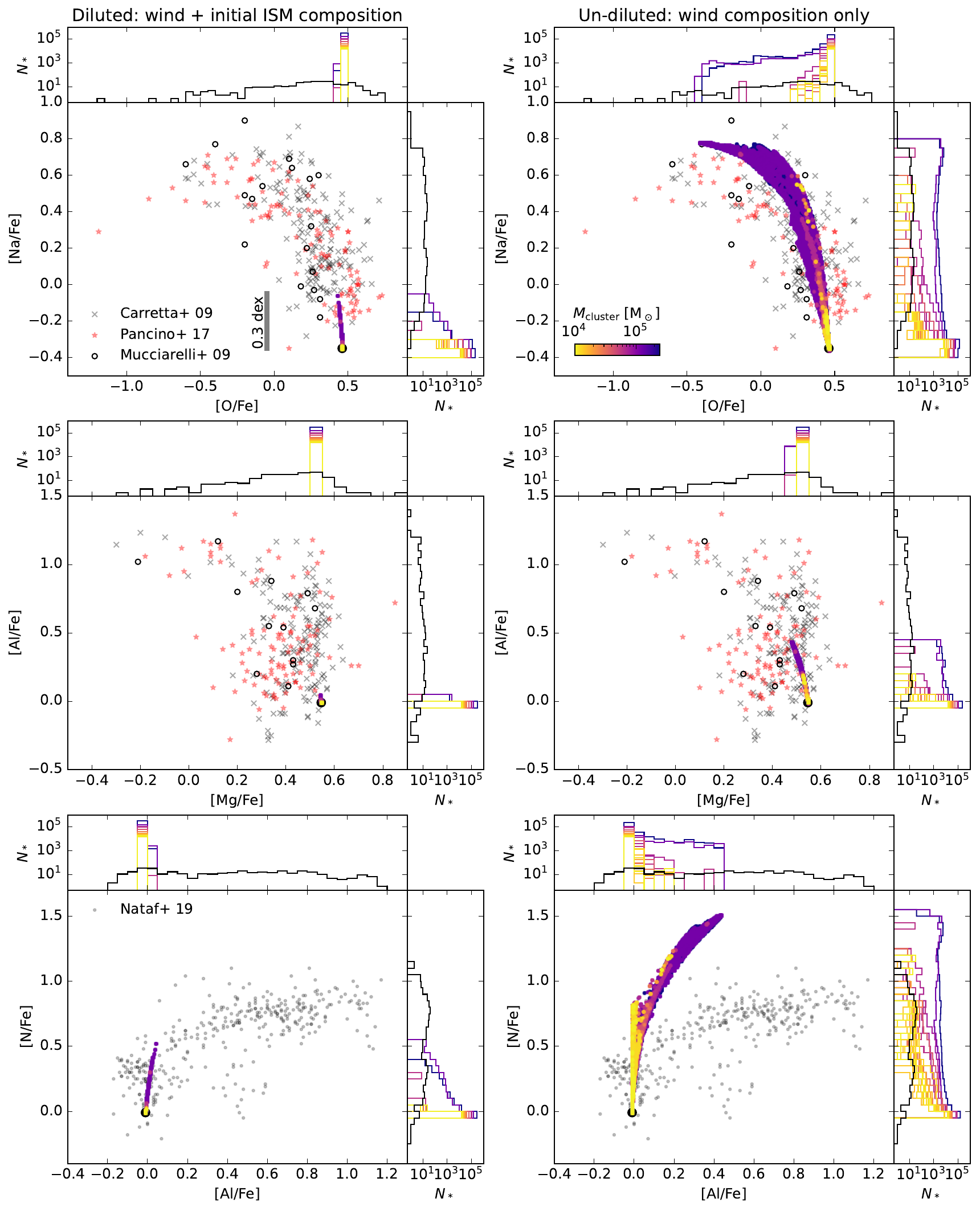}
\caption{The correlation between various abundance ratios in stars bound to clusters more massive than $10^4$ \mdot{} measured within $2r_\mathrm{1/2}$. The left column shows the diluted abundance ratios measured as the mix of stellar wind and low-metallicity ISM out of which each star formed. The right column shows the abundance ratios for the recycled wind-material only (see Fig. \ref{fig:tracks}), to showcase the most extreme spread that would be obtained if the majority of the mass of a star was composed of the wind-mixture. The histograms show the distribution of stars with each abundance ratio in 0.1 dex bins. The stars and the histograms have been colour-coded by the total mass of the host cluster. The black dot indicates the zero enrichment composition, as in Fig. \ref{fig:tracks}. The observed data include Galactic GC stars from \citet[gray crosses]{2009A&A...505..139C} and \citet[red stars]{2017A&A...601A.112P} and LMC GC stars from  \citet[empy circles]{2009ApJ...695L.134M} in the two uppermost rows, and from \citet[and references therein]{2019AJ....158...14N} for the bottom row. All observed data have been limited to clusters with $[\mathrm{Fe}/\mathrm{H}]<-0.9$, and in the bottom row to stars with signal-to-noise greater than 100. The 0.3 dex bar in the top left panel shows the typical [Na/Fe] threshold for stars considered to be part of the chemically peculiar 2P, which we reach in the most wind-enriched stars}. \label{fig:abundances}
\end{figure*}

Next, we inspect the variation of abundances often used to identify MPs in GCs: Na, O, Mg, Al and N.
Fig. \ref{fig:abundances} shows the abundance ratio distribution in [Na/Fe], [O/Fe], [Al/Fe], [Mg/Fe] and [N/Fe] in stars of the most massive clusters ($>10^4$ \mdot). The variations in these abundances are dominated in the simulated clusters by wind-enrichment: stellar winds are the source of at least $95\%$ of the recycled mass in all of these elements. We compute the \textit{diluted} composition of cluster-stars (left column) as given by the mixture of wind-material and ISM, inherited from the gas out of which the star was born. Since observed GC-stars are often unevolved low-mass stars, we do not here include the internal chemical evolution of the stars themselves. The diluted abundance ratios reflect the chemical composition of each star when they formed, hence any change from the initial pristine value will show the level of chemical pollution in the star-forming environment. The \textit{un-diluted} values (right column) exclude the ISM component locked in the stars, and thus shows only the composition of the recycled wind-mixture (see Fig. \ref{fig:tracks}). The un-diluted mixture would represent the most extreme chemical variations if the stars were 100\% made of such material at formation. Because the enrichment of these elements is driven by the stellar winds, we can safely exclude the SN-enriched stars whose chemical contents are artificially enhanced due to the discrete SN injection scheme and the lack of metal diffusion discussed in Sections \ref{section:feedback} and \ref{section:fe_per_h}. This excludes up to 1\% of both the stars and the wind-mass locked in those SN-enriched stars in each clusters.

We compare the wind-enriched star-to-star abundance ratios to single stars observed in Galactic GCs with detected light-element spreads. The data are from \citet{2009A&A...505..139C} and \citet{2017A&A...601A.112P} for [Na/Fe], [O/Fe], [Al/Fe] and [Mg/Fe] in the top two rows, and from \citet{2019AJ....158...14N} for [N/Fe] and [Al/Fe] in the bottom row. In the absence of large data sets for GCs in dwarf galaxies, we selected Galactic GCs with metallicity below $[\mathrm{Fe/H}]<-0.9$ to match the low metallicity of the simulated system. In addition, in the top two rows of Fig. \ref{fig:abundances}, we also show a few single stars observed in three GCs that host MPs in the Large Magellanic Clouds from \citet{2009ApJ...695L.134M}. According to these observations, old dwarf galaxy GCs seem to host similar star-to-star spreads in [Na/Fe]-[O/Fe] and [Al/Fe]-[Mg/Fe] as unambiguously observed in the low-metallicity GCs of the more massive Milky Way. Integrated-light measurements of dwarf galaxy GCs \citep{2018A&A...613A..56L} also indicate cluster-to-cluster abundance variations and differences between clusters and field stars, strengthening the possible universality of MPs in low-metallicity GCs. We show the observed data in both columns of Fig. \ref{fig:abundances} for reference, however the diluted measurements of the simulated stars (left column) offer a more direct comparison.

In the right hand column of Fig. \ref{fig:abundances}, the pure wind abundances show a strong anticorrelation between the enhanced [Na/Fe] and depleted [O/Fe], typical for nuclear reactions taking place at high temperatures as shown in Fig. \ref{fig:tracks}, and similar to the abundance ratios as measured in GCs. [N/Fe] is also strongly enhanced. However, when the stellar wind abundances are mixed with the ISM to form the bulk of the stellar mass within the cluster (left column), the spread in abundance ratios narrows down. The diluted [Na/Fe] and [N/Fe] are still enhanced by 0.3--0.5 dex. The enhanced [N/Fe] that persists even after dilution is particularly interesting in relation to recent observations of extreme star forming objects at higher redshifts \citep{2023ApJ...957...77P, 2023A&A...677A..88B}. Such N-enhanced systems have been suggested to represent young GCs observed during their formation \citep{2023A&A...673L...7C, 2023arXiv230304179S, 2024A&A...681A..30M}.

The [O/Fe] and [Mg/Fe] depletion and the [Al/Fe] enhancement, on the other hand, are almost completely washed out after dilution with the ISM. Even if the stellar wind is strongly enriched in light elements as shown on the right of Fig. \ref{fig:abundances}, the mass recycled in the cluster stars is not enough to produce a significant spread in abundance ratios, and falls short of the extent observed in ancient GCs. The 90\% of mass at unenriched ISM composition dilutes the enriched abundances, resulting in only a modest spread in abundances even in the most enriched, massive (\mbox{$>10^4$ \mdot}) clusters. This is a manifestation of the so-called mass-budget problem. As a possible caveat, we note that the stellar wind injection method injects the mass into existing gas particles. Hence, we always end up diluting the wind material by some amount. As a result, the absolute spread in abundance ratios can be viewed as a lower limit. For complementary examples of the upper limit abundance ratios in wind-material released and retained in stellar populations, accounting for radiation and semi-analytic hydrodynamics with no dilution, we refer the reader to \citet{2019ApJ...871...20S}. Having the possibility of stars made 100\% of wind material would increase the number of extremely peculiar stars, possibly even up to the extreme values shown on the right of Fig. \ref{fig:abundances}. Such redistribution of the wind mass would however not change the fact that the clusters return only up to a couple per cent of their mass in stellar winds, that is up to a few $10^3$ \mdot{} for a cluster with initial mass of $10^5$ \mdot.

The extent of the abundance ratio (anti-)correlations increases with cluster-mass in terms of absolute spread and standard deviation, as shown in the histograms of Fig. \ref{fig:abundances}, following Figs. \ref{fig:mean_enr} and \ref{fig:wind_fraction} where the increasing cluster-mass resulted in increasingly high wind-mass fractions. These correlations with cluster-mass are in qualitative agreement with observed GCs, where the maximum spread and standard deviation of light element abundances increase with the present-day GC mass \citep{2017A&A...601A.112P, 2019AJ....158...14N}. We will discuss the distribution of enriched stars star-by-star in the context of GCs in more detail below.  

\section{Discussion}\label{section:discussion}

\subsection{Implications for GC formation}

The simulated star clusters form with up to \mbox{$2\times10^5$ \mdot{}} of stellar mass, corresponding to the lower mass range of evolved GCs. In the simulated clusters, the chemically peculiar stars that differ significantly from the unenriched population in Fig. \ref{fig:abundances} comprise a minority of the stars. This is in contrast to observed ancient GCs where the peculiar 2P stars start dominating between present-day cluster-masses of a few \mbox{$10^4$ \mdot{}} and \mbox{$10^5$ \mdot}. What would it take to reach a dominant, strongly enriched population of stars, based on the results here so far? Using the nomenclature of observed GCs, we will refer to the enriched population of stars as the second population, 2P, and the unenriched population as the first population, 1P. The defining factor between 1P and 2P is traditionally based on the separation between the populations in the abundance ratio plane, e.g. as a threshold in [Na/Fe] and [O/Fe]. As can be seen in Fig. \ref{fig:abundances}, the stars that contain 10\% of wind-material are already quite separated from the unenriched composition and almost reach the typical 2P limit of $[\mathrm{Na/Fe]}\sim+0.3$ dex \citep{2009A&A...505..139C, 2017A&A...601A.112P} compared to the unenriched population. Based on the most extreme stars in Fig. \ref{fig:abundances}, we can conclude that the stars must contain at least $\sim10\%$ of enriched wind-material to be considered part of the 2P. The actual extent of the 2P across the abundance ratio plane of course depends on the detailed combination of abundances and ISM mixture from which the stars form, however we use here the wind-mass fraction as the first order proxy for enrichment when extrapolating our results to higher cluster-masses.

\begin{figure*}
\includegraphics[width=0.9\textwidth]{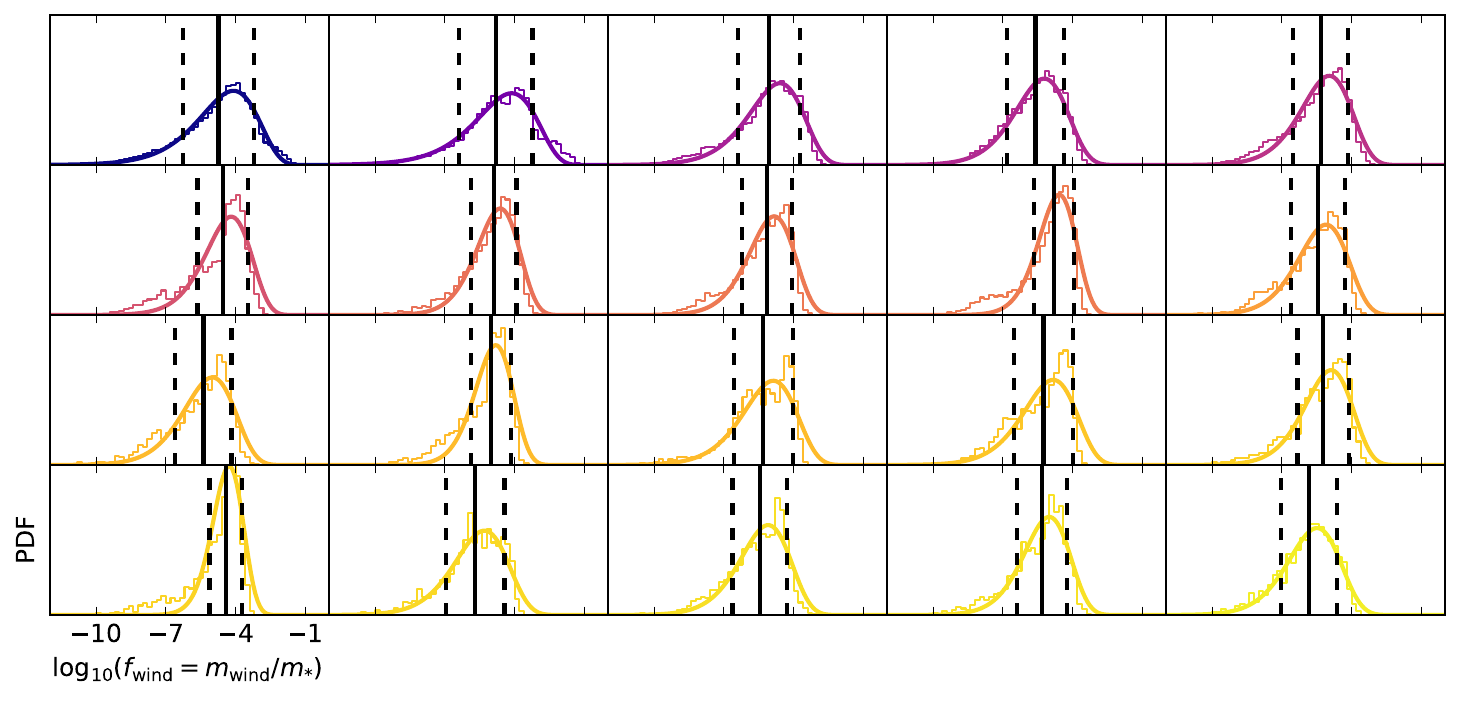}
\caption{The PDF and best-fit log-normal PDF (Eq. \ref{eq:log-normal}) of the wind-mass fraction in 20 most massive clusters, colour-coded as in Fig. \ref{fig:cluster_SFR}. The black solid and dashed vertical lines show the arithmetic mean and standard deviation of the best-fit profiles (Eqs. \ref{eq:mean} and \ref{eq:std}). \label{fig:fraction_stamps}}
\end{figure*}

\begin{figure*}
\includegraphics[width=\textwidth]{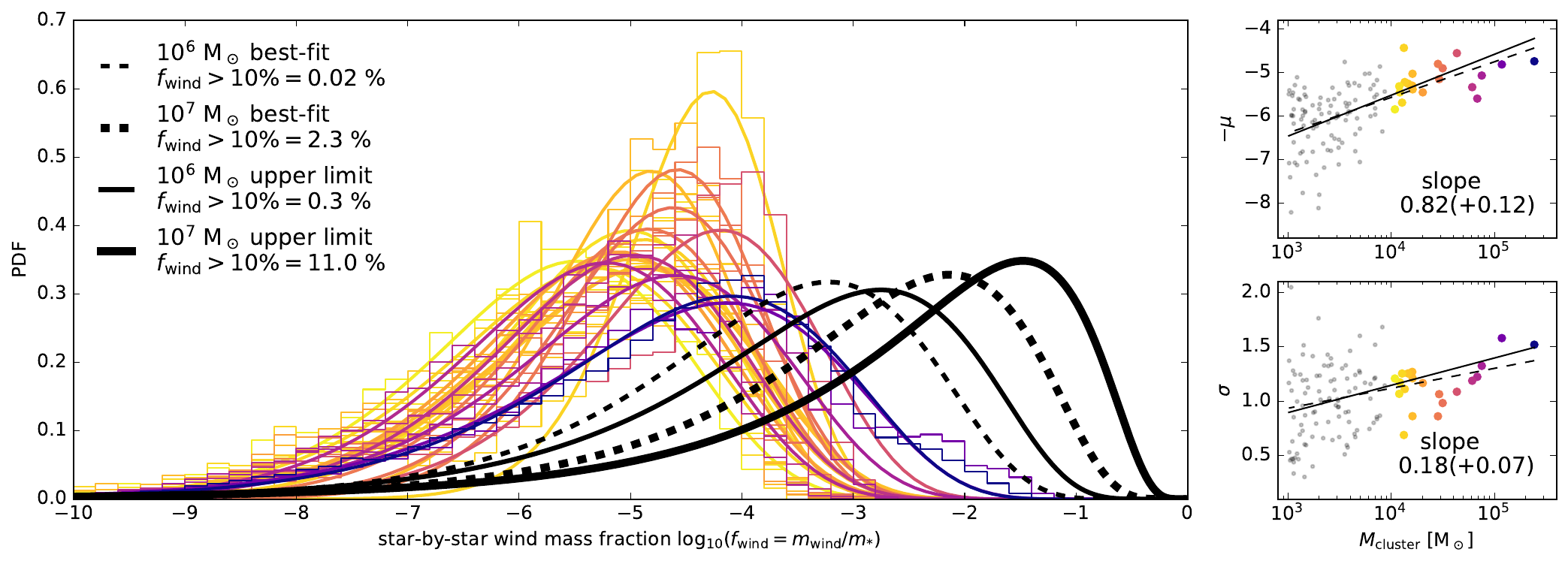}
\caption{\textit{Left:} the probability density function of the stellar mass composed of wind-material in stars of clusters more massive than $10^4$ \mdot{} within $2\times r_{1/2}$, as in Figs. \ref{fig:wind_fraction} and \ref{fig:fraction_stamps}. We only show the most massive clusters for clarity. The best-fit log-normal distributions (Eq. \ref{eq:log-normal}), using inverted $\log_{10}(f_\mathrm{wind})$ as the x-axis and zero point at $f_\mathrm{wind}=100\%$, are shown with the same colour coding as the simulated cluster data from Fig. \ref{fig:cluster_SFR}. \textit{Right:} the arithmetic mean $-\mu$ and standard deviation $\sigma$ (Eqs. \ref{eq:mean} and \ref{eq:std}) of the log-normal fits to $f_\mathrm{wind}$ in clusters more massive than $10^3$ \mdot, as well as the power-law scaling relations fit to $-\mu$ and $\sigma$. The lines show the best-fit power-law slopes ($\propto M_\mathrm{cluster}^\alpha$, dashed) and the slope upper limits (slope plus bootstrapped standard deviation, solid) , with values indicated in the panels. The black curves in the left panel show log-normal profiles extrapolated according to the best-fit (dashed) and the upper limit (solid) parameter-scaling in the right hand panels for clusters with mass of \mbox{$10^6$ \mdot{}} (thin) and \mbox{$10^7$ \mdot{}} (thick). The legend indicates the fraction of stars composed of more than 10\% of wind-material in the extrapolated distributions. \label{fig:fraction_extra}}
\end{figure*}

As an illustrative experiment, we fit the star-by-star wind-mass fraction distribution in Fig. \ref{fig:wind_fraction} with a distribution function and extrapolate the fit parameters to higher star cluster-masses. Fig. \ref{fig:fraction_stamps} shows the data in Fig. \ref{fig:wind_fraction} for 20 most massive clusters, this time with linear y-axis normalized to the area below each distribution to obtain the probability density function (PDF). Because the massive stars release most of their increasingly enriched winds toward the end of the cluster formation sequence, they result in a skewed distribution toward enriched stars with a tapered cut-off at high values of $f_\mathrm{wind}$ and a longer tail of low values of $f_\mathrm{wind}$. The distributions in Fig. \ref{fig:fraction_stamps} resemble a reversed log-normal or a gamma-distribution that have conveniently a natural lower limit of zero, corresponding to the upper limit of the wind-mass fraction of $\log_{10}(f_\mathrm{wind}=100\%)=0$ in log-10 scale. After experimenting with fitting the profiles with both distribution functions, the log-normal shape seems to yield more robust results across the cluster population. When we reverse $f_\mathrm{wind}$ and define $x=-\log_{10}(f_\mathrm{wind})$\footnote{The log-normal distribution is only defined for positive values.}, we can fit the wind-mass fraction PDFs with the log-normal distribution function \citep{univariate_distributions}
\begin{equation}\label{eq:log-normal}
    \frac{1}{ \sqrt{2\pi} \sigma_\mathrm{ln} x} \exp \left( -\frac{(\ln{x}-\mu_\mathrm{ln})^2}{2\sigma_\mathrm{ln}^2}\right)
\end{equation}
where $\mu_\mathrm{ln}$ and $\sigma_\mathrm{ln}$ are the mean and standard deviation of the natural logarithm of $x$. The arithmetic mean (expected value) and standard deviation of the log-normal distribution are then given by
\begin{equation}\label{eq:mean}
    \mu = e^{\mu_\mathrm{ln} +\frac{1}{2}\sigma_\mathrm{ln}^2}
\end{equation}
and
\begin{equation}\label{eq:std}
    \sigma = e^{\mu_\mathrm{ln} +\frac{1}{2}\sigma_\mathrm{ln}^2}\sqrt{e^{\sigma_\mathrm{ln}^2}-1} = \mu\sqrt{e^{\sigma_\mathrm{ln}^2}-1}.
\end{equation}

The best-fit log-normal distributions of wind-mass fraction locked in our simulated cluster stars are shown in Fig. \ref{fig:fraction_stamps}, for the 20 most massive clusters. The x-axis is shown in log-10 scale for simplicity, and the log-normal profiles have been reversed back by multiplying the x-axis by $-1$. The corresponding values of mean and standard deviation are indicated as well.
The $f_\mathrm{wind}$ profiles show, for the most parts, a good match to a log-normal distribution. In some cases, the distributions deviate in the tails, while some of the profiles have a more peaked distribution than the best-fit log-normal. The final profile is the convolution between the star formation and enrichment histories in the cluster forming region. Even though the SFRs (Fig. \ref{fig:cluster_SFR}) are often gaussian-like, sub-cluster mergers or irregular patterns in the star formation and enrichment histories may cause irregularities in the $f_\mathrm{wind}$ distribution.

Because the trend between the $f_\mathrm{wind}$ shape and the cluster-mass is perhaps not obvious in the individual panels of \ref{fig:fraction_stamps}, we show in Fig. \ref{fig:fraction_extra} all the $f_\mathrm{wind}$ profiles of clusters more massive than $10^4$ \mdot, together with their best-fit log-normal distributions. As seen already in Fig. \ref{fig:wind_fraction}, the distributions in Fig. \ref{fig:fraction_stamps} tend to shift to higher values of $f_\mathrm{wind}$ and broaden with increasing cluster-mass. To quantify the scaling of the $f_\mathrm{wind}$-distribution with cluster-mass, we also show the correlation of the arithmetic mean ($-\mu$) and standard deviation ($\sigma$) of the log-normal distributions with cluster-mass for all massive clusters in the right hand side of Fig. \ref{fig:fraction_extra}. We fit the log-normal parameters with a power-law relation between $-\mu$ and $\sigma$ against cluster-mass, and estimate the standard deviation of the best-fit power-law slopes by bootstrapping 1000 times. The mean and standard deviation of the $f_\mathrm{wind}$-distributions increase with cluster-mass with power-law slopes of $0.82\pm 0.12$ and $0.18\pm 0.07$, respectively. In the following, we use these power-law scaling relations to extrapolate the distribution of $f_\mathrm{wind}$ to higher cluster-masses.

Using the power-law fits to the parameters of the log-normal distributions shown in Fig. \ref{fig:fraction_extra}, we first extrapolate the wind-mass fraction distribution to \mbox{$10^6$ \mdot{}} and \mbox{$10^7$ \mdot{}} clusters. The extrapolated distributions are shown in Fig. \ref{fig:fraction_extra} with black, with thin lines for \mbox{$10^6$ \mdot{}} and thick lines for \mbox{$10^7$ \mdot} clusters, respectively. The dashed black lines show the distribution extrapolated using the best-fit relation of $\mu$ and $\sigma$, and the solid lines show the upper limit scenario, extrapolated according to the best-fit plus its standard deviation (as given in the legend of the right hand panels of Fig. \ref{fig:fraction_extra}). The legend in the left hand panel of Fig. \ref{fig:fraction_extra} indicates the fraction of stars consisting of more than 10\% of wind-material, obtained by integrating over the the extrapolated log-normal distributions. These stars are considered as the 2P of the extrapolated clusters, based on the most enriched stars that are on the verge of reaching the traditional 2P limit in the simulated clusters shown in Fig. \ref{fig:abundances}.

Assuming that the stellar wind released and captured in cluster stars scales with cluster-mass as given by the distribution of star-by-star wind-mass fractions, the conservative extrapolation would result in only up to $\sim0.02\%$ and $\sim2.3\%$ of strongly enriched stars in $10^6$ \mdot{} and $10^7$ \mdot{} clusters, respectively. The picture changes slightly when we consider the extrapolated distributions using the upper limits of $-\mu$ and $\sigma$: the \mbox{$10^6$ \mdot{}} and \mbox{$10^7$ \mdot{}} clusters would have $\sim0.3\%$ and $\sim11 \%$ of strongly enriched stars. We are therefore still not able to produce a cluster with a dominant 2P as observed in many GCs, even after extrapolating the enrichment properties of our simulated cluster to almost two orders of magnitude higher cluster masses.

\begin{figure}
\includegraphics[width=\columnwidth]{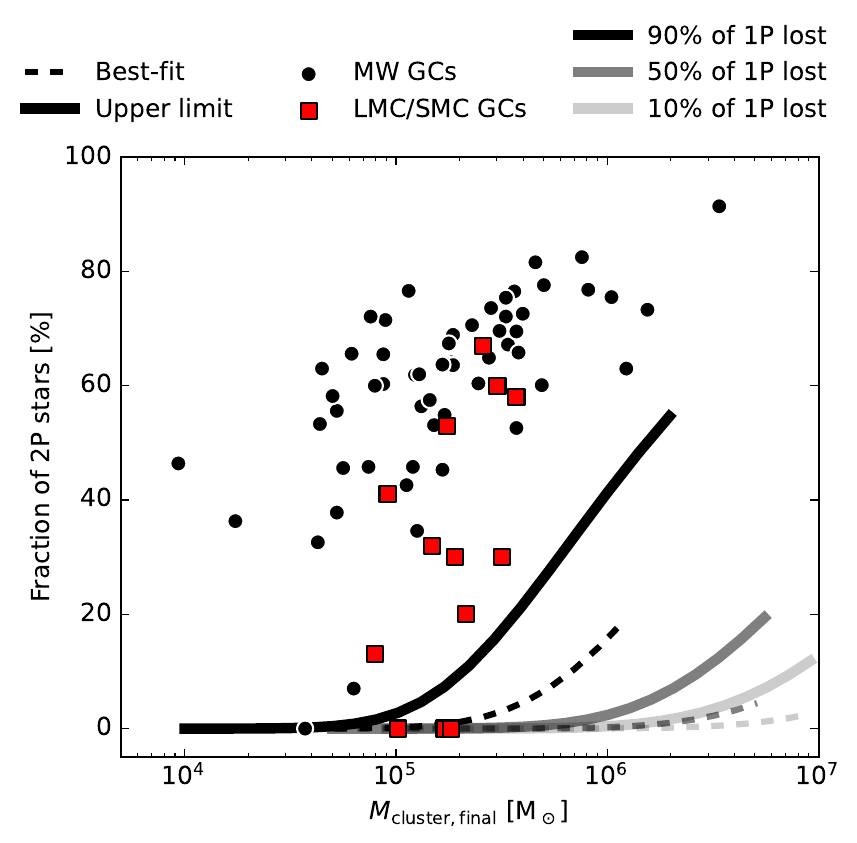}
\caption{The fraction of 2P stars as a function of the final cluster-mass, using the mass-dependent scaling of the log-normal parameters obtained in Fig. \ref{fig:fraction_extra} and factoring in the loss of stars due to cluster evolution in a galactic tidal field. The extrapolated relations assume that either $90\%$, $50\%$ or $0\%$ (from darkest to lightest opacity) of the 1P stars are lost. The dashed lines show the 2P fraction obtained by integrating over the best-fit log-normal wind-mass fraction distributions, and the solid lines show the upper limit using the best-fit parameters plus standard deviation, as illustrated in Fig. \ref{fig:fraction_extra}. The observed Milky Way and Magellanic GCs are from Tables 7, 8 and 9 in \citet{2019A&ARv..27....8G}, see text for details. \label{fig:2P_fraction}}
\end{figure}

\subsection{Evolution of 2P fraction due to GC mass-loss}

A $\sim10\%$ 2P fraction is low compared to the typical values in the Local Group where $10^6$--$10^7$ \mdot{} GCs show dominant 2P fractions of up to $90\%$. However, the 2P fractions estimated above describe the initial division between 1P and 2P stars immediately after the clusters have formed. Because the GCs observed in the Local Group are very old and orbit in a galactic potential, they must have lost some stars throughout their evolution in the galactic environment \citep{2010MNRAS.409..305L, 2011MNRAS.413.2297S, 2019MNRAS.482.5138B, 2016MNRAS.458.1450W, 2024MNRAS.527.2765M, 2024A&A...681A..45L}. The fraction of mass lost depends on both the initial properties of the cluster (mass, density profile, IMF) and the galactic tidal field, and estimates for the mass-loss of Milky Way GCs extend from half to more than 90\% of the initial GC mass \citep{2015MNRAS.453.3278W, 2019MNRAS.482.5138B}. Some of the mass-loss is due to stellar evolution, and the rest is caused by stars escaping the cluster. A detailed simulation of massive clusters through Gyrs of evolution is beyond the scope of the current study, especially since such modelling would need to account for the tidal field of the Milky Way in addition to that of our initial dwarf galaxy system. We instead approximate the \textit{evolved} 2P fraction in the extrapolated clusters by parametrizing the loss of stars over time. We assume that a fraction of the cluster stars are lost throughout the evolution of the clusters, with 1P dominating the mass-loss. This assumption is based on observations of the spatial distribution of the populations and our simulated radial profiles of enriched stars in Fig. \ref{fig:radial_enr}, wherein the 2P stars are often more centrally concentrated within GCs and therefore less susceptible to be lost to the tidal field. To compute the final extrapolated 2P fraction we use
\begin{equation}
    f_\mathrm{2P,f} = \frac{N_\mathrm{2P,f}}{N_\mathrm{1P,f}+N_\mathrm{2P,f}}=\frac{N_\mathrm{2P,i}}{\delta_\mathrm{1P}(1-N_\mathrm{2P,i})+N_\mathrm{2P,i}},
\end{equation}
assuming that the number of 2P stars initially $N_\mathrm{2P,i}$ does not evolve, i.e. $N_\mathrm{2P,f}=N_\mathrm{2P,i}$, while a fraction $\delta_\mathrm{1P}$ of the initial 1P stars $N_\mathrm{1P,i}$ is lost and results in the final 1P fraction of $N_\mathrm{1P,f}=\delta_\mathrm{1P}N_\mathrm{1P,i}$. We assume only two populations, therefore $N_\mathrm{1P,i}=1-N_\mathrm{2P,i}$. To gauge the range of $f_\mathrm{2P,f}$, we vary $\delta_\mathrm{1P}$ from zero to $90\%$, with $\delta_\mathrm{1P}=0$ equal to no evolution. For simplicity, we define the final cluster-mass as 
\begin{equation}
M_\mathrm{cluster,final}=\frac{N_\mathrm{1P,f}+N_\mathrm{2P,f}}{N_\mathrm{1P,i}+N_\mathrm{2P,i}}M_\mathrm{cluster}, 
\end{equation}
therefore neglecting mass loss through stellar evolution and dynamical effects that depend on stellar mass.

Fig. \ref{fig:2P_fraction} shows the final fraction of 2P stars in our extrapolated log-normal wind-mass fraction distributions obtained in Fig. \ref{fig:fraction_extra}, assuming $\delta_\mathrm{1P}=90\%$, $\delta_\mathrm{1P}=50\%$ or $\delta_\mathrm{1P}=0\%$. We refrain from changing the 2P fraction for simplicity, however since the 1P is always dominant initially, the loss of 2P stars will mostly result in only a minor change in the final 2P fraction. For instance, in the case where $90\%$ of the 1P is lost, the loss of half of the 2P stars would only decrease the final upper limit 2P fraction of a $10^7$ \mdot{} cluster from $\sim55\%$ to $\sim38\%$.  We compare in Fig. \ref{fig:2P_fraction} our extrapolated final 2P fractions to GCs in the Milky Way and in the Magellanic clouds collected in Tables 7, 8 and 9 in \citet{2019A&ARv..27....8G}. The masses and 2P fractions for clusters in the Milky Way are from \citet{2019MNRAS.482.5138B} and \citet{2017MNRAS.464.3636M, 2019MNRAS.487.3239Z}, respectively. The masses of SMC/LMC clusters are from \citet{2003MNRAS.338...85M, 2011AJ....142...36G}, and the 2P fraction are from \citet{2009ApJ...695L.134M, 2014ApJ...793L...6M, 2017MNRAS.464...94N, 2017MNRAS.465.4159N, 2017MNRAS.468.3150M, 2018MNRAS.476..114H, 2018ApJ...853..186Z, 2018MNRAS.477.4696M, 2019MNRAS.484.4718H}. As noted earlier, the observed clusters seem to indicate a correlation between the 2P fraction and cluster-mass, although we should stress that there are notable uncertainties in the observed 2P fractions that are at times based on only a handful of stars.

None of the extrapolated models result in 2P dominated clusters when we use the best-fit log-normal parameters from Fig. \ref{fig:fraction_extra}, even after most of the 1P is lost. Only when using the upper limits of the log-normal parameters are we able to reach a dominant 2P population in clusters with initial mass close to $10^7$ \mdot. Even then, a significant fraction ($\delta_\mathrm{1P}\gg 50\%$) of the 1P has to be lost. Compared to the observed GCs, the increase in the extrapolated 2P fraction occurs more gradually with cluster-mass, and the highest 2P fractions are still 20--30 percentage units too low compared to the typical 2P fractions in the local GC population with similar present-day masses. The strong increase in the 2P fraction of observed GCs therefore occurs at lower cluster-masses than in any of our extrapolated relations. This again demonstrated the mass-budget problem, in addition to the directly simulated modest chemical spreads in Fig. \ref{fig:abundances}: the self-enriching winds of massive stars self-consistently sampled from a standard Kroupa IMF may not alone produce enough material to result in observationally consistent 2P fractions in a low-metallicity starburst environment.

In addition to dynamical loss of stars, the 2P fraction might evolve also due to changes in stellar evolution induced by the different chemical composition of the 1P and the 2P stars. For the low-mass stars still alive in present-day GCs, the helium content affects the duration of the phases of stellar evolution \citep{2015A&A...578A.117C}. Because the chemical pollution from massive stars is produced through central hydrogen burning, the increasing light-element enrichment correlates with increased helium enrichment. \citet{2016A&A...592A.111C} argued that the most helium-enriched 2P stars may finish their nuclear burning faster than their helium-normal counterparts, decreasing the present-day 2P fraction in clusters that start with initially extremely helium-enriched stars. The maximum enhancement of star-to-star helium-mass fraction due to stellar winds in our simulated clusters is $\sim0.02$, not enough to drastically change the evolution of the stars.

\begin{figure}
\includegraphics[width=\columnwidth]{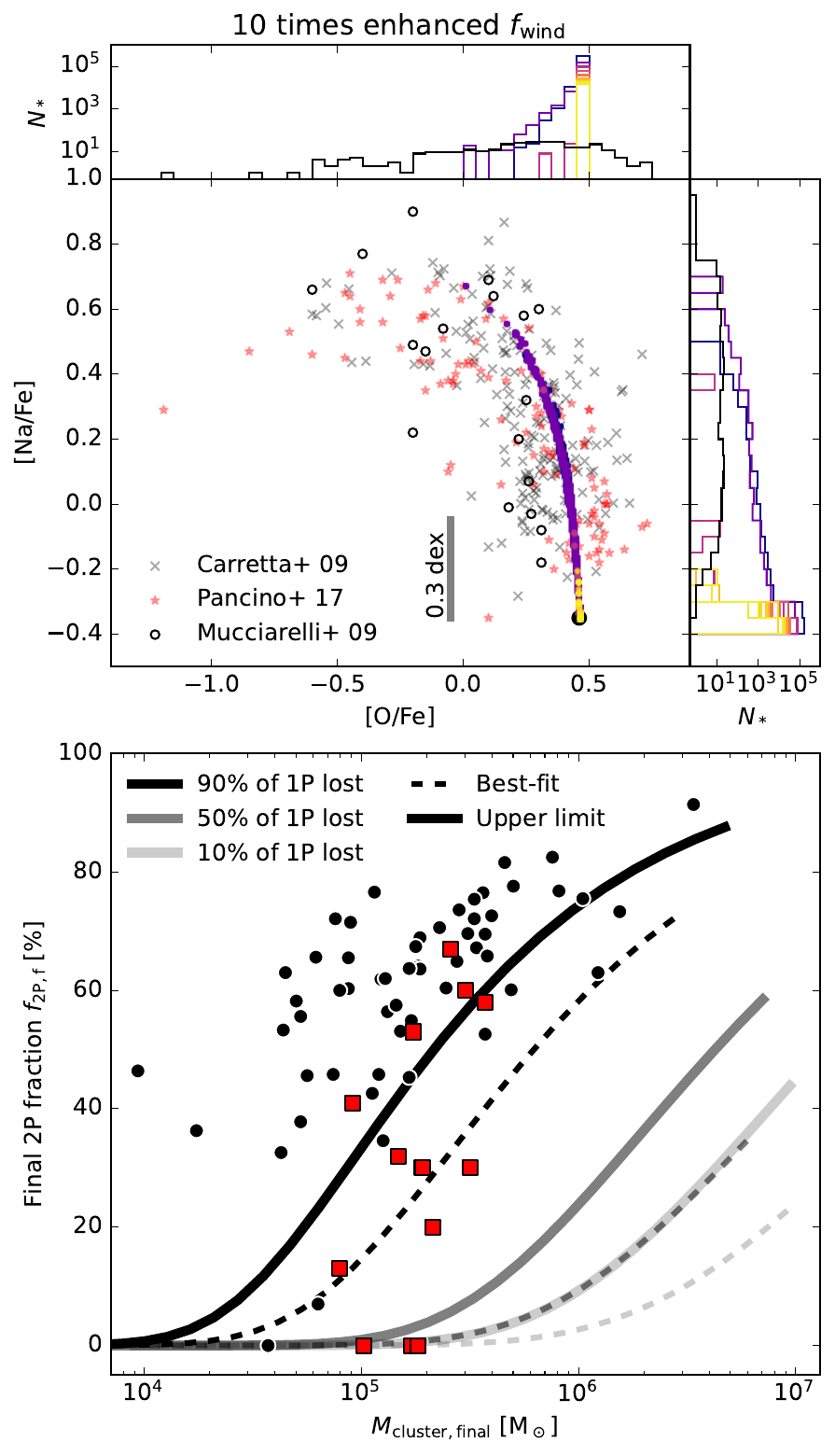}
\caption{The stellar [Na/Fe]-[O/Fe] abundance ratios measured at formation equivalent to those shown in the top left panel of Fig. \ref{fig:abundances} (top) and the final 2P fractions equivalent to Fig. \ref{fig:2P_fraction} (bottom) in an enhanced enrichment scenario: the wind-masses captured in cluster stars used to recover the log-normal fits in Fig. \ref{fig:fraction_extra} have been scaled uniformly up by a factor of ten while keeping the stellar mass the same. Such enhanced wind-output can be expected e.g. from a top-heavy IMF. The colour-coding in the top panel is the same as in Fig. \ref{fig:cluster_SFR} and the observed data points are the same as in the respective Figs. \ref{fig:abundances} and \ref{fig:2P_fraction}. \label{fig:enhanced}}
\end{figure}

\subsection{Avenues to enhance the final 2P fraction}

Our extrapolations are based on the simulated clusters that still, even at $10^5$ \mdot, do not fully realise the input IMF. If all of the winds from a fully realised IMF up to 500 \mdot{} in a very massive cluster would be retained in the 2P, there should potentially be up to a factor of ten more material to be recycled (compare Figs. \ref{fig:theor_output} and \ref{fig:mean_enr}).
A simple exercise of multiplying the wind-material captured in the cluster stars by a factor of two, four and eight results in highest extrapolated 2P fractions of 68\%, 79\% and 86\%, instead of the $55\%$ estimated directly based on the simulated clusters. As an extreme example, we show in the Fig. \ref{fig:enhanced} how the [Na/Fe] abundances and the extrapolated 2P fractions would better agree with the observed distributions if we had ten times more wind-material recycled in the cluster stars. Given ten times more captured wind-material, the extrapolated clusters with initial mass of $10^7$ \mdot{} would host a $73\%$ and $88\%$ 2P fraction based on the the best-fit and upper limit log-normal parameters, at final cluster-mass of \mbox{$2.9\times 10^6$ \mdot{}} and \mbox{$4.8\times10^6$ \mdot}, respectively. In terms of abundance variations, a boost by a factor of ten would increase the largest $f_\mathrm{wind}$ values to $\sim80\%$. Such enhancement in wind-material would shift the most enriched stars up to values in excess of [Na/Fe]$\sim0.6$, resulting in a [Na/Fe] spread of 1 dex that is in agreement with observed GCs. Even then, more than 50\% of the 1P stars would still have to be lost.
 
To reach a dominant 2P we would therefore require more winds to be captured in the cluster stars. As long as the clusters experience mass-loss dominated by the number of 1P stars, the clusters do not necessarily need to be dominated by 2P stars at birth. An initially higher 2P fraction would help in producing a higher final 2P fraction. Based on Fig. \ref{fig:mean_enr}, more massive clusters may be able to retain a larger fraction of the released stellar wind, aiding the formation of strongly wind-enriched stars. Dedicated hydrodynamical simulations with similar methodology as presented here would be required to assess whether the initial fraction of 2P stars in more massive clusters would be even more enhanced than what we are able to infer based on clusters up to $2\times 10^5 $\mdot{} in Fig. \ref{fig:fraction_extra}. 

\begin{figure}
\includegraphics[width=\columnwidth]{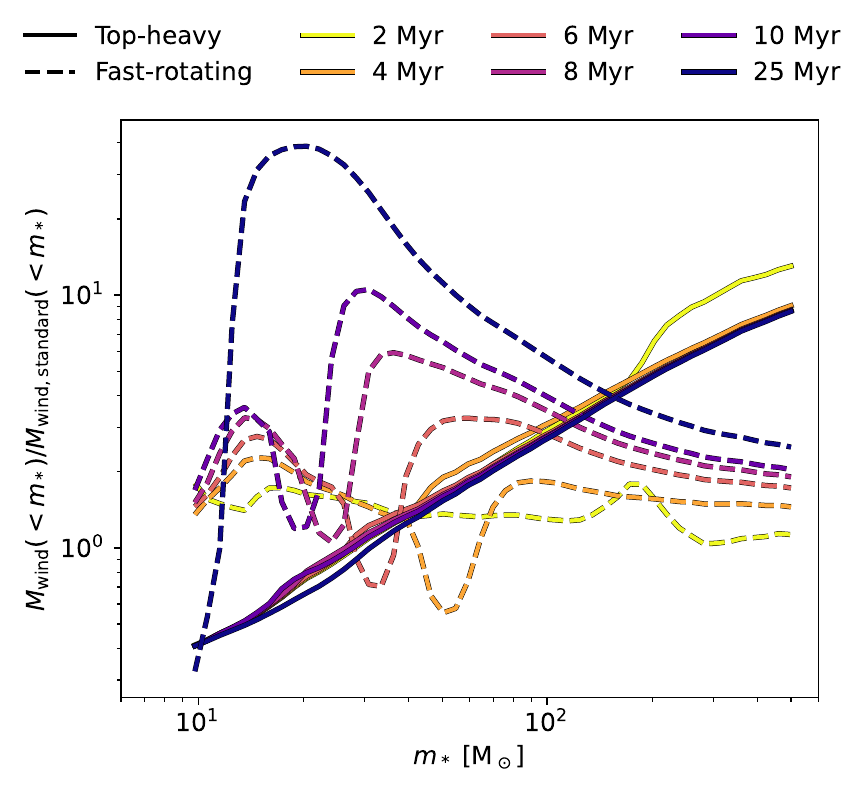}
\caption{The cumulative wind-mass released by a $10^6$ \mdot{} star cluster that either has a top-heavy IMF with high-mass slope $-1.3$ (solid lines) or consists of fast-rotating massive stars that evolve chemically homogeneously (dashed lines). The stellar winds have been integrated using the appropriate slow and fast-rotating BoOST stellar tracks at 0.016 \zdot{} metallicity, see text for details. The wind-masses have been normalized to the output of the standard IMF shown in the bottom right panel of Fig. \ref{fig:theor_output} and the lines are coloured by cluster age up to 25 Myr. The values indicate the mean of a Monte Carlo sample of 10 clusters. \label{fig:relative_wind}}
\end{figure}

The other option is to enhance the wind production, as long as the added enrichment can be captured within the star forming enrivonment. In addition to actually sampling the IMF in full, enhancing the wind production by a top-heavy IMF \citep{2006A&A...458..135P} has been suggested as a particular solution to the mass-budget problem by the semi-analytic computations using similar wind tables to ours presented in \citet{2019ApJ...871...20S}. Other avenues for enhanced enrichment could be the addition of other sources, such as interacting binaries, rapidly rotating stars or supermassive stars, occurring coevally at such short timescales before the clusters are completely vacuated of star-forming gas. 

We show in Fig. \ref{fig:relative_wind} examples of wind production by a top-heavy IMF and fast-rotating, chemically homogeneously evolving stars computed using the BoOST stellar tracks at our fiducial metallicity. For the fast-rotating stars, we use the IZw18CHE tracks from \citet{2022A&A...658A.125S} where the stars have an initial rotation velocity of \mbox{$500$ km s$^{-1}$}. The wind production of the two alternative models is shown relative to the fiducial model shown at the bottom right in Fig. \ref{fig:theor_output}.
The output of the top-heavy IMF is computed by replacing the Kroupa or \citet{1955ApJ...121..161S}-like high-mass slope of $-2.3$ by a relatively extreme $-1.3$.
Such a top-heavy IMF would result in the production of ten times more wind-material, which according to Fig. \ref{fig:2P_fraction} would in theory be enough to result in a dominant 2P after significant cluster mass-loss ($\delta_\mathrm{1P}> 50\%$). Such enhanced wind production would occur already during the first couple of million years, driven by the increased number of very massive stars. 

The integrated stellar wind produced by a population of fast-rotating stars in Fig. \ref{fig:relative_wind} is also enhanced compared to the fiducial BoOST models, in total by a factor of 2--3. The enhancement increases as the cluster ages due to the increased contribution of massive stars in the 10--100 \mdot{} range. Fast-rotating stars with a few tens of \mdot{} contribute tens of times more winds compared to their slowly rotating counterparts. The capture of the extra wind-material released at later ages may require an extended star formation period, or consecutive episodes of star formation similar to the AGB enrichment scenario. The chemical abundances on the surface of the fast-rotating stars may however result in different abundance patterns (as in Fig. \ref{fig:abundances}), and the stellar winds of such stars are faster, possibly complicating their capture in the cluster centre (see \citealt{2022A&A...658A.125S}). 

Lastly, the wind mass-loss rate could be increased by introducing even more massive stars. The line-driven wind mass-loss rate is expected to increase with stellar mass \citep{2018A&A...615A.119V}, thus stellar mass-growth through collisions or further gas-accretion might lead to enhanced production of enriched material.

The enhanced wind-enrichment would result in enhanced helium-enrichment. Multiplying the recycled wind-masses in the stars by a factor of ten, the helium mass-fraction would increase by 0.1--0.2, which might already start impacting stellar evolution. The response of the ISM and the final 2P fraction to changes in the IMF and/or stellar feedback models may therefore be highly non-linear and would require self-consistent testing in a hydrodynamical setting. The scenarios to alleviate the mass-budget problem outlined above offer intriguing avenues for future studies.

\subsection{Numerical limitations}

Because the stellar wind increments released by single massive stars at a rate of up to \mbox{$10^{-3}$ \mdot{} yr$^{-1}$} would require a significant improvement to the gas mass resolution to allow their injection as a continuous stream of new particles, the winds are here injected into existing gas particles. For the stellar wind to be retained in the cluster forming region, the wind injection scheme adopted here relies on there being gas particles to be injected with. Once a cluster has expelled all gas within its vicinity, the stellar winds cannot by construction remain in the clusters, nor can they cool down or contribute to further star formation without the presence of gas particles. The results presented here for enrichment within massive star clusters can therefore be viewed as lower limits. If the stellar wind could be retained and cool into star forming conditions without mixing with the ISM, it may be possible to form stars consisting of 100\% wind-material in the later stages of cluster formation. As noted in Fig. \ref{fig:theor_output}, the maximum returned mass via winds would still be only a couple per cent.

The protostar phase, especially for low-mass stars, may last for a significant duration up to Myrs \citep{2014prpl.conf..195D}. Our simulations do not resolve the protostellar phase with gas accretion and jet-like outflows (see e.g. \citealt{2021MNRAS.506.2199G}). To remain conservative, we only inject energy and metal-enriched material into the gaseous component within and surrounding the star-forming regions. Once a gas particle is defined star-forming, we do not allow further enrichment. If we were to consider the inert time as the protostar phase, the metal-content of the star formation reservoir particles (Section \ref{section:SF}) could be enhanced by allowing them to accrete wind and supernova-material.

As indicated in Fig. \ref{fig:theor_output}, stochastic sampling of the IMF can also have a strong impact on the integrated feedback of massive stars in clusters as massive as $10^5$ \mdot. The massive clusters realised in the current study reach stellar masses up to $2\times 10^5$ \mdot, and remain susceptible to stochasticity. This is best illustrated by concentrating on the most massive clusters in Figs. \ref{fig:mean_enr}--\ref{fig:fraction_extra}: the clusters host very massive stars of up to 379 \mdot, but their evolution within the cluster formation sequence governs how well their feedback can actually be recycled. As a result, the most massive cluster is less enriched in wind-material compared to the second most massive one. Simulations reaching higher cluster-masses, from $10^6$ \mdot{} upwards, would be required to guarantee the realisation of a wealth of very massive stars spread across the cluster formation sequence.

Regardless of how the massive stars are sampled and how their stellar winds are implemented in the simulations, the stars need to be present within the cluster forming regions during the period when the wind is released to be able to enrich their surroundings. Due to gravitational softening, our simulations do not take into account the dynamical impact of short-range interactions or stellar multiples. Binary interactions are especially relevant for massive stars which are predominantly found in systems of multiple stars \citep{2012Sci...337..444S, 2012MNRAS.424.1925C}. Processes including binary stars may result in the ejection of massive stars out of young star clusters through binary dissolution due to exploding companion stars \citep{1961BAN....15..265B} or through strong gravitational encounters \citep{1967BOTT....4...86P}. Observed runaway and walkaway stars have indeed been traced back to their cluster origins \citep{2000ApJ...544L.133H, 2005A&A...437..247D} even at very young cluster ages (e.g. \citealt{2008A&A...490.1071G, 2023A&A...670A.108S}). The removal of the massive stars during the early stages of cluster formation may have a direct impact on their role in enrichment and gas dispersal, however direct N-body simulations indicate that not all massive stars are necessarily ejected \citep{2022MNRAS.514.2513F}. Molecular cloud scale simulations that model both star formation and stellar two-body dynamics \citep{2021PASJ...73.1057F, 2021MNRAS.506.2199G, 2022MNRAS.509.6155R, 2023MNRAS.521.1338C} have paved the way and we will hopefully in the future be able to study the impact of accurate gravitational dynamics in cluster forming regions within a full galactic environment.

\section{Conclusions}\label{section:summary}

We have examined star formation and stellar feedback in a low-metallicity dwarf galaxy starburst, focusing on the stellar self-enrichment during the formation of massive star clusters. The hydrodynamical simulations employ a Jeans mass dependent star formation threshold, a stellar IMF realised with single stars between $0.08$ M$_\odot$ and $\sim400$ M$_\odot$, and detailed prescriptions for stellar winds, radiation, and SNe of individual stars. The evolution models for massive ($>9$ \mdot) and very massive ($>100$ \mdot) stars include the chemical composition of the stellar winds an SNe, which allows us to trace the propagation of individual elements throughout the starburst.

Massive star clusters up to $\sim 2\times 10^5$ \mdot{} form on timescales of less than $\sim5$ Myr. Massive stars evolve on a few-Myr timescale, allowing their light-element enriched winds to be released and recycled in new stars while the clusters are still forming. Following the individual elements released in stellar winds and SNe of massive stars, we showed how the wind-material is on average ten times more efficiently recycled within clusters compared to SN-material. Very massive stars, when present, contribute most of the stellar wind-mass and die either as direct collapse black holes or very energetic PISNe, inhibiting significant self-enrichment by heavy elements in the vicinity of the rapidly forming star clusters. 

Some of the simulated clusters show small amounts of heavy element enrichment, with the cluster-to-cluster mean [Fe/H] varying by 0.01 dex. Because the measured mass fraction of SN-material locked in cluster stars is independent of the total cluster-mass, the iron enrichment has to occur on galaxy-scale, rather than locally. The iron variations reflect the global mixing of SN-material within the star forming regions throughout the starburst, rather than localized self-enrichment. These results are consistent with detailed spectral observation of GC stars that indicate small but non-zero spreads in iron, especially in the primordial 1P, in iron-normal\footnote{Type I GCs, as opposed to iron-complex Type II GCs such as $\omega$ Cen that host multiple sequences in [Fe/H] \citep{2017MNRAS.469..800M}.} GCs that were originally thought to have uniform heavy element contents.

We have taken a detailed look at the light-element enrichment through stellar winds in the most massive star clusters that host very massive stars. The stellar wind-material is radially concentrated at the cluster centre in many of the clusters more massive than a few \mbox{$10^4$ \mdot}, consistent with some of the observed GCs that host a centrally concentrated chemically peculiar 2P. The total mass of recycled wind-material within the simulated clusters increases with the cluster-mass, reflecting the enhanced wind production and possibly easier capture of wind-material during the formation phase of massive star clusters. This is conceptually consistent with observed GCs, where the number fraction and elemental abundance spreads of enriched 2P stars increase with cluster-mass.

The abundance ratios in the pure, un-diluted stellar winds captured in the simulated 2P cluster stars are consistent with the star-to-star [Na/Fe]--[O/Fe] and [Al/Fe]--[Mg/Fe] anticorrelation and the enhanced [N/Fe] observed in practically all GCs. After dilution with the star-forming ISM, the spread of the simulated light-element abundance ratios decreases and we run into a manifestation of the mass budget problem with many of the abundance ratios. Even the most enriched stars show barely extreme enough abundance variations in [Na/Fe], [O/Fe], [Al/Fe] and [Mg/Fe] to be considered as observationally selected 2P stars. The largest fraction of stellar mass composed of wind-material reaches 10\% per star, resulting in star-to-star spread of up to \mbox{$\sim0.3$ dex} in [Na/Fe]. [N/Fe] shows the largest spread of up to  $\sim0.5$ dex and N-enhancement in particular has been observed in young high-redshift stellar objects that may represent young GCs in formation.

The wind-mass captured in the stars of each simulated cluster is reasonably well fit with a reversed log-normal PDF. Using the best-fit parameters, we obtained scaling relations for the simulated $f_\mathrm{wind}$ distribution as a function of cluster-mass. We then extrapolated the parameters of the $f_\mathrm{wind}$ distributions to obtain an observationally motivated 2P fraction ($\Delta[\mathrm{Na/Fe}]>0.3,\, f_\mathrm{wind}>10\%$) of clusters that represent the initial mass range of present-day observed GCs. After accounting for significant mass loss (up to 90\% of the 1P) via two-body dynamics and tidal effects, we expect winds of evolved massive stars to result in a final 2P fraction of $\sim55\%$ at most in clusters up to an initial mass of $10^7$ \mdot. Observed GCs, on the other hand, show 2P fractions of up to $90 \%$ and an onset of the increase in 2P fraction already at $10^5$ \mdot. Such high 2P fractions at so low cluster masses are at first hand inaccessible to our current model because of the mass budget problem. However, we showed that an increase by a factor of ten in enriched material recycled within the simulated clusters could result in an observationally consistent relation between the 2P fraction and final cluster-mass, as well as observationally consistent spreads in star-to-star [Na/Fe] and [O/Fe]. Such enhancement in enriched material may be provided by stellar wind output of stars with a top-heavy IMF, or by a combination of various enrichment sources such as interacting binaries, rotating massive stars or supermassive stars. For maximal wind-production, the high-mass end of the IMF would also need to be well populated. In terms of the wind injection method, the recycled wind mass and the most extreme abundance ratios might change by implementing a method that spawns wind-particles instead of relying on existing gas particles that inevitably dilute the chemical composition of the 2P stars. Even then, an enhanced wind-production rate would still be needed if an initially dominant 2P was to be achieved. 

Despite the expected mass-budget problem encountered in our study, we can conclude that the observationally consistent, rapidly generated centrally concentrated 2P distribution and increase of recycled wind-mass with cluster-mass support self-enrichment by pre-SN feedback in cluster forming regions as the source of the 2P. 

\section*{Acknowledgements}

The authors thank Guinevere Kauffmann for insightful comments and discussions that helped improve the manuscript. NL is grateful to Angela Adamo and Corinne Charbonnel for discussions on observational aspects of globular clusters.
NL and TN acknowledge the computing time granted by the LRZ (Leibniz-Rechenzentrum) on SuperMUC-NG under
project numbers pn49qi and pn72bu. TN acknowledges support from the Deutsche Forschungsgemeinschaft (DFG, German Research Foundation) under Germany's Excellence Strategy - EXC-2094 - 390783311 from the DFG Cluster of Excellence "ORIGINS". This research was funded in part by the National Science Center (NCN), Poland under grant number OPUS 2021/41/B/ST9/00757. For the  purpose of Open Access, the author has applied a CC-BY public copyright license to any Author Accepted Manuscript (AAM) version arising from this submission. 
The computations were carried out at SuperMUC-NG hosted by the LRZ and the COBRA and FREYA clusters hosted by The Max Planck Computing and Data Facility (MPCDF) in Garching, Germany.

This research made use of \textsc{python} packages \textsc{scipy} \citep{2020SciPy-NMeth}, \textsc{numpy} \citep{2020NumPy-Array}, \textsc{matplotlib} \citep{Hunter:2007}, \textsc{pygad} \citep{2020MNRAS.496..152R} and \textsc{h5py} \citep{collette_python_hdf5_2014}.

\section*{Data Availability}

The data will be made available on reasonable request to the corresponding author.



\bibliographystyle{mnras}

\begin{thebibliography}{}
\makeatletter
\relax
\def\mn@urlcharsother{\let\do\@makeother \do\$\do\&\do\#\do\^\do\_\do\%\do\~}
\def\mn@doi{\begingroup\mn@urlcharsother \@ifnextchar [ {\mn@doi@}
  {\mn@doi@[]}}
\def\mn@doi@[#1]#2{\def\@tempa{#1}\ifx\@tempa\@empty \href
  {http://dx.doi.org/#2} {doi:#2}\else \href {http://dx.doi.org/#2} {#1}\fi
  \endgroup}
\def\mn@eprint#1#2{\mn@eprint@#1:#2::\@nil}
\def\mn@eprint@arXiv#1{\href {http://arxiv.org/abs/#1} {{\tt arXiv:#1}}}
\def\mn@eprint@dblp#1{\href {http://dblp.uni-trier.de/rec/bibtex/#1.xml}
  {dblp:#1}}
\def\mn@eprint@#1:#2:#3:#4\@nil{\def\@tempa {#1}\def\@tempb {#2}\def\@tempc
  {#3}\ifx \@tempc \@empty \let \@tempc \@tempb \let \@tempb \@tempa \fi \ifx
  \@tempb \@empty \def\@tempb {arXiv}\fi \@ifundefined
  {mn@eprint@\@tempb}{\@tempb:\@tempc}{\expandafter \expandafter \csname
  mn@eprint@\@tempb\endcsname \expandafter{\@tempc}}}

\bibitem[\protect\citeauthoryear{{Adamo} et~al.,}{{Adamo}
  et~al.}{2020}]{2020SSRv..216...69A}
{Adamo} A.,  et~al., 2020, \mn@doi [\ssr] {10.1007/s11214-020-00690-x}, \href
  {https://ui.adsabs.harvard.edu/abs/2020SSRv..216...69A} {216, 69}

\bibitem[\protect\citeauthoryear{{Adamo} et~al.,}{{Adamo}
  et~al.}{2024}]{2024arXiv240103224A}
{Adamo} A.,  et~al., 2024, \mn@doi [arXiv e-prints]
  {10.48550/arXiv.2401.03224}, \href
  {https://ui.adsabs.harvard.edu/abs/2024arXiv240103224A} {p. arXiv:2401.03224}

\bibitem[\protect\citeauthoryear{{Andersson}, {Mac Low}, {Agertz}, {Renaud}  \&
  {Li}}{{Andersson} et~al.}{2024}]{2024A&A...681A..28A}
{Andersson} E.~P.,  {Mac Low} M.-M.,  {Agertz} O.,  {Renaud} F.,   {Li} H.,
  2024, \mn@doi [\aap] {10.1051/0004-6361/202347792}, \href
  {https://ui.adsabs.harvard.edu/abs/2024A&A...681A..28A} {681, A28}

\bibitem[\protect\citeauthoryear{{Arnould}, {Goriely}  \& {Jorissen}}{{Arnould}
  et~al.}{1999}]{1999A&A...347..572A}
{Arnould} M.,  {Goriely} S.,   {Jorissen} A.,  1999, \mn@doi [\aap]
  {10.48550/arXiv.astro-ph/9904407}, \href
  {https://ui.adsabs.harvard.edu/abs/1999A&A...347..572A} {347, 572}

\bibitem[\protect\citeauthoryear{{Bastian} \& {Lardo}}{{Bastian} \&
  {Lardo}}{2018}]{2018ARA&A..56...83B}
{Bastian} N.,  {Lardo} C.,  2018, \mn@doi [\araa]
  {10.1146/annurev-astro-081817-051839}, \href
  {https://ui.adsabs.harvard.edu/abs/2018ARA&A..56...83B} {56, 83}

\bibitem[\protect\citeauthoryear{{Bastian}, {Lamers}, {de Mink}, {Longmore},
  {Goodwin}  \& {Gieles}}{{Bastian} et~al.}{2013}]{2013MNRAS.436.2398B}
{Bastian} N.,  {Lamers} H.~J.~G.~L.~M.,  {de Mink} S.~E.,  {Longmore} S.~N.,
  {Goodwin} S.~P.,   {Gieles} M.,  2013, \mn@doi [\mnras]
  {10.1093/mnras/stt1745}, \href
  {https://ui.adsabs.harvard.edu/abs/2013MNRAS.436.2398B} {436, 2398}

\bibitem[\protect\citeauthoryear{{Bastian}, {Hollyhead}  \&
  {Cabrera-Ziri}}{{Bastian} et~al.}{2014}]{2014MNRAS.445..378B}
{Bastian} N.,  {Hollyhead} K.,   {Cabrera-Ziri} I.,  2014, \mn@doi [\mnras]
  {10.1093/mnras/stu1775}, \href
  {https://ui.adsabs.harvard.edu/abs/2014MNRAS.445..378B} {445, 378}

\bibitem[\protect\citeauthoryear{{Bastian}, {Cabrera-Ziri}  \&
  {Salaris}}{{Bastian} et~al.}{2015}]{2015MNRAS.449.3333B}
{Bastian} N.,  {Cabrera-Ziri} I.,   {Salaris} M.,  2015, \mn@doi [\mnras]
  {10.1093/mnras/stv543}, \href
  {https://ui.adsabs.harvard.edu/abs/2015MNRAS.449.3333B} {449, 3333}

\bibitem[\protect\citeauthoryear{{Baumgardt}, {Hilker}, {Sollima}  \&
  {Bellini}}{{Baumgardt} et~al.}{2019}]{2019MNRAS.482.5138B}
{Baumgardt} H.,  {Hilker} M.,  {Sollima} A.,   {Bellini} A.,  2019, \mn@doi
  [\mnras] {10.1093/mnras/sty2997}, \href
  {https://ui.adsabs.harvard.edu/abs/2019MNRAS.482.5138B} {482, 5138}

\bibitem[\protect\citeauthoryear{{Bekki}}{{Bekki}}{2011}]{2011MNRAS.412.2241B}
{Bekki} K.,  2011, \mn@doi [\mnras] {10.1111/j.1365-2966.2010.18047.x}, \href
  {https://ui.adsabs.harvard.edu/abs/2011MNRAS.412.2241B} {412, 2241}

\bibitem[\protect\citeauthoryear{{Bekki}}{{Bekki}}{2019}]{2019MNRAS.486.2570B}
{Bekki} K.,  2019, \mn@doi [\mnras] {10.1093/mnras/stz999}, \href
  {https://ui.adsabs.harvard.edu/abs/2019MNRAS.486.2570B} {486, 2570}

\bibitem[\protect\citeauthoryear{{Bisbas} et~al.,}{{Bisbas}
  et~al.}{2022}]{2022ApJ...934..115B}
{Bisbas} T.~G.,  et~al., 2022, \mn@doi [\apj] {10.3847/1538-4357/ac7960}, \href
  {https://ui.adsabs.harvard.edu/abs/2022ApJ...934..115B} {934, 115}

\bibitem[\protect\citeauthoryear{{Blaauw}}{{Blaauw}}{1961}]{1961BAN....15..265B}
{Blaauw} A.,  1961, \bain, \href
  {https://ui.adsabs.harvard.edu/abs/1961BAN....15..265B} {15, 265}

\bibitem[\protect\citeauthoryear{{Bunker} et~al.,}{{Bunker}
  et~al.}{2023}]{2023A&A...677A..88B}
{Bunker} A.~J.,  et~al., 2023, \mn@doi [\aap] {10.1051/0004-6361/202346159},
  \href {https://ui.adsabs.harvard.edu/abs/2023A&A...677A..88B} {677, A88}

\bibitem[\protect\citeauthoryear{{Cabrera-Ziri}, {Lardo}, {Davies}, {Bastian},
  {Beccari}, {Larsen}  \& {Hernandez}}{{Cabrera-Ziri}
  et~al.}{2016}]{2016MNRAS.460.1869C}
{Cabrera-Ziri} I.,  {Lardo} C.,  {Davies} B.,  {Bastian} N.,  {Beccari} G.,
  {Larsen} S.~S.,   {Hernandez} S.,  2016, \mn@doi [\mnras]
  {10.1093/mnras/stw1090}, \href
  {https://ui.adsabs.harvard.edu/abs/2016MNRAS.460.1869C} {460, 1869}

\bibitem[\protect\citeauthoryear{{Calura}, {D'Ercole}, {Vesperini}, {Vanzella}
  \& {Sollima}}{{Calura} et~al.}{2019}]{2019MNRAS.489.3269C}
{Calura} F.,  {D'Ercole} A.,  {Vesperini} E.,  {Vanzella} E.,   {Sollima} A.,
  2019, \mn@doi [\mnras] {10.1093/mnras/stz2055}, \href
  {https://ui.adsabs.harvard.edu/abs/2019MNRAS.489.3269C} {489, 3269}

\bibitem[\protect\citeauthoryear{{Cannon}, {Croke}, {Bell}, {Hesser}  \&
  {Stathakis}}{{Cannon} et~al.}{1998}]{1998MNRAS.298..601C}
{Cannon} R.~D.,  {Croke} B.~F.~W.,  {Bell} R.~A.,  {Hesser} J.~E.,
  {Stathakis} R.~A.,  1998, \mn@doi [\mnras]
  {10.1046/j.1365-8711.1998.01671.x}, \href
  {https://ui.adsabs.harvard.edu/abs/1998MNRAS.298..601C} {298, 601}

\bibitem[\protect\citeauthoryear{{Carretta}}{{Carretta}}{2014}]{2014ApJ...795L..28C}
{Carretta} E.,  2014, \mn@doi [\apjl] {10.1088/2041-8205/795/2/L28}, \href
  {https://ui.adsabs.harvard.edu/abs/2014ApJ...795L..28C} {795, L28}

\bibitem[\protect\citeauthoryear{{Carretta} et~al.,}{{Carretta}
  et~al.}{2009a}]{2009A&A...505..117C}
{Carretta} E.,  et~al., 2009a, \mn@doi [\aap] {10.1051/0004-6361/200912096},
  \href {https://ui.adsabs.harvard.edu/abs/2009A&A...505..117C} {505, 117}

\bibitem[\protect\citeauthoryear{{Carretta}, {Bragaglia}, {Gratton}  \&
  {Lucatello}}{{Carretta} et~al.}{2009b}]{2009A&A...505..139C}
{Carretta} E.,  {Bragaglia} A.,  {Gratton} R.,   {Lucatello} S.,  2009b,
  \mn@doi [\aap] {10.1051/0004-6361/200912097}, \href
  {https://ui.adsabs.harvard.edu/abs/2009A&A...505..139C} {505, 139}

\bibitem[\protect\citeauthoryear{{Carretta}, {Bragaglia}, {Gratton},
  {Recio-Blanco}, {Lucatello}, {D'Orazi}  \& {Cassisi}}{{Carretta}
  et~al.}{2010}]{2010A&A...516A..55C}
{Carretta} E.,  {Bragaglia} A.,  {Gratton} R.~G.,  {Recio-Blanco} A.,
  {Lucatello} S.,  {D'Orazi} V.,   {Cassisi} S.,  2010, \mn@doi [\aap]
  {10.1051/0004-6361/200913451}, \href
  {https://ui.adsabs.harvard.edu/abs/2010A&A...516A..55C} {516, A55}

\bibitem[\protect\citeauthoryear{{Carretta}, {Bragaglia}, {Gratton},
  {Lucatello}  \& {D'Orazi}}{{Carretta} et~al.}{2012}]{2012ApJ...750L..14C}
{Carretta} E.,  {Bragaglia} A.,  {Gratton} R.~G.,  {Lucatello} S.,   {D'Orazi}
  V.,  2012, \mn@doi [\apjl] {10.1088/2041-8205/750/1/L14}, \href
  {https://ui.adsabs.harvard.edu/abs/2012ApJ...750L..14C} {750, L14}

\bibitem[\protect\citeauthoryear{{Chandar} et~al.,}{{Chandar}
  et~al.}{2023}]{2023ApJ...949..116C}
{Chandar} R.,  et~al., 2023, \mn@doi [\apj] {10.3847/1538-4357/acc93b}, \href
  {https://ui.adsabs.harvard.edu/abs/2023ApJ...949..116C} {949, 116}

\bibitem[\protect\citeauthoryear{{Chantereau}, {Charbonnel}  \&
  {Decressin}}{{Chantereau} et~al.}{2015}]{2015A&A...578A.117C}
{Chantereau} W.,  {Charbonnel} C.,   {Decressin} T.,  2015, \mn@doi [\aap]
  {10.1051/0004-6361/201525929}, \href
  {https://ui.adsabs.harvard.edu/abs/2015A&A...578A.117C} {578, A117}

\bibitem[\protect\citeauthoryear{{Chantereau}, {Charbonnel}  \&
  {Meynet}}{{Chantereau} et~al.}{2016}]{2016A&A...592A.111C}
{Chantereau} W.,  {Charbonnel} C.,   {Meynet} G.,  2016, \mn@doi [\aap]
  {10.1051/0004-6361/201628418}, \href
  {https://ui.adsabs.harvard.edu/abs/2016A&A...592A.111C} {592, A111}

\bibitem[\protect\citeauthoryear{{Charbonnel}}{{Charbonnel}}{2016}]{2016EAS....80..177C}
{Charbonnel} C.,  2016, in {Moraux} E.,  {Lebreton} Y.,   {Charbonnel} C.,
  eds,  EAS Publications Series Vol. 80-81, EAS Publications Series. pp
  177--226 (\mn@eprint {arXiv} {1611.08855}), \mn@doi{10.1051/eas/1680006}

\bibitem[\protect\citeauthoryear{{Charbonnel}, {Schaerer}, {Prantzos},
  {Ram{\'\i}rez-Galeano}, {Fragos}, {Kuruvanthodi}, {Marques-Chaves}  \&
  {Gieles}}{{Charbonnel} et~al.}{2023}]{2023A&A...673L...7C}
{Charbonnel} C.,  {Schaerer} D.,  {Prantzos} N.,  {Ram{\'\i}rez-Galeano} L.,
  {Fragos} T.,  {Kuruvanthodi} A.,  {Marques-Chaves} R.,   {Gieles} M.,  2023,
  \mn@doi [\aap] {10.1051/0004-6361/202346410}, \href
  {https://ui.adsabs.harvard.edu/abs/2023A&A...673L...7C} {673, L7}

\bibitem[\protect\citeauthoryear{{Chieffi} \& {Limongi}}{{Chieffi} \&
  {Limongi}}{2004}]{2004ApJ...608..405C}
{Chieffi} A.,  {Limongi} M.,  2004, \mn@doi [\apj] {10.1086/392523}, \href
  {https://ui.adsabs.harvard.edu/abs/2004ApJ...608..405C} {608, 405}

\bibitem[\protect\citeauthoryear{{Chini}, {Hoffmeister}, {Nasseri}, {Stahl}  \&
  {Zinnecker}}{{Chini} et~al.}{2012}]{2012MNRAS.424.1925C}
{Chini} R.,  {Hoffmeister} V.~H.,  {Nasseri} A.,  {Stahl} O.,   {Zinnecker} H.,
   2012, \mn@doi [\mnras] {10.1111/j.1365-2966.2012.21317.x}, \href
  {https://ui.adsabs.harvard.edu/abs/2012MNRAS.424.1925C} {424, 1925}

\bibitem[\protect\citeauthoryear{{Clark}, {Glover}  \& {Klessen}}{{Clark}
  et~al.}{2012}]{2012MNRAS.420..745C}
{Clark} P.~C.,  {Glover} S. C.~O.,   {Klessen} R.~S.,  2012, \mn@doi [\mnras]
  {10.1111/j.1365-2966.2011.20087.x}, \href
  {https://ui.adsabs.harvard.edu/abs/2012MNRAS.420..745C} {420, 745}

\bibitem[\protect\citeauthoryear{{Cohen}, {Huang}  \& {Kirby}}{{Cohen}
  et~al.}{2011}]{2011ApJ...740...60C}
{Cohen} J.~G.,  {Huang} W.,   {Kirby} E.~N.,  2011, \mn@doi [\apj]
  {10.1088/0004-637X/740/2/60}, \href
  {https://ui.adsabs.harvard.edu/abs/2011ApJ...740...60C} {740, 60}

\bibitem[\protect\citeauthoryear{Collette}{Collette}{2013}]{collette_python_hdf5_2014}
Collette A.,  2013, Python and HDF5.
O'Reilly Media, Inc., Sebostopol, USA

\bibitem[\protect\citeauthoryear{{Cook} et~al.,}{{Cook}
  et~al.}{2023}]{2023MNRAS.519.3749C}
{Cook} D.~O.,  et~al., 2023, \mn@doi [\mnras] {10.1093/mnras/stac3748}, \href
  {https://ui.adsabs.harvard.edu/abs/2023MNRAS.519.3749C} {519, 3749}

\bibitem[\protect\citeauthoryear{{Cottrell} \& {Da Costa}}{{Cottrell} \& {Da
  Costa}}{1981}]{1981ApJ...245L..79C}
{Cottrell} P.~L.,  {Da Costa} G.~S.,  1981, \mn@doi [\apjl] {10.1086/183527},
  \href {https://ui.adsabs.harvard.edu/abs/1981ApJ...245L..79C} {245, L79}

\bibitem[\protect\citeauthoryear{{Cournoyer-Cloutier}
  et~al.,}{{Cournoyer-Cloutier} et~al.}{2023}]{2023MNRAS.521.1338C}
{Cournoyer-Cloutier} C.,  et~al., 2023, \mn@doi [\mnras]
  {10.1093/mnras/stad568}, \href
  {https://ui.adsabs.harvard.edu/abs/2023MNRAS.521.1338C} {521, 1338}

\bibitem[\protect\citeauthoryear{{D'Ercole}, {Vesperini}, {D'Antona},
  {McMillan}  \& {Recchi}}{{D'Ercole} et~al.}{2008}]{2008MNRAS.391..825D}
{D'Ercole} A.,  {Vesperini} E.,  {D'Antona} F.,  {McMillan} S. L.~W.,
  {Recchi} S.,  2008, \mn@doi [\mnras] {10.1111/j.1365-2966.2008.13915.x},
  \href {https://ui.adsabs.harvard.edu/abs/2008MNRAS.391..825D} {391, 825}

\bibitem[\protect\citeauthoryear{{D'Ercole}, {D'Antona}, {Ventura}, {Vesperini}
   \& {McMillan}}{{D'Ercole} et~al.}{2010}]{2010MNRAS.407..854D}
{D'Ercole} A.,  {D'Antona} F.,  {Ventura} P.,  {Vesperini} E.,   {McMillan} S.
  L.~W.,  2010, \mn@doi [\mnras] {10.1111/j.1365-2966.2010.16996.x}, \href
  {https://ui.adsabs.harvard.edu/abs/2010MNRAS.407..854D} {407, 854}

\bibitem[\protect\citeauthoryear{{Dalessandro} et~al.,}{{Dalessandro}
  et~al.}{2014}]{2014ApJ...791L...4D}
{Dalessandro} E.,  et~al., 2014, \mn@doi [\apjl] {10.1088/2041-8205/791/1/L4},
  \href {https://ui.adsabs.harvard.edu/abs/2014ApJ...791L...4D} {791, L4}

\bibitem[\protect\citeauthoryear{{Dalessandro} et~al.,}{{Dalessandro}
  et~al.}{2019}]{2019ApJ...884L..24D}
{Dalessandro} E.,  et~al., 2019, \mn@doi [\apjl] {10.3847/2041-8213/ab45f7},
  \href {https://ui.adsabs.harvard.edu/abs/2019ApJ...884L..24D} {884, L24}

\bibitem[\protect\citeauthoryear{{Decressin}, {Meynet}, {Charbonnel},
  {Prantzos}  \& {Ekstr{\"o}m}}{{Decressin} et~al.}{2007}]{2007A&A...464.1029D}
{Decressin} T.,  {Meynet} G.,  {Charbonnel} C.,  {Prantzos} N.,   {Ekstr{\"o}m}
  S.,  2007, \mn@doi [\aap] {10.1051/0004-6361:20066013}, \href
  {https://ui.adsabs.harvard.edu/abs/2007A&A...464.1029D} {464, 1029}

\bibitem[\protect\citeauthoryear{{Decressin}, {Baumgardt}, {Charbonnel}  \&
  {Kroupa}}{{Decressin} et~al.}{2010}]{2010A&A...516A..73D}
{Decressin} T.,  {Baumgardt} H.,  {Charbonnel} C.,   {Kroupa} P.,  2010,
  \mn@doi [\aap] {10.1051/0004-6361/200913703}, \href
  {https://ui.adsabs.harvard.edu/abs/2010A&A...516A..73D} {516, A73}

\bibitem[\protect\citeauthoryear{{Denisenkov} \& {Denisenkova}}{{Denisenkov} \&
  {Denisenkova}}{1989}]{1989ATsir1538...11D}
{Denisenkov} P.~A.,  {Denisenkova} S.~N.,  1989, Astronomicheskij Tsirkulyar,
  \href {https://ui.adsabs.harvard.edu/abs/1989ATsir1538...11D} {1538, 11}

\bibitem[\protect\citeauthoryear{{Denissenkov} \& {Hartwick}}{{Denissenkov} \&
  {Hartwick}}{2014}]{2014MNRAS.437L..21D}
{Denissenkov} P.~A.,  {Hartwick} F.~D.~A.,  2014, \mn@doi [\mnras]
  {10.1093/mnrasl/slt133}, \href
  {https://ui.adsabs.harvard.edu/abs/2014MNRAS.437L..21D} {437, L21}

\bibitem[\protect\citeauthoryear{{Dolag}, {Borgani}, {Murante}  \&
  {Springel}}{{Dolag} et~al.}{2009}]{2009MNRAS.399..497D}
{Dolag} K.,  {Borgani} S.,  {Murante} G.,   {Springel} V.,  2009, \mn@doi
  [\mnras] {10.1111/j.1365-2966.2009.15034.x}, \href
  {https://ui.adsabs.harvard.edu/abs/2009MNRAS.399..497D} {399, 497}

\bibitem[\protect\citeauthoryear{{Dunham} et~al.,}{{Dunham}
  et~al.}{2014}]{2014prpl.conf..195D}
{Dunham} M.~M.,  et~al., 2014, in {Beuther} H.,  {Klessen} R.~S.,  {Dullemond}
  C.~P.,   {Henning} T.,  eds, Protostars and Planets VI. pp 195--218
  (\mn@eprint {arXiv} {1401.1809}),
  \mn@doi{10.2458/azu_uapress_9780816531240-ch009}

\bibitem[\protect\citeauthoryear{{Elmegreen} \& {Hunter}}{{Elmegreen} \&
  {Hunter}}{2015}]{2015ApJ...805..145E}
{Elmegreen} B.~G.,  {Hunter} D.~A.,  2015, \mn@doi [\apj]
  {10.1088/0004-637X/805/2/145}, \href
  {https://ui.adsabs.harvard.edu/abs/2015ApJ...805..145E} {805, 145}

\bibitem[\protect\citeauthoryear{{Fujii}, {Saitoh}, {Wang}  \& {Hirai}}{{Fujii}
  et~al.}{2021}]{2021PASJ...73.1057F}
{Fujii} M.~S.,  {Saitoh} T.~R.,  {Wang} L.,   {Hirai} Y.,  2021, \mn@doi
  [\pasj] {10.1093/pasj/psab037}, \href
  {https://ui.adsabs.harvard.edu/abs/2021PASJ...73.1057F} {73, 1057}

\bibitem[\protect\citeauthoryear{{Fujii}, {Wang}, {Hirai}, {Shimajiri},
  {Kumamoto}  \& {Saitoh}}{{Fujii} et~al.}{2022}]{2022MNRAS.514.2513F}
{Fujii} M.~S.,  {Wang} L.,  {Hirai} Y.,  {Shimajiri} Y.,  {Kumamoto} J.,
  {Saitoh} T.,  2022, \mn@doi [\mnras] {10.1093/mnras/stac1496}, \href
  {https://ui.adsabs.harvard.edu/abs/2022MNRAS.514.2513F} {514, 2513}

\bibitem[\protect\citeauthoryear{{Genzel} et~al.,}{{Genzel}
  et~al.}{2010}]{2010MNRAS.407.2091G}
{Genzel} R.,  et~al., 2010, \mn@doi [\mnras]
  {10.1111/j.1365-2966.2010.16969.x}, \href
  {https://ui.adsabs.harvard.edu/abs/2010MNRAS.407.2091G} {407, 2091}

\bibitem[\protect\citeauthoryear{{Georgy} et~al.,}{{Georgy}
  et~al.}{2013}]{2013A&A...558A.103G}
{Georgy} C.,  et~al., 2013, \mn@doi [\aap] {10.1051/0004-6361/201322178}, \href
  {https://ui.adsabs.harvard.edu/abs/2013A&A...558A.103G} {558, A103}

\bibitem[\protect\citeauthoryear{{Gieles} et~al.,}{{Gieles}
  et~al.}{2018}]{2018MNRAS.478.2461G}
{Gieles} M.,  et~al., 2018, \mn@doi [\mnras] {10.1093/mnras/sty1059}, \href
  {https://ui.adsabs.harvard.edu/abs/2018MNRAS.478.2461G} {478, 2461}

\bibitem[\protect\citeauthoryear{{Gilmer}, {Kozyreva}, {Hirschi},
  {Fr{\"o}hlich}  \& {Yusof}}{{Gilmer} et~al.}{2017}]{2017ApJ...846..100G}
{Gilmer} M.~S.,  {Kozyreva} A.,  {Hirschi} R.,  {Fr{\"o}hlich} C.,   {Yusof}
  N.,  2017, \mn@doi [\apj] {10.3847/1538-4357/aa8461}, \href
  {https://ui.adsabs.harvard.edu/abs/2017ApJ...846..100G} {846, 100}

\bibitem[\protect\citeauthoryear{{Glatt} et~al.,}{{Glatt}
  et~al.}{2011}]{2011AJ....142...36G}
{Glatt} K.,  et~al., 2011, \mn@doi [\aj] {10.1088/0004-6256/142/2/36}, \href
  {https://ui.adsabs.harvard.edu/abs/2011AJ....142...36G} {142, 36}

\bibitem[\protect\citeauthoryear{{G{\'o}rski} \& {Hivon}}{{G{\'o}rski} \&
  {Hivon}}{2011}]{2011ascl.soft07018G}
{G{\'o}rski} K.~M.,  {Hivon} E.,  2011, {HEALPix: Hierarchical Equal Area
  isoLatitude Pixelization of a sphere}, Astrophysics Source Code Library
  (\mn@eprint {ascl} {1107.018})

\bibitem[\protect\citeauthoryear{{Gratton} et~al.,}{{Gratton}
  et~al.}{2001}]{2001A&A...369...87G}
{Gratton} R.~G.,  et~al., 2001, \mn@doi [\aap] {10.1051/0004-6361:20010144},
  \href {https://ui.adsabs.harvard.edu/abs/2001A&A...369...87G} {369, 87}

\bibitem[\protect\citeauthoryear{{Gratton}, {Bragaglia}, {Carretta}, {D'Orazi},
  {Lucatello}  \& {Sollima}}{{Gratton} et~al.}{2019}]{2019A&ARv..27....8G}
{Gratton} R.,  {Bragaglia} A.,  {Carretta} E.,  {D'Orazi} V.,  {Lucatello} S.,
   {Sollima} A.,  2019, \mn@doi [\aapr] {10.1007/s00159-019-0119-3}, \href
  {https://ui.adsabs.harvard.edu/abs/2019A&ARv..27....8G} {27, 8}

\bibitem[\protect\citeauthoryear{{Groh} et~al.,}{{Groh}
  et~al.}{2019}]{2019A&A...627A..24G}
{Groh} J.~H.,  et~al., 2019, \mn@doi [\aap] {10.1051/0004-6361/201833720},
  \href {https://ui.adsabs.harvard.edu/abs/2019A&A...627A..24G} {627, A24}

\bibitem[\protect\citeauthoryear{{Grudi{\'c}}, {Guszejnov}, {Hopkins}, {Offner}
   \& {Faucher-Gigu{\`e}re}}{{Grudi{\'c}} et~al.}{2021}]{2021MNRAS.506.2199G}
{Grudi{\'c}} M.~Y.,  {Guszejnov} D.,  {Hopkins} P.~F.,  {Offner} S. S.~R.,
  {Faucher-Gigu{\`e}re} C.-A.,  2021, \mn@doi [\mnras]
  {10.1093/mnras/stab1347}, \href
  {https://ui.adsabs.harvard.edu/abs/2021MNRAS.506.2199G} {506, 2199}

\bibitem[\protect\citeauthoryear{{Gutcke}, {Pakmor}, {Naab}  \&
  {Springel}}{{Gutcke} et~al.}{2021}]{2021MNRAS.501.5597G}
{Gutcke} T.~A.,  {Pakmor} R.,  {Naab} T.,   {Springel} V.,  2021, \mn@doi
  [\mnras] {10.1093/mnras/staa3875}, \href
  {https://ui.adsabs.harvard.edu/abs/2021MNRAS.501.5597G} {501, 5597}

\bibitem[\protect\citeauthoryear{{Gvaramadze} \& {Bomans}}{{Gvaramadze} \&
  {Bomans}}{2008}]{2008A&A...490.1071G}
{Gvaramadze} V.~V.,  {Bomans} D.~J.,  2008, \mn@doi [\aap]
  {10.1051/0004-6361:200810411}, \href
  {https://ui.adsabs.harvard.edu/abs/2008A&A...490.1071G} {490, 1071}

\bibitem[\protect\citeauthoryear{Harris et~al.,}{Harris
  et~al.}{2020}]{2020NumPy-Array}
Harris C.~R.,  et~al., 2020, \mn@doi [Nature] {10.1038/s41586-020-2649-2}, 585,
  357–362

\bibitem[\protect\citeauthoryear{{Heger} \& {Woosley}}{{Heger} \&
  {Woosley}}{2002}]{2002ApJ...567..532H}
{Heger} A.,  {Woosley} S.~E.,  2002, \mn@doi [\apj] {10.1086/338487}, \href
  {https://ui.adsabs.harvard.edu/abs/2002ApJ...567..532H} {567, 532}

\bibitem[\protect\citeauthoryear{{H{\'e}nault-Brunet}, {Gieles}, {Agertz}  \&
  {Read}}{{H{\'e}nault-Brunet} et~al.}{2015}]{2015MNRAS.450.1164H}
{H{\'e}nault-Brunet} V.,  {Gieles} M.,  {Agertz} O.,   {Read} J.~I.,  2015,
  \mn@doi [\mnras] {10.1093/mnras/stv675}, \href
  {https://ui.adsabs.harvard.edu/abs/2015MNRAS.450.1164H} {450, 1164}

\bibitem[\protect\citeauthoryear{{Hibbard}, {van der Hulst}, {Barnes}  \&
  {Rich}}{{Hibbard} et~al.}{2001}]{2001AJ....122.2969H}
{Hibbard} J.~E.,  {van der Hulst} J.~M.,  {Barnes} J.~E.,   {Rich} R.~M.,
  2001, \mn@doi [\aj] {10.1086/324102}, \href
  {https://ui.adsabs.harvard.edu/abs/2001AJ....122.2969H} {122, 2969}

\bibitem[\protect\citeauthoryear{{Hirai}, {Fujii}  \& {Saitoh}}{{Hirai}
  et~al.}{2021}]{2021PASJ...73.1036H}
{Hirai} Y.,  {Fujii} M.~S.,   {Saitoh} T.~R.,  2021, \mn@doi [\pasj]
  {10.1093/pasj/psab038}, \href
  {https://ui.adsabs.harvard.edu/abs/2021PASJ...73.1036H} {73, 1036}

\bibitem[\protect\citeauthoryear{{Hislop}, {Naab}, {Steinwandel}, {Lah{\'e}n},
  {Irodotou}, {Johansson}  \& {Walch}}{{Hislop}
  et~al.}{2022}]{2022MNRAS.509.5938H}
{Hislop} J.~M.,  {Naab} T.,  {Steinwandel} U.~P.,  {Lah{\'e}n} N.,  {Irodotou}
  D.,  {Johansson} P.~H.,   {Walch} S.,  2022, \mn@doi [\mnras]
  {10.1093/mnras/stab3347}, \href
  {https://ui.adsabs.harvard.edu/abs/2022MNRAS.509.5938H} {509, 5938}

\bibitem[\protect\citeauthoryear{{Hollyhead} et~al.,}{{Hollyhead}
  et~al.}{2018}]{2018MNRAS.476..114H}
{Hollyhead} K.,  et~al., 2018, \mn@doi [\mnras] {10.1093/mnras/sty230}, \href
  {https://ui.adsabs.harvard.edu/abs/2018MNRAS.476..114H} {476, 114}

\bibitem[\protect\citeauthoryear{{Hollyhead} et~al.,}{{Hollyhead}
  et~al.}{2019}]{2019MNRAS.484.4718H}
{Hollyhead} K.,  et~al., 2019, \mn@doi [\mnras] {10.1093/mnras/stz317}, \href
  {https://ui.adsabs.harvard.edu/abs/2019MNRAS.484.4718H} {484, 4718}

\bibitem[\protect\citeauthoryear{{Hong} et~al.,}{{Hong}
  et~al.}{2017}]{2017MNRAS.472...67H}
{Hong} J.,  et~al., 2017, \mn@doi [\mnras] {10.1093/mnras/stx1954}, \href
  {https://ui.adsabs.harvard.edu/abs/2017MNRAS.472...67H} {472, 67}

\bibitem[\protect\citeauthoryear{{Hoogerwerf}, {de Bruijne}  \& {de
  Zeeuw}}{{Hoogerwerf} et~al.}{2000}]{2000ApJ...544L.133H}
{Hoogerwerf} R.,  {de Bruijne} J.~H.~J.,   {de Zeeuw} P.~T.,  2000, \mn@doi
  [\apjl] {10.1086/317315}, \href
  {https://ui.adsabs.harvard.edu/abs/2000ApJ...544L.133H} {544, L133}

\bibitem[\protect\citeauthoryear{{Howard}, {Pudritz}, {Sills}  \&
  {Harris}}{{Howard} et~al.}{2019}]{2019MNRAS.486.1146H}
{Howard} C.~S.,  {Pudritz} R.~E.,  {Sills} A.,   {Harris} W.~E.,  2019, \mn@doi
  [\mnras] {10.1093/mnras/stz924}, \href
  {https://ui.adsabs.harvard.edu/abs/2019MNRAS.486.1146H} {486, 1146}

\bibitem[\protect\citeauthoryear{{Hu}}{{Hu}}{2019}]{2019MNRAS.483.3363H}
{Hu} C.-Y.,  2019, \mn@doi [\mnras] {10.1093/mnras/sty3252}, \href
  {https://ui.adsabs.harvard.edu/abs/2019MNRAS.483.3363H} {483, 3363}

\bibitem[\protect\citeauthoryear{{Hu}, {Naab}, {Walch}, {Moster}  \&
  {Oser}}{{Hu} et~al.}{2014}]{2014MNRAS.443.1173H}
{Hu} C.-Y.,  {Naab} T.,  {Walch} S.,  {Moster} B.~P.,   {Oser} L.,  2014,
  \mn@doi [\mnras] {10.1093/mnras/stu1187}, \href
  {https://ui.adsabs.harvard.edu/abs/2014MNRAS.443.1173H} {443, 1173}

\bibitem[\protect\citeauthoryear{{Hu}, {Naab}, {Walch}, {Glover}  \&
  {Clark}}{{Hu} et~al.}{2016}]{2016MNRAS.458.3528H}
{Hu} C.-Y.,  {Naab} T.,  {Walch} S.,  {Glover} S. C.~O.,   {Clark} P.~C.,
  2016, \mn@doi [\mnras] {10.1093/mnras/stw544}, \href
  {https://ui.adsabs.harvard.edu/abs/2016MNRAS.458.3528H} {458, 3528}

\bibitem[\protect\citeauthoryear{{Hu}, {Naab}, {Glover}, {Walch}  \&
  {Clark}}{{Hu} et~al.}{2017}]{2017MNRAS.471.2151H}
{Hu} C.-Y.,  {Naab} T.,  {Glover} S. C.~O.,  {Walch} S.,   {Clark} P.~C.,
  2017, \mn@doi [\mnras] {10.1093/mnras/stx1773}, \href
  {https://ui.adsabs.harvard.edu/abs/2017MNRAS.471.2151H} {471, 2151}

\bibitem[\protect\citeauthoryear{Hunter}{Hunter}{2007}]{Hunter:2007}
Hunter J.~D.,  2007, \mn@doi [Computing in Science \& Engineering]
  {10.1109/MCSE.2007.55}, 9, 90

\bibitem[\protect\citeauthoryear{{Hypki}, {Giersz}, {Hong}, {Leveque}, {Askar},
  {Belloni}  \& {Otulakowska-Hypka}}{{Hypki}
  et~al.}{2022}]{2022MNRAS.517.4768H}
{Hypki} A.,  {Giersz} M.,  {Hong} J.,  {Leveque} A.,  {Askar} A.,  {Belloni}
  D.,   {Otulakowska-Hypka} M.,  2022, \mn@doi [\mnras]
  {10.1093/mnras/stac2815}, \href
  {https://ui.adsabs.harvard.edu/abs/2022MNRAS.517.4768H} {517, 4768}

\bibitem[\protect\citeauthoryear{{Johnson}, {Kotz}  \&
  {Balakrishnan}}{{Johnson} et~al.}{1994}]{univariate_distributions}
{Johnson} N.~L.,  {Kotz} S.,   {Balakrishnan} N.,  1994, Continuous Univariate
  Distributions, Volume 1, 2 edn.
Wiley Series in Probability and Mathematical Statistics: Applied Probability
  and Statistics, John Wiley \& Sons, Inc., New York

\bibitem[\protect\citeauthoryear{{Johnson}, {Rich}, {Pilachowski}, {Caldwell},
  {Mateo}, {Bailey}  \& {Crane}}{{Johnson} et~al.}{2015}]{2015AJ....150...63J}
{Johnson} C.~I.,  {Rich} R.~M.,  {Pilachowski} C.~A.,  {Caldwell} N.,  {Mateo}
  M.,  {Bailey} John~I. I.,   {Crane} J.~D.,  2015, \mn@doi [\aj]
  {10.1088/0004-6256/150/2/63}, \href
  {https://ui.adsabs.harvard.edu/abs/2015AJ....150...63J} {150, 63}

\bibitem[\protect\citeauthoryear{{Johnson}, {Caldwell}, {Michael Rich}, {Mateo}
   \& {Bailey}}{{Johnson} et~al.}{2019}]{2019MNRAS.485.4311J}
{Johnson} C.~I.,  {Caldwell} N.,  {Michael Rich} R.,  {Mateo} M.,   {Bailey}
  J.~I.,  2019, \mn@doi [\mnras] {10.1093/mnras/stz587}, \href
  {https://ui.adsabs.harvard.edu/abs/2019MNRAS.485.4311J} {485, 4311}

\bibitem[\protect\citeauthoryear{{Karakas}}{{Karakas}}{2010}]{2010MNRAS.403.1413K}
{Karakas} A.~I.,  2010, \mn@doi [\mnras] {10.1111/j.1365-2966.2009.16198.x},
  \href {https://ui.adsabs.harvard.edu/abs/2010MNRAS.403.1413K} {403, 1413}

\bibitem[\protect\citeauthoryear{{Karl}, {Naab}, {Johansson}, {Kotarba},
  {Boily}, {Renaud}  \& {Theis}}{{Karl} et~al.}{2010}]{2010ApJ...715L..88K}
{Karl} S.~J.,  {Naab} T.,  {Johansson} P.~H.,  {Kotarba} H.,  {Boily} C.~M.,
  {Renaud} F.,   {Theis} C.,  2010, \mn@doi [\apjl]
  {10.1088/2041-8205/715/2/L88}, \href
  {https://ui.adsabs.harvard.edu/abs/2010ApJ...715L..88K} {715, L88}

\bibitem[\protect\citeauthoryear{{Kennicutt}}{{Kennicutt}}{1998}]{1998ApJ...498..541K}
{Kennicutt} Jr. R.~C.,  1998, \mn@doi [\apj] {10.1086/305588}, \href
  {http://adsabs.harvard.edu/abs/1998ApJ...498..541K} {498, 541}

\bibitem[\protect\citeauthoryear{{Kennicutt} \& {Evans}}{{Kennicutt} \&
  {Evans}}{2012}]{2012ARA&A..50..531K}
{Kennicutt} R.~C.,  {Evans} N.~J.,  2012, \mn@doi [\araa]
  {10.1146/annurev-astro-081811-125610}, \href
  {https://ui.adsabs.harvard.edu/abs/2012ARA&A..50..531K} {50, 531}

\bibitem[\protect\citeauthoryear{{Khalaj} \& {Baumgardt}}{{Khalaj} \&
  {Baumgardt}}{2015}]{2015MNRAS.452..924K}
{Khalaj} P.,  {Baumgardt} H.,  2015, \mn@doi [\mnras] {10.1093/mnras/stv1356},
  \href {https://ui.adsabs.harvard.edu/abs/2015MNRAS.452..924K} {452, 924}

\bibitem[\protect\citeauthoryear{{Kimbro}, {Reines}, {Molina}, {Deller}  \&
  {Stern}}{{Kimbro} et~al.}{2021}]{2021ApJ...912...89K}
{Kimbro} E.,  {Reines} A.~E.,  {Molina} M.,  {Deller} A.~T.,   {Stern} D.,
  2021, \mn@doi [\apj] {10.3847/1538-4357/abec6a}, \href
  {https://ui.adsabs.harvard.edu/abs/2021ApJ...912...89K} {912, 89}

\bibitem[\protect\citeauthoryear{{Kochanek}}{{Kochanek}}{2016}]{2016MNRAS.458..127K}
{Kochanek} C.~S.,  2016, \mn@doi [\mnras] {10.1093/mnras/stw267}, \href
  {https://ui.adsabs.harvard.edu/abs/2016MNRAS.458..127K} {458, 127}

\bibitem[\protect\citeauthoryear{{Kozyreva}, {Yoon}  \& {Langer}}{{Kozyreva}
  et~al.}{2014}]{2014A&A...566A.146K}
{Kozyreva} A.,  {Yoon} S.~C.,   {Langer} N.,  2014, \mn@doi [\aap]
  {10.1051/0004-6361/201423641}, \href
  {https://ui.adsabs.harvard.edu/abs/2014A&A...566A.146K} {566, A146}

\bibitem[\protect\citeauthoryear{{Krause}, {Charbonnel}, {Bastian}  \&
  {Diehl}}{{Krause} et~al.}{2016}]{2016A&A...587A..53K}
{Krause} M. G.~H.,  {Charbonnel} C.,  {Bastian} N.,   {Diehl} R.,  2016,
  \mn@doi [\aap] {10.1051/0004-6361/201526685}, \href
  {https://ui.adsabs.harvard.edu/abs/2016A&A...587A..53K} {587, A53}

\bibitem[\protect\citeauthoryear{{Kravtsov}, {Alca{\'\i}no}, {Marconi}  \&
  {Alvarado}}{{Kravtsov} et~al.}{2011}]{2011A&A...527L...9K}
{Kravtsov} V.,  {Alca{\'\i}no} G.,  {Marconi} G.,   {Alvarado} F.,  2011,
  \mn@doi [\aap] {10.1051/0004-6361/201015975}, \href
  {https://ui.adsabs.harvard.edu/abs/2011A&A...527L...9K} {527, L9}

\bibitem[\protect\citeauthoryear{{Kroupa}}{{Kroupa}}{2001}]{2001MNRAS.322..231K}
{Kroupa} P.,  2001, \mn@doi [\mnras] {10.1046/j.1365-8711.2001.04022.x}, \href
  {https://ui.adsabs.harvard.edu/abs/2001MNRAS.322..231K} {322, 231}

\bibitem[\protect\citeauthoryear{{Lacchin}, {Mastrobuono-Battisti}, {Calura},
  {Nipoti}, {Milone}, {Meneghetti}  \& {Vanzella}}{{Lacchin}
  et~al.}{2024}]{2024A&A...681A..45L}
{Lacchin} E.,  {Mastrobuono-Battisti} A.,  {Calura} F.,  {Nipoti} C.,  {Milone}
  A.~P.,  {Meneghetti} M.,   {Vanzella} E.,  2024, \mn@doi [\aap]
  {10.1051/0004-6361/202347268}, \href
  {https://ui.adsabs.harvard.edu/abs/2024A&A...681A..45L} {681, A45}

\bibitem[\protect\citeauthoryear{{Lah{\'e}n}, {Naab}, {Johansson}, {Elmegreen},
  {Hu}  \& {Walch}}{{Lah{\'e}n} et~al.}{2019}]{2019ApJ...879L..18L}
{Lah{\'e}n} N.,  {Naab} T.,  {Johansson} P.~H.,  {Elmegreen} B.,  {Hu} C.-Y.,
  {Walch} S.,  2019, \mn@doi [\apjl] {10.3847/2041-8213/ab2a13}, \href
  {https://ui.adsabs.harvard.edu/abs/2019ApJ...879L..18L} {879, L18}

\bibitem[\protect\citeauthoryear{{Lah{\'e}n}, {Naab}, {Johansson}, {Elmegreen},
  {Hu}, {Walch}, {Steinwandel}  \& {Moster}}{{Lah{\'e}n}
  et~al.}{2020a}]{2020ApJ...891....2L}
{Lah{\'e}n} N.,  {Naab} T.,  {Johansson} P.~H.,  {Elmegreen} B.,  {Hu} C.-Y.,
  {Walch} S.,  {Steinwandel} U.~P.,   {Moster} B.~P.,  2020a, \mn@doi [\apj]
  {10.3847/1538-4357/ab7190}, \href
  {https://ui.adsabs.harvard.edu/abs/2020ApJ...891....2L} {891, 2}

\bibitem[\protect\citeauthoryear{{Lah{\'e}n}, {Naab}, {Johansson}, {Elmegreen},
  {Hu}  \& {Walch}}{{Lah{\'e}n} et~al.}{2020b}]{2020ApJ...904...71L}
{Lah{\'e}n} N.,  {Naab} T.,  {Johansson} P.~H.,  {Elmegreen} B.,  {Hu} C.-Y.,
  {Walch} S.,  2020b, \mn@doi [\apj] {10.3847/1538-4357/abc001}, \href
  {https://ui.adsabs.harvard.edu/abs/2020ApJ...904...71L} {904, 71}

\bibitem[\protect\citeauthoryear{{Lah{\'e}n}, {Naab}  \&
  {Kauffmann}}{{Lah{\'e}n} et~al.}{2022}]{2022MNRAS.514.4560L}
{Lah{\'e}n} N.,  {Naab} T.,   {Kauffmann} G.,  2022, \mn@doi [\mnras]
  {10.1093/mnras/stac1594}, \href
  {https://ui.adsabs.harvard.edu/abs/2022MNRAS.514.4560L} {514, 4560}

\bibitem[\protect\citeauthoryear{{Lah{\'e}n} et~al.,}{{Lah{\'e}n}
  et~al.}{2023}]{2023MNRAS.522.3092L}
{Lah{\'e}n} N.,  et~al., 2023, \mn@doi [\mnras] {10.1093/mnras/stad1147}, \href
  {https://ui.adsabs.harvard.edu/abs/2023MNRAS.522.3092L} {522, 3092}

\bibitem[\protect\citeauthoryear{{Lamers}, {Baumgardt}  \& {Gieles}}{{Lamers}
  et~al.}{2010}]{2010MNRAS.409..305L}
{Lamers} H. J.~G.~L.~M.,  {Baumgardt} H.,   {Gieles} M.,  2010, \mn@doi
  [\mnras] {10.1111/j.1365-2966.2010.17309.x}, \href
  {https://ui.adsabs.harvard.edu/abs/2010MNRAS.409..305L} {409, 305}

\bibitem[\protect\citeauthoryear{{Lancaster}, {Ostriker}, {Kim}  \&
  {Kim}}{{Lancaster} et~al.}{2021a}]{2021ApJ...914...89L}
{Lancaster} L.,  {Ostriker} E.~C.,  {Kim} J.-G.,   {Kim} C.-G.,  2021a, \mn@doi
  [\apj] {10.3847/1538-4357/abf8ab}, \href
  {https://ui.adsabs.harvard.edu/abs/2021ApJ...914...89L} {914, 89}

\bibitem[\protect\citeauthoryear{{Lancaster}, {Ostriker}, {Kim}  \&
  {Kim}}{{Lancaster} et~al.}{2021b}]{2021ApJ...914...90L}
{Lancaster} L.,  {Ostriker} E.~C.,  {Kim} J.-G.,   {Kim} C.-G.,  2021b, \mn@doi
  [\apj] {10.3847/1538-4357/abf8ac}, \href
  {https://ui.adsabs.harvard.edu/abs/2021ApJ...914...90L} {914, 90}

\bibitem[\protect\citeauthoryear{{Lancaster}, {Ostriker}, {Kim}  \&
  {Kim}}{{Lancaster} et~al.}{2021c}]{2021ApJ...922L...3L}
{Lancaster} L.,  {Ostriker} E.~C.,  {Kim} J.-G.,   {Kim} C.-G.,  2021c, \mn@doi
  [\apjl] {10.3847/2041-8213/ac3333}, \href
  {https://ui.adsabs.harvard.edu/abs/2021ApJ...922L...3L} {922, L3}

\bibitem[\protect\citeauthoryear{{Langer}, {Hoffman}  \& {Sneden}}{{Langer}
  et~al.}{1993}]{1993PASP..105..301L}
{Langer} G.~E.,  {Hoffman} R.,   {Sneden} C.,  1993, \mn@doi [\pasp]
  {10.1086/133147}, \href
  {https://ui.adsabs.harvard.edu/abs/1993PASP..105..301L} {105, 301}

\bibitem[\protect\citeauthoryear{{Lardo}, {Bellazzini}, {Pancino}, {Carretta},
  {Bragaglia}  \& {Dalessandro}}{{Lardo} et~al.}{2011}]{2011A&A...525A.114L}
{Lardo} C.,  {Bellazzini} M.,  {Pancino} E.,  {Carretta} E.,  {Bragaglia} A.,
  {Dalessandro} E.,  2011, \mn@doi [\aap] {10.1051/0004-6361/201015662}, \href
  {https://ui.adsabs.harvard.edu/abs/2011A&A...525A.114L} {525, A114}

\bibitem[\protect\citeauthoryear{{Lardo}, {Cabrera-Ziri}, {Davies}  \&
  {Bastian}}{{Lardo} et~al.}{2017}]{2017MNRAS.468.2482L}
{Lardo} C.,  {Cabrera-Ziri} I.,  {Davies} B.,   {Bastian} N.,  2017, \mn@doi
  [\mnras] {10.1093/mnras/stx628}, \href
  {https://ui.adsabs.harvard.edu/abs/2017MNRAS.468.2482L} {468, 2482}

\bibitem[\protect\citeauthoryear{{Lardo}, {Salaris}, {Cassisi}, {Bastian},
  {Mucciarelli}, {Cabrera-Ziri}  \& {Dalessandro}}{{Lardo}
  et~al.}{2023}]{2023A&A...669A..19L}
{Lardo} C.,  {Salaris} M.,  {Cassisi} S.,  {Bastian} N.,  {Mucciarelli} A.,
  {Cabrera-Ziri} I.,   {Dalessandro} E.,  2023, \mn@doi [\aap]
  {10.1051/0004-6361/202245090}, \href
  {https://ui.adsabs.harvard.edu/abs/2023A&A...669A..19L} {669, A19}

\bibitem[\protect\citeauthoryear{{Larsen}, {Baumgardt}, {Bastian}, {Brodie},
  {Grundahl}  \& {Strader}}{{Larsen} et~al.}{2015}]{2015ApJ...804...71L}
{Larsen} S.~S.,  {Baumgardt} H.,  {Bastian} N.,  {Brodie} J.~P.,  {Grundahl}
  F.,   {Strader} J.,  2015, \mn@doi [\apj] {10.1088/0004-637X/804/1/71}, \href
  {https://ui.adsabs.harvard.edu/abs/2015ApJ...804...71L} {804, 71}

\bibitem[\protect\citeauthoryear{{Larsen}, {Brodie}, {Wasserman}  \&
  {Strader}}{{Larsen} et~al.}{2018}]{2018A&A...613A..56L}
{Larsen} S.~S.,  {Brodie} J.~P.,  {Wasserman} A.,   {Strader} J.,  2018,
  \mn@doi [\aap] {10.1051/0004-6361/201731909}, \href
  {https://ui.adsabs.harvard.edu/abs/2018A&A...613A..56L} {613, A56}

\bibitem[\protect\citeauthoryear{{Leitinger}, {Baumgardt}, {Cabrera-Ziri},
  {Hilker}  \& {Pancino}}{{Leitinger} et~al.}{2023}]{2023MNRAS.520.1456L}
{Leitinger} E.,  {Baumgardt} H.,  {Cabrera-Ziri} I.,  {Hilker} M.,   {Pancino}
  E.,  2023, \mn@doi [\mnras] {10.1093/mnras/stad093}, \href
  {https://ui.adsabs.harvard.edu/abs/2023MNRAS.520.1456L} {520, 1456}

\bibitem[\protect\citeauthoryear{{Lejeune}, {Cuisinier}  \& {Buser}}{{Lejeune}
  et~al.}{1997}]{1997A&AS..125..229L}
{Lejeune} T.,  {Cuisinier} F.,   {Buser} R.,  1997, \mn@doi [\aaps]
  {10.1051/aas:1997373}, \href
  {http://adsabs.harvard.edu/abs/1997A%26AS..125..229L} {125, 229}

\bibitem[\protect\citeauthoryear{{Lejeune}, {Cuisinier}  \& {Buser}}{{Lejeune}
  et~al.}{1998}]{1998A&AS..130...65L}
{Lejeune} T.,  {Cuisinier} F.,   {Buser} R.,  1998, \mn@doi [\aaps]
  {10.1051/aas:1998405}, \href
  {http://adsabs.harvard.edu/abs/1998A%26AS..130...65L} {130, 65}

\bibitem[\protect\citeauthoryear{{Leroy}, {Walter}, {Brinks}, {Bigiel}, {de
  Blok}, {Madore}  \& {Thornley}}{{Leroy} et~al.}{2008}]{2008AJ....136.2782L}
{Leroy} A.~K.,  {Walter} F.,  {Brinks} E.,  {Bigiel} F.,  {de Blok} W.~J.~G.,
  {Madore} B.,   {Thornley} M.~D.,  2008, \mn@doi [\aj]
  {10.1088/0004-6256/136/6/2782}, \href
  {http://adsabs.harvard.edu/abs/2008AJ....136.2782L} {136, 2782}

\bibitem[\protect\citeauthoryear{{Li}, {Vogelsberger}, {Bryan}, {Marinacci},
  {Sales}  \& {Torrey}}{{Li} et~al.}{2022}]{2022MNRAS.514..265L}
{Li} H.,  {Vogelsberger} M.,  {Bryan} G.~L.,  {Marinacci} F.,  {Sales} L.~V.,
  {Torrey} P.,  2022, \mn@doi [\mnras] {10.1093/mnras/stac1136}, \href
  {https://ui.adsabs.harvard.edu/abs/2022MNRAS.514..265L} {514, 265}

\bibitem[\protect\citeauthoryear{{Lochhaas} \& {Thompson}}{{Lochhaas} \&
  {Thompson}}{2017}]{2017MNRAS.470..977L}
{Lochhaas} C.,  {Thompson} T.~A.,  2017, \mn@doi [\mnras]
  {10.1093/mnras/stx1289}, \href
  {https://ui.adsabs.harvard.edu/abs/2017MNRAS.470..977L} {470, 977}

\bibitem[\protect\citeauthoryear{{Mackey} \& {Gilmore}}{{Mackey} \&
  {Gilmore}}{2003}]{2003MNRAS.338...85M}
{Mackey} A.~D.,  {Gilmore} G.~F.,  2003, \mn@doi [\mnras]
  {10.1046/j.1365-8711.2003.06021.x}, \href
  {https://ui.adsabs.harvard.edu/abs/2003MNRAS.338...85M} {338, 85}

\bibitem[\protect\citeauthoryear{{Maeder} \& {Meynet}}{{Maeder} \&
  {Meynet}}{2006}]{2006A&A...448L..37M}
{Maeder} A.,  {Meynet} G.,  2006, \mn@doi [\aap] {10.1051/0004-6361:200600012},
  \href {https://ui.adsabs.harvard.edu/abs/2006A&A...448L..37M} {448, L37}

\bibitem[\protect\citeauthoryear{{Marino} et~al.,}{{Marino}
  et~al.}{2015}]{2015MNRAS.450..815M}
{Marino} A.~F.,  et~al., 2015, \mn@doi [\mnras] {10.1093/mnras/stv420}, \href
  {https://ui.adsabs.harvard.edu/abs/2015MNRAS.450..815M} {450, 815}

\bibitem[\protect\citeauthoryear{{Marino} et~al.,}{{Marino}
  et~al.}{2019}]{2019MNRAS.487.3815M}
{Marino} A.~F.,  et~al., 2019, \mn@doi [\mnras] {10.1093/mnras/stz1415}, \href
  {https://ui.adsabs.harvard.edu/abs/2019MNRAS.487.3815M} {487, 3815}

\bibitem[\protect\citeauthoryear{{Marino} et~al.,}{{Marino}
  et~al.}{2023}]{2023ApJ...958...31M}
{Marino} A.~F.,  et~al., 2023, \mn@doi [\apj] {10.3847/1538-4357/acfca3}, \href
  {https://ui.adsabs.harvard.edu/abs/2023ApJ...958...31M} {958, 31}

\bibitem[\protect\citeauthoryear{{Marques-Chaves} et~al.,}{{Marques-Chaves}
  et~al.}{2024}]{2024A&A...681A..30M}
{Marques-Chaves} R.,  et~al., 2024, \mn@doi [\aap]
  {10.1051/0004-6361/202347411}, \href
  {https://ui.adsabs.harvard.edu/abs/2024A&A...681A..30M} {681, A30}

\bibitem[\protect\citeauthoryear{{Martocchia} et~al.,}{{Martocchia}
  et~al.}{2017}]{2017MNRAS.468.3150M}
{Martocchia} S.,  et~al., 2017, \mn@doi [\mnras] {10.1093/mnras/stx660}, \href
  {https://ui.adsabs.harvard.edu/abs/2017MNRAS.468.3150M} {468, 3150}

\bibitem[\protect\citeauthoryear{{Martocchia} et~al.,}{{Martocchia}
  et~al.}{2018a}]{2018MNRAS.473.2688M}
{Martocchia} S.,  et~al., 2018a, \mn@doi [\mnras] {10.1093/mnras/stx2556},
  \href {https://ui.adsabs.harvard.edu/abs/2018MNRAS.473.2688M} {473, 2688}

\bibitem[\protect\citeauthoryear{{Martocchia} et~al.,}{{Martocchia}
  et~al.}{2018b}]{2018MNRAS.477.4696M}
{Martocchia} S.,  et~al., 2018b, \mn@doi [\mnras] {10.1093/mnras/sty916}, \href
  {https://ui.adsabs.harvard.edu/abs/2018MNRAS.477.4696M} {477, 4696}

\bibitem[\protect\citeauthoryear{{McKenzie} \& {Bekki}}{{McKenzie} \&
  {Bekki}}{2021}]{2021MNRAS.500.4578M}
{McKenzie} M.,  {Bekki} K.,  2021, \mn@doi [\mnras] {10.1093/mnras/staa3376},
  \href {https://ui.adsabs.harvard.edu/abs/2021MNRAS.500.4578M} {500, 4578}

\bibitem[\protect\citeauthoryear{{Mi{\'c}i{\'c}}, {Holmes}, {Wells}  \&
  {Irwin}}{{Mi{\'c}i{\'c}} et~al.}{2023}]{2023ApJ...944..160M}
{Mi{\'c}i{\'c}} M.,  {Holmes} O.~J.,  {Wells} B.~N.,   {Irwin} J.~A.,  2023,
  \mn@doi [\apj] {10.3847/1538-4357/aca1bb}, \href
  {https://ui.adsabs.harvard.edu/abs/2023ApJ...944..160M} {944, 160}

\bibitem[\protect\citeauthoryear{{Milone} et~al.,}{{Milone}
  et~al.}{2017a}]{2017MNRAS.464.3636M}
{Milone} A.~P.,  et~al., 2017a, \mn@doi [\mnras] {10.1093/mnras/stw2531}, \href
  {https://ui.adsabs.harvard.edu/abs/2017MNRAS.464.3636M} {464, 3636}

\bibitem[\protect\citeauthoryear{{Milone} et~al.,}{{Milone}
  et~al.}{2017b}]{2017MNRAS.469..800M}
{Milone} A.~P.,  et~al., 2017b, \mn@doi [\mnras] {10.1093/mnras/stx836}, \href
  {https://ui.adsabs.harvard.edu/abs/2017MNRAS.469..800M} {469, 800}

\bibitem[\protect\citeauthoryear{{Milone} et~al.,}{{Milone}
  et~al.}{2020}]{2020MNRAS.491..515M}
{Milone} A.~P.,  et~al., 2020, \mn@doi [\mnras] {10.1093/mnras/stz2999}, \href
  {https://ui.adsabs.harvard.edu/abs/2020MNRAS.491..515M} {491, 515}

\bibitem[\protect\citeauthoryear{{Moreno-Hilario}, {Martinez-Medina}, {Li},
  {Souza}  \& {P{\'e}rez-Villegas}}{{Moreno-Hilario}
  et~al.}{2024}]{2024MNRAS.527.2765M}
{Moreno-Hilario} E.,  {Martinez-Medina} L.~A.,  {Li} H.,  {Souza} S.~O.,
  {P{\'e}rez-Villegas} A.,  2024, \mn@doi [\mnras] {10.1093/mnras/stad3306},
  \href {https://ui.adsabs.harvard.edu/abs/2024MNRAS.527.2765M} {527, 2765}

\bibitem[\protect\citeauthoryear{{Mucciarelli}, {Origlia}, {Ferraro}  \&
  {Pancino}}{{Mucciarelli} et~al.}{2009}]{2009ApJ...695L.134M}
{Mucciarelli} A.,  {Origlia} L.,  {Ferraro} F.~R.,   {Pancino} E.,  2009,
  \mn@doi [\apjl] {10.1088/0004-637X/695/2/L134}, \href
  {https://ui.adsabs.harvard.edu/abs/2009ApJ...695L.134M} {695, L134}

\bibitem[\protect\citeauthoryear{{Mucciarelli}, {Dalessandro}, {Ferraro},
  {Origlia}  \& {Lanzoni}}{{Mucciarelli} et~al.}{2014}]{2014ApJ...793L...6M}
{Mucciarelli} A.,  {Dalessandro} E.,  {Ferraro} F.~R.,  {Origlia} L.,
  {Lanzoni} B.,  2014, \mn@doi [\apjl] {10.1088/2041-8205/793/1/L6}, \href
  {https://ui.adsabs.harvard.edu/abs/2014ApJ...793L...6M} {793, L6}

\bibitem[\protect\citeauthoryear{{Mucciarelli}, {Bellazzini}, {Merle}, {Plez},
  {Dalessandro}  \& {Ibata}}{{Mucciarelli} et~al.}{2015}]{2015ApJ...801...68M}
{Mucciarelli} A.,  {Bellazzini} M.,  {Merle} T.,  {Plez} B.,  {Dalessandro} E.,
    {Ibata} R.,  2015, \mn@doi [\apj] {10.1088/0004-637X/801/1/68}, \href
  {https://ui.adsabs.harvard.edu/abs/2015ApJ...801...68M} {801, 68}

\bibitem[\protect\citeauthoryear{{Mullan} et~al.,}{{Mullan}
  et~al.}{2011}]{2011ApJ...731...93M}
{Mullan} B.,  et~al., 2011, \mn@doi [\apj] {10.1088/0004-637X/731/2/93}, \href
  {https://ui.adsabs.harvard.edu/abs/2011ApJ...731...93M} {731, 93}

\bibitem[\protect\citeauthoryear{{Naab} \& {Ostriker}}{{Naab} \&
  {Ostriker}}{2017}]{2017ARA&A..55...59N}
{Naab} T.,  {Ostriker} J.~P.,  2017, \mn@doi [\araa]
  {10.1146/annurev-astro-081913-040019}, \href
  {https://ui.adsabs.harvard.edu/abs/2017ARA&A..55...59N} {55, 59}

\bibitem[\protect\citeauthoryear{{Nataf} et~al.,}{{Nataf}
  et~al.}{2019}]{2019AJ....158...14N}
{Nataf} D.~M.,  et~al., 2019, \mn@doi [\aj] {10.3847/1538-3881/ab1a27}, \href
  {https://ui.adsabs.harvard.edu/abs/2019AJ....158...14N} {158, 14}

\bibitem[\protect\citeauthoryear{{Niederhofer} et~al.,}{{Niederhofer}
  et~al.}{2017a}]{2017MNRAS.464...94N}
{Niederhofer} F.,  et~al., 2017a, \mn@doi [\mnras] {10.1093/mnras/stw2269},
  \href {https://ui.adsabs.harvard.edu/abs/2017MNRAS.464...94N} {464, 94}

\bibitem[\protect\citeauthoryear{{Niederhofer} et~al.,}{{Niederhofer}
  et~al.}{2017b}]{2017MNRAS.465.4159N}
{Niederhofer} F.,  et~al., 2017b, \mn@doi [\mnras] {10.1093/mnras/stw3084},
  \href {https://ui.adsabs.harvard.edu/abs/2017MNRAS.465.4159N} {465, 4159}

\bibitem[\protect\citeauthoryear{{Norris}}{{Norris}}{2004}]{2004ApJ...612L..25N}
{Norris} J.~E.,  2004, \mn@doi [\apjl] {10.1086/423986}, \href
  {https://ui.adsabs.harvard.edu/abs/2004ApJ...612L..25N} {612, L25}

\bibitem[\protect\citeauthoryear{{Pancino} et~al.,}{{Pancino}
  et~al.}{2017}]{2017A&A...601A.112P}
{Pancino} E.,  et~al., 2017, \mn@doi [\aap] {10.1051/0004-6361/201730474},
  \href {https://ui.adsabs.harvard.edu/abs/2017A&A...601A.112P} {601, A112}

\bibitem[\protect\citeauthoryear{{Pascale}, {Dai}, {McKee}  \&
  {Tsang}}{{Pascale} et~al.}{2023}]{2023ApJ...957...77P}
{Pascale} M.,  {Dai} L.,  {McKee} C.~F.,   {Tsang} B. T.~H.,  2023, \mn@doi
  [\apj] {10.3847/1538-4357/acf75c}, \href
  {https://ui.adsabs.harvard.edu/abs/2023ApJ...957...77P} {957, 77}

\bibitem[\protect\citeauthoryear{{Piotto} et~al.,}{{Piotto}
  et~al.}{2015}]{2015AJ....149...91P}
{Piotto} G.,  et~al., 2015, \mn@doi [\aj] {10.1088/0004-6256/149/3/91}, \href
  {https://ui.adsabs.harvard.edu/abs/2015AJ....149...91P} {149, 91}

\bibitem[\protect\citeauthoryear{{Poveda}, {Ruiz}  \& {Allen}}{{Poveda}
  et~al.}{1967}]{1967BOTT....4...86P}
{Poveda} A.,  {Ruiz} J.,   {Allen} C.,  1967, Boletin de los Observatorios
  Tonantzintla y Tacubaya, \href
  {https://ui.adsabs.harvard.edu/abs/1967BOTT....4...86P} {4, 86}

\bibitem[\protect\citeauthoryear{{Prantzos} \& {Charbonnel}}{{Prantzos} \&
  {Charbonnel}}{2006}]{2006A&A...458..135P}
{Prantzos} N.,  {Charbonnel} C.,  2006, \mn@doi [\aap]
  {10.1051/0004-6361:20065374}, \href
  {https://ui.adsabs.harvard.edu/abs/2006A&A...458..135P} {458, 135}

\bibitem[\protect\citeauthoryear{{Prantzos}, {Charbonnel}  \&
  {Iliadis}}{{Prantzos} et~al.}{2007}]{2007A&A...470..179P}
{Prantzos} N.,  {Charbonnel} C.,   {Iliadis} C.,  2007, \mn@doi [\aap]
  {10.1051/0004-6361:20077205}, \href
  {https://ui.adsabs.harvard.edu/abs/2007A&A...470..179P} {470, 179}

\bibitem[\protect\citeauthoryear{{Prantzos}, {Charbonnel}  \&
  {Iliadis}}{{Prantzos} et~al.}{2017}]{2017A&A...608A..28P}
{Prantzos} N.,  {Charbonnel} C.,   {Iliadis} C.,  2017, \mn@doi [\aap]
  {10.1051/0004-6361/201731528}, \href
  {https://ui.adsabs.harvard.edu/abs/2017A&A...608A..28P} {608, A28}

\bibitem[\protect\citeauthoryear{{Renzini} et~al.,}{{Renzini}
  et~al.}{2015}]{2015MNRAS.454.4197R}
{Renzini} A.,  et~al., 2015, \mn@doi [\mnras] {10.1093/mnras/stv2268}, \href
  {https://ui.adsabs.harvard.edu/abs/2015MNRAS.454.4197R} {454, 4197}

\bibitem[\protect\citeauthoryear{{Rieder}, {Dobbs}, {Bending}, {Liow}  \&
  {Wurster}}{{Rieder} et~al.}{2022}]{2022MNRAS.509.6155R}
{Rieder} S.,  {Dobbs} C.,  {Bending} T.,  {Liow} K.~Y.,   {Wurster} J.,  2022,
  \mn@doi [\mnras] {10.1093/mnras/stab3425}, \href
  {https://ui.adsabs.harvard.edu/abs/2022MNRAS.509.6155R} {509, 6155}

\bibitem[\protect\citeauthoryear{{R{\"o}ttgers}, {Naab}, {Cernetic},
  {Dav{\'e}}, {Kauffmann}, {Borthakur}  \& {Foidl}}{{R{\"o}ttgers}
  et~al.}{2020}]{2020MNRAS.496..152R}
{R{\"o}ttgers} B.,  {Naab} T.,  {Cernetic} M.,  {Dav{\'e}} R.,  {Kauffmann} G.,
   {Borthakur} S.,   {Foidl} H.,  2020, \mn@doi [\mnras]
  {10.1093/mnras/staa1490}, \href
  {https://ui.adsabs.harvard.edu/abs/2020MNRAS.496..152R} {496, 152}

\bibitem[\protect\citeauthoryear{{Roychowdhury}, {Huang}, {Kauffmann}, {Wang}
  \& {Chengalur}}{{Roychowdhury} et~al.}{2015}]{2015MNRAS.449.3700R}
{Roychowdhury} S.,  {Huang} M.-L.,  {Kauffmann} G.,  {Wang} J.,   {Chengalur}
  J.~N.,  2015, \mn@doi [\mnras] {10.1093/mnras/stv515}, \href
  {https://ui.adsabs.harvard.edu/abs/2015MNRAS.449.3700R} {449, 3700}

\bibitem[\protect\citeauthoryear{{Salpeter}}{{Salpeter}}{1955}]{1955ApJ...121..161S}
{Salpeter} E.~E.,  1955, \mn@doi [\apj] {10.1086/145971}, \href
  {https://ui.adsabs.harvard.edu/abs/1955ApJ...121..161S} {121, 161}

\bibitem[\protect\citeauthoryear{{Sana} et~al.,}{{Sana}
  et~al.}{2012}]{2012Sci...337..444S}
{Sana} H.,  et~al., 2012, \mn@doi [Science] {10.1126/science.1223344}, \href
  {https://ui.adsabs.harvard.edu/abs/2012Sci...337..444S} {337, 444}

\bibitem[\protect\citeauthoryear{{Schaerer} \& {Charbonnel}}{{Schaerer} \&
  {Charbonnel}}{2011}]{2011MNRAS.413.2297S}
{Schaerer} D.,  {Charbonnel} C.,  2011, \mn@doi [\mnras]
  {10.1111/j.1365-2966.2011.18304.x}, \href
  {https://ui.adsabs.harvard.edu/abs/2011MNRAS.413.2297S} {413, 2297}

\bibitem[\protect\citeauthoryear{{Senchyna}, {Plat}, {Stark}  \&
  {Rudie}}{{Senchyna} et~al.}{2023}]{2023arXiv230304179S}
{Senchyna} P.,  {Plat} A.,  {Stark} D.~P.,   {Rudie} G.~C.,  2023, \mn@doi
  [arXiv e-prints] {10.48550/arXiv.2303.04179}, \href
  {https://ui.adsabs.harvard.edu/abs/2023arXiv230304179S} {p. arXiv:2303.04179}

\bibitem[\protect\citeauthoryear{{Sills} \& {Glebbeek}}{{Sills} \&
  {Glebbeek}}{2010}]{2010MNRAS.407..277S}
{Sills} A.,  {Glebbeek} E.,  2010, \mn@doi [\mnras]
  {10.1111/j.1365-2966.2010.16876.x}, \href
  {https://ui.adsabs.harvard.edu/abs/2010MNRAS.407..277S} {407, 277}

\bibitem[\protect\citeauthoryear{{Simioni}, {Milone}, {Bedin}, {Aparicio},
  {Piotto}, {Vesperini}  \& {Hong}}{{Simioni}
  et~al.}{2016}]{2016MNRAS.463..449S}
{Simioni} M.,  {Milone} A.~P.,  {Bedin} L.~R.,  {Aparicio} A.,  {Piotto} G.,
  {Vesperini} E.,   {Hong} J.,  2016, \mn@doi [\mnras] {10.1093/mnras/stw2003},
  \href {https://ui.adsabs.harvard.edu/abs/2016MNRAS.463..449S} {463, 449}

\bibitem[\protect\citeauthoryear{{Springel}}{{Springel}}{2005}]{2005MNRAS.364.1105S}
{Springel} V.,  2005, \mn@doi [\mnras] {10.1111/j.1365-2966.2005.09655.x},
  \href {https://ui.adsabs.harvard.edu/abs/2005MNRAS.364.1105S} {364, 1105}

\bibitem[\protect\citeauthoryear{{Springel}, {White}, {Tormen}  \&
  {Kauffmann}}{{Springel} et~al.}{2001}]{2001MNRAS.328..726S}
{Springel} V.,  {White} S. D.~M.,  {Tormen} G.,   {Kauffmann} G.,  2001,
  \mn@doi [\mnras] {10.1046/j.1365-8711.2001.04912.x}, \href
  {https://ui.adsabs.harvard.edu/abs/2001MNRAS.328..726S} {328, 726}

\bibitem[\protect\citeauthoryear{{Steinwandel} \& {Goldberg}}{{Steinwandel} \&
  {Goldberg}}{2023}]{2023arXiv231011495S}
{Steinwandel} U.~P.,  {Goldberg} J.~A.,  2023, \mn@doi [arXiv e-prints]
  {10.48550/arXiv.2310.11495}, \href
  {https://ui.adsabs.harvard.edu/abs/2023arXiv231011495S} {p. arXiv:2310.11495}

\bibitem[\protect\citeauthoryear{{Steinwandel}, {Bryan}, {Somerville},
  {Hayward}  \& {Burkhart}}{{Steinwandel} et~al.}{2023}]{2023MNRAS.526.1408S}
{Steinwandel} U.~P.,  {Bryan} G.~L.,  {Somerville} R.~S.,  {Hayward} C.~C.,
  {Burkhart} B.,  2023, \mn@doi [\mnras] {10.1093/mnras/stad2744}, \href
  {https://ui.adsabs.harvard.edu/abs/2023MNRAS.526.1408S} {526, 1408}

\bibitem[\protect\citeauthoryear{{Stoop}, {Kaper}, {de Koter}, {Guo}, {Lamers}
  \& {Rieder}}{{Stoop} et~al.}{2023}]{2023A&A...670A.108S}
{Stoop} M.,  {Kaper} L.,  {de Koter} A.,  {Guo} D.,  {Lamers} H.~J.~G.~L.~M.,
  {Rieder} S.,  2023, \mn@doi [\aap] {10.1051/0004-6361/202244511}, \href
  {https://ui.adsabs.harvard.edu/abs/2023A&A...670A.108S} {670, A108}

\bibitem[\protect\citeauthoryear{{Szakacs}, {P{\'e}roux}, {Zwaan}, {Nelson},
  {Schinnerer}, {Lah{\'e}n}, {Weng}  \& {Fresco}}{{Szakacs}
  et~al.}{2022}]{2022MNRAS.512.4736S}
{Szakacs} R.,  {P{\'e}roux} C.,  {Zwaan} M.~A.,  {Nelson} D.,  {Schinnerer} E.,
   {Lah{\'e}n} N.,  {Weng} S.,   {Fresco} A.~Y.,  2022, \mn@doi [\mnras]
  {10.1093/mnras/stac510}, \href
  {https://ui.adsabs.harvard.edu/abs/2022MNRAS.512.4736S} {512, 4736}

\bibitem[\protect\citeauthoryear{{Sz{\'e}csi} \& {W{\"u}nsch}}{{Sz{\'e}csi} \&
  {W{\"u}nsch}}{2019}]{2019ApJ...871...20S}
{Sz{\'e}csi} D.,  {W{\"u}nsch} R.,  2019, \mn@doi [\apj]
  {10.3847/1538-4357/aaf4be}, \href
  {https://ui.adsabs.harvard.edu/abs/2019ApJ...871...20S} {871, 20}

\bibitem[\protect\citeauthoryear{{Sz{\'e}csi}, {Mackey}  \&
  {Langer}}{{Sz{\'e}csi} et~al.}{2018}]{2018A&A...612A..55S}
{Sz{\'e}csi} D.,  {Mackey} J.,   {Langer} N.,  2018, \mn@doi [\aap]
  {10.1051/0004-6361/201731500}, \href
  {https://ui.adsabs.harvard.edu/abs/2018A&A...612A..55S} {612, A55}

\bibitem[\protect\citeauthoryear{{Sz{\'e}csi}, {Agrawal}, {W{\"u}nsch}  \&
  {Langer}}{{Sz{\'e}csi} et~al.}{2022}]{2022A&A...658A.125S}
{Sz{\'e}csi} D.,  {Agrawal} P.,  {W{\"u}nsch} R.,   {Langer} N.,  2022, \mn@doi
  [\aap] {10.1051/0004-6361/202141536}, \href
  {https://ui.adsabs.harvard.edu/abs/2022A&A...658A.125S} {658, A125}

\bibitem[\protect\citeauthoryear{{Tacconi} et~al.,}{{Tacconi}
  et~al.}{2013}]{2013ApJ...768...74T}
{Tacconi} L.~J.,  et~al., 2013, \mn@doi [\apj] {10.1088/0004-637X/768/1/74},
  \href {https://ui.adsabs.harvard.edu/abs/2013ApJ...768...74T} {768, 74}

\bibitem[\protect\citeauthoryear{{Teyssier}, {Chapon}  \&
  {Bournaud}}{{Teyssier} et~al.}{2010}]{2010ApJ...720L.149T}
{Teyssier} R.,  {Chapon} D.,   {Bournaud} F.,  2010, \mn@doi [\apjl]
  {10.1088/2041-8205/720/2/L149}, \href
  {https://ui.adsabs.harvard.edu/abs/2010ApJ...720L.149T} {720, L149}

\bibitem[\protect\citeauthoryear{{Tiongco}, {Vesperini}  \& {Varri}}{{Tiongco}
  et~al.}{2019}]{2019MNRAS.487.5535T}
{Tiongco} M.~A.,  {Vesperini} E.,   {Varri} A.~L.,  2019, \mn@doi [\mnras]
  {10.1093/mnras/stz1595}, \href
  {https://ui.adsabs.harvard.edu/abs/2019MNRAS.487.5535T} {487, 5535}

\bibitem[\protect\citeauthoryear{{Vanzella} et~al.,}{{Vanzella}
  et~al.}{2023}]{2023ApJ...945...53V}
{Vanzella} E.,  et~al., 2023, \mn@doi [\apj] {10.3847/1538-4357/acb59a}, \href
  {https://ui.adsabs.harvard.edu/abs/2023ApJ...945...53V} {945, 53}

\bibitem[\protect\citeauthoryear{{Ventura}, {D'Antona}, {Mazzitelli}  \&
  {Gratton}}{{Ventura} et~al.}{2001}]{2001ApJ...550L..65V}
{Ventura} P.,  {D'Antona} F.,  {Mazzitelli} I.,   {Gratton} R.,  2001, \mn@doi
  [\apjl] {10.1086/319496}, \href
  {https://ui.adsabs.harvard.edu/abs/2001ApJ...550L..65V} {550, L65}

\bibitem[\protect\citeauthoryear{{Vesperini}, {McMillan}, {D'Antona}  \&
  {D'Ercole}}{{Vesperini} et~al.}{2013}]{2013MNRAS.429.1913V}
{Vesperini} E.,  {McMillan} S. L.~W.,  {D'Antona} F.,   {D'Ercole} A.,  2013,
  \mn@doi [\mnras] {10.1093/mnras/sts434}, \href
  {https://ui.adsabs.harvard.edu/abs/2013MNRAS.429.1913V} {429, 1913}

\bibitem[\protect\citeauthoryear{{Vesperini}, {Hong}, {Giersz}  \&
  {Hypki}}{{Vesperini} et~al.}{2021}]{2021MNRAS.502.4290V}
{Vesperini} E.,  {Hong} J.,  {Giersz} M.,   {Hypki} A.,  2021, \mn@doi [\mnras]
  {10.1093/mnras/stab223}, \href
  {https://ui.adsabs.harvard.edu/abs/2021MNRAS.502.4290V} {502, 4290}

\bibitem[\protect\citeauthoryear{{Vink}}{{Vink}}{2018}]{2018A&A...615A.119V}
{Vink} J.~S.,  2018, \mn@doi [\aap] {10.1051/0004-6361/201832773}, \href
  {https://ui.adsabs.harvard.edu/abs/2018A&A...615A.119V} {615, A119}

\bibitem[\protect\citeauthoryear{Virtanen et~al.,}{Virtanen
  et~al.}{2020}]{2020SciPy-NMeth}
Virtanen P.,  et~al., 2020, \mn@doi [Nature Methods]
  {10.1038/s41592-019-0686-2}, \href {https://rdcu.be/b08Wh} {17, 261}

\bibitem[\protect\citeauthoryear{{Wang} et~al.,}{{Wang}
  et~al.}{2016}]{2016MNRAS.458.1450W}
{Wang} L.,  et~al., 2016, \mn@doi [\mnras] {10.1093/mnras/stw274}, \href
  {https://ui.adsabs.harvard.edu/abs/2016MNRAS.458.1450W} {458, 1450}

\bibitem[\protect\citeauthoryear{{Wang}, {Kroupa}, {Takahashi}  \&
  {Jerabkova}}{{Wang} et~al.}{2020}]{2020MNRAS.491..440W}
{Wang} L.,  {Kroupa} P.,  {Takahashi} K.,   {Jerabkova} T.,  2020, \mn@doi
  [\mnras] {10.1093/mnras/stz3033}, \href
  {https://ui.adsabs.harvard.edu/abs/2020MNRAS.491..440W} {491, 440}

\bibitem[\protect\citeauthoryear{{Webb} \& {Leigh}}{{Webb} \&
  {Leigh}}{2015}]{2015MNRAS.453.3278W}
{Webb} J.~J.,  {Leigh} N. W.~C.,  2015, \mn@doi [\mnras]
  {10.1093/mnras/stv1780}, \href
  {https://ui.adsabs.harvard.edu/abs/2015MNRAS.453.3278W} {453, 3278}

\bibitem[\protect\citeauthoryear{{Westera}, {Lejeune}, {Buser}, {Cuisinier}  \&
  {Bruzual}}{{Westera} et~al.}{2002}]{2002A&A...381..524W}
{Westera} P.,  {Lejeune} T.,  {Buser} R.,  {Cuisinier} F.,   {Bruzual} G.,
  2002, \mn@doi [\aap] {10.1051/0004-6361:20011493}, \href
  {https://ui.adsabs.harvard.edu/abs/2002A&A...381..524W} {381, 524}

\bibitem[\protect\citeauthoryear{{Whitmore}, {Zhang}, {Leitherer}, {Fall},
  {Schweizer}  \& {Miller}}{{Whitmore} et~al.}{1999}]{1999AJ....118.1551W}
{Whitmore} B.~C.,  {Zhang} Q.,  {Leitherer} C.,  {Fall} S.~M.,  {Schweizer} F.,
    {Miller} B.~W.,  1999, \mn@doi [\aj] {10.1086/301041}, \href
  {https://ui.adsabs.harvard.edu/abs/1999AJ....118.1551W} {118, 1551}

\bibitem[\protect\citeauthoryear{{Wiersma}, {Schaye}  \& {Smith}}{{Wiersma}
  et~al.}{2009}]{2009MNRAS.393...99W}
{Wiersma} R. P.~C.,  {Schaye} J.,   {Smith} B.~D.,  2009, \mn@doi [\mnras]
  {10.1111/j.1365-2966.2008.14191.x}, \href
  {https://ui.adsabs.harvard.edu/abs/2009MNRAS.393...99W} {393, 99}

\bibitem[\protect\citeauthoryear{{W{\"u}nsch}, {Palou{\v{s}}}, {Tenorio-Tagle}
  \& {Ehlerov{\'a}}}{{W{\"u}nsch} et~al.}{2017}]{2017ApJ...835...60W}
{W{\"u}nsch} R.,  {Palou{\v{s}}} J.,  {Tenorio-Tagle} G.,   {Ehlerov{\'a}} S.,
  2017, \mn@doi [\apj] {10.3847/1538-4357/835/1/60}, \href
  {https://ui.adsabs.harvard.edu/abs/2017ApJ...835...60W} {835, 60}

\bibitem[\protect\citeauthoryear{{Yong}, {Grundahl}, {Lambert}, {Nissen}  \&
  {Shetrone}}{{Yong} et~al.}{2003}]{2003A&A...402..985Y}
{Yong} D.,  {Grundahl} F.,  {Lambert} D.~L.,  {Nissen} P.~E.,   {Shetrone}
  M.~D.,  2003, \mn@doi [\aap] {10.1051/0004-6361:20030296}, \href
  {https://ui.adsabs.harvard.edu/abs/2003A&A...402..985Y} {402, 985}

\bibitem[\protect\citeauthoryear{{Zennaro}, {Milone}, {Marino}, {Cordoni},
  {Lagioia}  \& {Tailo}}{{Zennaro} et~al.}{2019}]{2019MNRAS.487.3239Z}
{Zennaro} M.,  {Milone} A.~P.,  {Marino} A.~F.,  {Cordoni} G.,  {Lagioia}
  E.~P.,   {Tailo} M.,  2019, \mn@doi [\mnras] {10.1093/mnras/stz1477}, \href
  {https://ui.adsabs.harvard.edu/abs/2019MNRAS.487.3239Z} {487, 3239}

\bibitem[\protect\citeauthoryear{{Zhang}, {de Grijs}, {Li}  \& {Wu}}{{Zhang}
  et~al.}{2018}]{2018ApJ...853..186Z}
{Zhang} H.,  {de Grijs} R.,  {Li} C.,   {Wu} X.,  2018, \mn@doi [\apj]
  {10.3847/1538-4357/aaa428}, \href
  {https://ui.adsabs.harvard.edu/abs/2018ApJ...853..186Z} {853, 186}

\bibitem[\protect\citeauthoryear{{de Mink}, {Pols}, {Langer}  \& {Izzard}}{{de
  Mink} et~al.}{2009}]{2009A&A...507L...1D}
{de Mink} S.~E.,  {Pols} O.~R.,  {Langer} N.,   {Izzard} R.~G.,  2009, \mn@doi
  [\aap] {10.1051/0004-6361/200913205}, \href
  {https://ui.adsabs.harvard.edu/abs/2009A&A...507L...1D} {507, L1}

\bibitem[\protect\citeauthoryear{{de Wit}, {Testi}, {Palla}  \&
  {Zinnecker}}{{de Wit} et~al.}{2005}]{2005A&A...437..247D}
{de Wit} W.~J.,  {Testi} L.,  {Palla} F.,   {Zinnecker} H.,  2005, \mn@doi
  [\aap] {10.1051/0004-6361:20042489}, \href
  {https://ui.adsabs.harvard.edu/abs/2005A&A...437..247D} {437, 247}

\makeatother
\end{thebibliography}




\appendix


\bsp	
\label{lastpage}
\end{document}